\newcommand{\printstyle}{reprint}

\newcommand{\thirdheight}{4.14cm}
\newcommand{\figsep}{\qquad}

\documentclass[aps,pra,superscriptaddress,floatfix,\printstyle ,nofootinbib]{revtex4-2}

\usepackage{color}

\usepackage[caption=false]{subfig}
\usepackage{graphicx,booktabs,array}
\usepackage{multirow}
\usepackage{tabularx} 
\DeclareGraphicsExtensions{.pdf,.png,.jpg,.eps,.ps}
\usepackage{epsfig}
\usepackage{epstopdf}
\usepackage{amssymb}
\usepackage{amsmath}
\usepackage{amsfonts}
\usepackage{mathrsfs}
\usepackage{amsthm}
\usepackage{float}
\usepackage{amsbsy}
\usepackage{bm}
\usepackage{grffile}
\usepackage{hyperref}
\hypersetup{
    colorlinks = true,
    citecolor = blue,
    urlcolor = blue,
    linkcolor = blue,
}
\usepackage{courier}
\usepackage{upgreek}
\usepackage{xspace}
\usepackage{xcolor}
\usepackage{calc} 
\usepackage{soul} 

\newcommand{\eg}{\textit{e}.\textit{g}. }
\newcommand{\etc}{\textit{etc.}\xspace} 

\newcommand{\Alfven}{Alfv\'{e}n\xspace}
\newcommand{\Alfvenic}{Alfv\'{e}nic\xspace}

\newcommand{\va}{v_A}
\newcommand{\vs}{v_S}

\newcommand{\fci}{f_{ci}}

\newcommand{\db}{\delta B}

\newcommand{\dbpar}{\db_\parallel}

\newcommand{\dbperp}{\db_\perp}

\newcommand{\betae}{\beta_e}

\newcommand{\W}{\mathcal{E}}

\newcommand{\Te}{T_e}
\newcommand{\Ti}{T_i}

\newcommand{\betap}{\beta_p}
\newcommand{\exb}{E \cross B}
\newcommand{\nustar}{\nu_*}
\newcommand{\ptsolver}{PT\_SOLVER\xspace}

\newcommand{\xtrue}{X_T}
\newcommand{\xpred}{X_P}
\newcommand{\xtruej}{X_{Tj}}
\newcommand{\xpredj}{X_{Pj}}
\newcommand{\epsj}{\epsilon_j}
\newcommand{\betan}{\beta_N}

\newcommand{\Teo}{T_{e0}}
\newcommand{\Tio}{T_{i0}}
\newcommand{\teti}{\Te/\Ti}
\newcommand{\ktheta}{k_\theta}

\newcommand{\qkiz}{QuaLiKiz\xspace}

\newcommand{\abs}[1]{\left|#1\right|}
\let\overdot\dot
\renewcommand{\dot}{\cdot}
\newcommand{\cross}{\times}
\renewcommand{\div}{\nabla\dot}
\newcommand{\grad}{\nabla}

\newcommand{\approptoinn}[2]{\mathrel{\vcenter{
  \offinterlineskip\halign{\hfil$##$\cr
    #1\propto\cr\noalign{\kern2pt}#1\sim\cr\noalign{\kern-2pt}}}}}

\newcommand{\ddfrac}[2]{\frac{\displaystyle #1}{\displaystyle #2}}

\newcommand{\pderiv}[2]{\frac{\partial #1}{\partial #2}}

\newcommand{\figref}[1]{Fig.\xspace\ref{#1}}
\renewcommand{\eqref}[1]{Eq.\xspace\ref{#1}}
\newcommand{\secref}[1]{Sec.\xspace\ref{#1}}
\newcommand{\citeref}[1]{Ref.\xspace\onlinecite{#1}}

\newcommand{\tabref}[1]{Table\xspace\ref{#1}}

\newcommand{\myname}{J.B. Lestz}

\newcommand{\Stan}{S.M. Kaye}
\newcommand{\Kathreen}{K.E. Thome}
\newcommand{\Galina}{G. Avdeeva}
\newcommand{\Joey}{J. McClenaghan}
\newcommand{\Federico}{F.D. Halpern}
\newcommand{\Alexei}{A.Y. Pankin}
\newcommand{\Marina}{M.V. Gorelenkova}
\newcommand{\Tom}{T.F. Neiser}

\newcommand{\PPPL}{Princeton Plasma Physics Lab, Princeton, NJ 08543, USA}

\newcommand{\GA}{General Atomics, San Diego, CA, 92121, USA}

\newcommand{\Podesta}{Podest{\`a}}

\definecolor{darkgreen}{rgb}{0,0.5,0}

\newcommand{\gadisclaimer}{This report was prepared as an account of work sponsored by an agency of the United States Government. Neither the United States Government nor any agency thereof, nor any of their employees, makes any warranty, express or implied, or assumes any legal liability or responsibility for the accuracy, completeness, or usefulness of any information, apparatus, product, or process disclosed, or represents that its use would not infringe privately owned rights. Reference herein to any specific commercial product, process, or service by trade name, trademark, manufacturer, or otherwise does not necessarily constitute or imply its endorsement, recommendation, or favoring by the United States Government or any agency thereof. The views and opinions of authors expressed herein do not necessarily state or reflect those of the United States Government or any agency thereof.}
\renewcommand{\vs}{{\xspace}versus\xspace}

\newcommand{\ktext}[1]{\textcolor{black}{#1}}
\newcommand{\rev}[1]{\ktext{#1}}

\begin{document}

\title{Assessing time-dependent temperature profile predictions using reduced transport models for high performing NSTX plasmas}

\author{\myname}
\email{lestzj@fusion.gat.com}
\affiliation{\GA}
\author{\Galina}
\affiliation{\GA}
\author{\Tom}
\affiliation{\GA}
\author{\Marina}
\affiliation{\PPPL}
\author{\Federico}
\affiliation{\GA}
\author{\Stan}
\affiliation{\PPPL}
\author{\Joey}
\affiliation{\GA}
\author{\Alexei}
\affiliation{\PPPL}
\author{\Kathreen}
\affiliation{\GA}
\date{\today}
\begin{abstract}

Time-dependent, predictive simulations were performed with the 1.5D tokamak integrated modeling code TRANSP on a large set of well-analyzed, high performing discharges from the National Spherical Torus Experiment (NSTX) in order to evaluate how well modern reduced transport models can reproduce experimentally observed temperature profiles in spherical tokamaks. Overall, it is found that simulations using the Multi-Mode Model (MMM) more consistently agree with the NSTX observations than those using the Trapped Gyro-Landau Fluid (TGLF) model, despite TGLF requiring orders of magnitude greater computational cost. When considering all examined discharges, MMM has median overpredictions of electron temperature ($\Te$) and ion temperature ($\Ti$) profiles of 28\% and 27\%, respectively, relative to the experiment. TGLF overpredicts $\Te$ by 46\%, with much larger variance than MMM, and underpredicts $\Ti$ by 25\%. As the ratio of kinetic to magnetic field pressure ($\beta$) is increased across NSTX discharges, TGLF predicts lower $\Te$ and significant flattening of the $\Ti$ profile, conflicting with NSTX observations. When using an electrostatic version of TGLF, both $\Te$ and $\Ti$ are substantially overpredicted, underscoring the importance of electromagnetic turbulence in the high $\beta$ spherical tokamak regime. Additionally, calculations with neural net surrogate models for TGLF were performed outside of TRANSP with a time slice flux matching transport solver, finding better agreement with experiment than the TRANSP simulations, highlighting the impact of different transport solvers and simulation techniques. Altogether, the reasonable agreement with experiment of temperature profiles predicted by MMM motivates a more detailed examination of the sensitivities of the TRANSP simulations with MMM to different NSTX plasma regimes in a companion paper \cite{Lestz2025pre2}, in preparation for self-consistent, time-dependent predictive modeling of NSTX-U scenarios. 

\end{abstract}
\maketitle

\section{Introduction}
\label{sec:intro}

Low aspect ratio tokamaks, such as the National Spherical Torus Experiment (NSTX) \cite{Ono2000NF,Sabbagh2013NF} and its successor NSTX-U \cite{Menard2012NF,Menard2017NF,Berkery2024NF}, operate in a qualitatively different transport regime from conventional large aspect ratio tokamaks \cite{Kaye2021PPCF}. In particular, spherical tokamaks can achieve toroidal $\beta = 2\mu_0 P/B_T^2 \approx 10 - 40\%$ ($P$ is the kinetic plasma pressure and $B_T$ is the toroidal magnetic field strength), much higher than the $\beta \approx 3 - 10\%$ for conventional tokamaks \cite{Ono2015POP}. The high $\beta$ conditions common in spherical tokamaks increases the importance of electromagnetic fluctuations for microinstabilities that drive turbulent transport. Favorable energy confinement has been experimentally observed in low aspect ratio tokamaks including NSTX \cite{Kaye2007NF,Kaye2013NF}, MAST \cite{Valovic2009NF,Valovic2011NF}, and Globus-M(2) \cite{Kurskiev2019NF,Kurskiev2022NF} at low normalized collisionality $\nustar = \nu_e/\omega_b \propto q R n_e/\Te^2\epsilon^{3/2}$, where $\nu_e$ is the electron collision frequency, $\omega_b$ is the bounce frequency, $q$ is the safety factor, $R$ is the major radius, $n_e$ is the electron density, $T_e$ is the electron temperature, and $\epsilon = a/R$ is the inverse aspect ratio between the major and minor radius ($a$). This property makes the spherical tokamak regime an attractive design point if the scaling extrapolates to even lower collisionality, as will be explored on NSTX-U. Improved confinement occurs in low aspect ratio tokamaks due to a relatively larger region of ``good curvature'' \cite{Rewoldt1996POP,Kinsey2007POP} and enhanced $\exb$ flow shear suppression of turbulent transport from ion temperature gradient modes (ITGs) and trapped electron modes (TEMs) \cite{Roach2009PPCF}, among other factors \cite{Kaye2021PPCF}. Previous studies with gyrokinetic simulations and reduced models have found that $\beta$ and $\nustar$ strongly influence the character of turbulent transport in spherical tokamaks such as NSTX(-U), with kinetic ballooning modes (KBMs) and microtearing modes (MTMs) playing a much larger role than in conventional tokamaks \cite{Guttenfelder2012POP,Guttenfelder2013NF,Kaye2014POP,Clauser2022POP,Patel2022NF,McClenaghan2023POP,Kennedy2023NF,Kennedy2024NF,
Giacomin2024PPCF,Dominski2024POP,McClenaghan2025PPCF,Singh2025NF}. 
\rev{Additionally, low aspect ratio devices have a higher trapped particle fraction than conventional tokamaks, which can impact the transport induced by many instabilities \cite{Kaye2021PPCF}.} 

NSTX had a major radius of 0.85 m and aspect ratio $\epsilon^{-1} \approx 1.4$, operating with on-axis toroidal magnetic fields of up to 0.5 T and plasma currents of up to 1.3 MA. 6 MW of on-axis neutral beam power and up to 6 MW of high harmonic fast wave (HHFW) power were available for auxiliary heating, though all of the discharges analyzed in this work were purely beam-heated. NSTX plasmas typically achieved densities of $10^{19} - 10^{20}\text{ m}^{-3}$ and $\Te \approx 1 - 2$ keV. Major components of the upgrade to NSTX-U in 2015 included significantly increasing the available neutral beam power (installing 6 MW of off-axis beams), the maximum magnetic field (up to 1 T), the maximum plasma current (up to 2 MA), and the anticipated maximum pulse length for advanced scenarios (up to 5 s, from 1 s on NSTX). Together, these and other upgrades are expected to enable significant progress towards demonstrating the viability of the spherical tokamak reactor concept \cite{Menard2011NF,Menard2016NF,Menard2019RSA,Menard2022NF}. Modern efforts include the STAR design from the Princeton Plasma Physics Lab (PPPL) \cite{Menard2023IAEA}, the STEP program being actively pursued by the United Kingdom government \cite{Wilson2020book,Waldon2024PTRSA,Meyer2024PTRA}, and a sequence of machines being constructed by the private company Tokamak Energy \cite{Buxton2019PPCF,Kingham2024POP}.

Reduced transport models are a key ingredient of integrated modeling for scenario development and reactor design studies. By sacrificing physics fidelity relative to first principles simulations, reduced models boast impressive computational efficiency, making possible large parameter scans and full pulse simulations that would otherwise be computationally infeasible. Despite being governed by different physics, the development of transport models for spherical tokamaks has historically not received as much attention as models which focus on large aspect ratio tokamaks \cite{Staebler2024NF}. Hence, validation studies and related investigations are needed in order to evaluate the reliability of such transport models in regimes of interest for spherical tokamaks. 

This work focuses on assessing the agreement between time-dependent predictive simulations of the electron and ion temperature profiles and experimental measurements for a large set of well-analyzed and high performing NSTX discharges. Specifically, the integrated modeling code TRANSP \cite{Hawryluk1980transp,Goldston1981JCP,transp2018,Grierson2018FST,Pankin2025CPC} is used to compare two mature reduced turbulent transport models, the Trapped Gyro-Landau Fluid (TGLF) model and the Multi-Mode Model (MMM). Both TGLF and MMM incorporate electromagnetic effects and $\exb$ shear. TRANSP will be described in greater detail in \secref{sec:methods}. 

TGLF is a trapped gyro-Landau fluid drift wave model for calculating linear eigenmodes and the quasilinear transport that they induce \cite{Staebler2007POP,Kinsey2008POP}. The linear TGLF system includes the following microinstabilities: electron temperature gradient (ETG), ion temperature gradient, trapped electron, trapped ion (TIM), and kinetic ballooning modes. 
\rev{Notably, microtearing modes are not yet accurately captured since TGLF represents eigenmodes with Hermite polynomials of similar widths in ballooning space, which poorly resolves MTMs that are known to have broader structure in their electrostatic fluctuations $\delta\phi$ than in the parallel component of the vector potential $\delta A_\parallel$.} 
For a given eigenmode spectrum, quasilinear fluxes are calculated via physics-based saturation rules (labeled as SAT0, SAT1, \etc) with parameters that were fitted to a database of nonlinear gyrokinetic simulations \cite{Staebler2007POP,Kinsey2008POP,Staebler2013NF,Staebler2016POP,Staebler2021PPCF,Staebler2021NF,Dudding2022NF}. Pedagogical descriptions of the progression of TGLF saturation rules can be found in a comprehensive review article \cite{Staebler2024NF} and recent PhD theses \cite{Dudding2022thesis,Pratt2024thesis}. 
\rev{The gyrofluid model is sytematically derived to include geometric effects, parallel dynamics, and a sophisticated computation and calibration of the quasilinear fluxes. Its ITG, TEM, ETG, and KBM linear physics are reasonably accurate when compared to gyrokinetic simulations with the CGYRO code \cite{Candy2016JCP,Hall2022APS}. Its main potential limitations for modeling spherical tokamaks are its limited treatment of MTMs and that the quasilinear fluxes have been calibrated to nonlinear gyrokinetic simulations in the large aspect ratio regime, which may not cover the unique transport physics of high $\beta$ spherical tokamaks. Additionally, there are indications that that the electron thermal transport channel can be extremely sensitive to nonlinear $\exb$ shear effects in spherical tokamaks \cite{Patel2025NF,Avdeeva2025pre}, requiring careful calibration against gyrokinetic simulations and posing a signficant modeling challenge for any quasilinear model.}

Two versions of TGLF are chosen to use for time-dependent TRANSP simulations on the full database of NSTX discharges -- fully electromagnetic (EM) SAT0 and electrostatic (ES) SAT1. Here, electrostatic \vs electromagnetic refers to the fields included in the eigenvalue calculation, with the electromagnetic settings including magnetic field fluctuations both perpendicular ($\dbperp$) and parallel ($\dbpar$) to the background field in addition to the electrostatic potential $\delta\phi$. SAT0 and SAT1 refer to the TGLF saturation rule used for calculating fluxes. Electromagnetic TGLF with SAT0 was chosen to evaluate based on the database study in \citeref{Neiser2024APS}, which found that it was the best performing TGLF model when tested against a large database of MAST-U time slices. Electrostatic TGLF with SAT1 was chosen based on its recent application to modeling two relatively low $\beta$ NSTX discharges, which were known to be dominated by electrostatic turbulence \cite{Avdeeva2023NF}. In that work, a careful time slice flux matching analysis found reasonable agreement with the observed temperature profiles. Comparison of these two different TGLF models is useful for isolating the influence of electromagnetic effects (theoretically stronger for larger $\beta$) on the model's ability to reproduce experimental observations. 

Physically, SAT0 calculates saturation amplitudes by balancing the linear growth rate with the $\exb$ zonal flow shearing rate, treating each poloidal wavenumber $\ktheta$ independently \cite{Staebler2007POP,Kinsey2008POP}. The SAT1 model goes beyond the SAT0 physics by introducing zonal flow mixing to include multiscale mixing with respect to $\ktheta$ \cite{Staebler2016POP}. Additionally, SAT1 employs a spectral shift model that modifies the 2D eigenmode spectrum with respect to the poloidal and radial wavenumbers $(\ktheta,k_r)$ \cite{Staebler2013PRL,Staebler2013NF}. Due to the computational expense of TGLF within TRANSP, the exploration of other TGLF physics models in this work is restricted to a smaller subset of discharges in the time-dependent TRANSP simulations and complemented by time slice flux matching analysis on all plasmas. Historically, the development and application of TGLF has focused much more on conventional tokamaks than those with low aspect ratio such as NSTX(-U). However, a detailed study of ETG instabilities in NSTX with gyrokinetic simulations, TGLF, and MMM has recently been performed in a few plasmas \cite{Clauser2025POP}. TGLF has also recently been used within time-dependent simulations of hot ion plasmas in the high field spherical tokamak ST40 \cite{AnastopoulosTzanis2025} \rev{and to evaluate the effect of different gas pre-fill levels in VEST \cite{Lee2022JKPS}}. While extensive validation studies have been completed in both time slice analysis \cite{Neiser2020APS} and time-dependent simulations \cite{Abbate2024POP} of the DIII-D conventional aspect ratio tokamak, TGLF has not previously been validated against a large database of NSTX discharges with either time slice analysis or time-dependent integrated modeling, to our knowledge.  

MMM is a multi-mode model which combines four different submodels, each for a different class of instabilities \cite{Rafiq2013POP,Luo2013CPC}. These are the Weiland model (TEM, ITG, KBM, and also high-$n$ MHD modes) \cite{Weiland2012text}, an ETG model \cite{Rafiq2022POP}, an MTM model \cite{Rafiq2016POP}, and a model for drift resistive inertial ballooning modes (DRIBM) \cite{Rafiq2010POP}. All of these submodels are enabled in the NSTX simulations presented here, except for the DRIBM model, which is disabled since previous studies show that these modes should not be unstable in NSTX \cite{Rafiq2024NF}. Each submodel of MMM uses a quasilinear approximation in order to calculate fluxes from the linear eigenspectrum. 
\rev{The ETG model was calibrated against NSTX data during its development in order to compensate for simplifying assumptions made in its derivation \cite{Rafiq2022POP,Rafiq2024NF}. Note that after that calibration was incorporated, it has remained unchanged when applying MMM to study different NSTX discharges and different devices. Notable strengths of MMM for modeling spherical tokamaks with high $\beta$ are its dedicated MTM model, capturing instabilities that are not well-described by the present formulation of TGLF, and very fast solution time due to its simple eigenvalue solver. Potential limitations include its limited use of flux-surface geometry (which is mostly $s-\alpha$) and its assumption that the total diffusivity can be modeled as the sum of the diffusivities calculated independently for different groups of instabilities.} 

MMM has been applied often to study transport physics in NSTX discharges, including improved profile predictions when including MTMs \cite{Rafiq2021POP}, the influence of $\beta$ on ETGs \cite{Rafiq2022POP}, and comparisons of low and high collisionality regimes \cite{Rafiq2024NF}. A companion paper to this manuscript investigates sensitivities of the agreement between MMM temperature profile predictions and experimental profile fits in NSTX in greater depth than presented here \cite{Lestz2025pre2}. Time-dependent TRANSP simulations using MMM have also been used to develop advanced current ramp-up trajectories in preparation for NSTX-U \cite{Poli2015NF,Lopez2018PPCF} and for neutral beam optimization to inform scenario development for the newly commissioned SMART low aspect ratio tokamak \cite{Podesta2024PPCF}. Performance scoping studies of SMART explored a range of different transport models within TRANSP, including both MMM and TGLF \cite{CruzZabala2024NF}. Beyond spherical tokamaks, MMM has been validated against a database of discharges from several different conventional tokamaks, finding agreement within experimental uncertainty \cite{Rafiq2023plasma}, and has also been used to investigate transport characteristics in the high $\betap$ (kinetic plasma pressure normalized to poloidal field pressure) advanced tokamak scenario on DIII-D \cite{Pankin2018POP}.   

\rev{Though not used in this work, it is worth briefly describing the \qkiz model to provide additional context for the TGLF and MMM approaches, as these three models are some of the most widely used quasilinear codes. \qkiz is a gyrokinetic transport model addressing ITG, TEM, and ETG turbulence derived using simple $s-\alpha$ geometry \cite{Bourdelle2016PPCF}. Since \qkiz does not consider electromagnetic fluctuations, it does not capture transport due to KBMs or MTMs, nor does it include electromagnetic modifications of the ITG, TEM, or ETG branches \cite{Citrin2022NF}. Consequently, \qkiz is usually applied to conventional tokamaks instead of spherical tokamaks. A detailed discussion comparing the \qkiz and TGLF models can be found in \citeref{Bourdelle2016PPCF}, with additional description of the historical progression of quasilinear models including MMM, TGLF, and \qkiz given in \citeref{Staebler2024NF}.} 

As will be elaborated in this paper, reasonable agreement is found between the predicted electron and ion temperature profiles and the experimentally measured profiles. Relative to the experimental profiles, MMM tends to predict the temperature profiles more reliably than TGLF. In particular, TGLF's profile predictions were more strongly influenced by $\beta$ than those from MMM. The characteristics of the profile predictions are examined in this work, providing insight into the relative capabilities of each reduced turbulent transport model in the spherical tokamak transport regime.  

The rest of the paper is organized as follows. The time-dependent predictive TRANSP simulation scheme is described in \secref{sec:methods}, along with a discussion of the approach used to quantify the degree of agreement with the experiment. \secref{sec:te} compares the $\Te$ profile predictions of TGLF and MMM. The ion temperature profile ($\Ti$) predictions are compared for the two models in \secref{sec:ti}, including significant differences found in the $\teti$ ratio and profile shapes. Improved agreement for the prediction of the stored energy is discussed in \secref{sec:storedenergy}. 
\secref{sec:cputime} compares the computational cost of TGLF and MMM when used within TRANSP. \secref{sec:tglf_other} explores the sensitivity of TGLF's predictions to electromagnetic effects, different quasilinear saturation rules, and a spectral shift model for a smaller subset of NSTX discharges. \secref{sec:tglf_nn} discusses initial results of surrogate models for TGLF that were trained on the same NSTX database used for the TRANSP simulations.  A summary of the different trends found for TGLF and MMM is given in \secref{sec:tglfmmm_summary} and a discussion within the context of related modeling studies is presented in \secref{sec:discussion}. Lastly, remarks on avenues for suggested future work are made in \secref{sec:future}.  

\section{Time-Dependent Predictive TRANSP Simulations and Analysis Approach}
\label{sec:methods}

\subsection{Description of the TRANSP code}
\label{sec:transp}

TRANSP is a 1.5D tokamak power balance and transport code which has become a commonly used platform for integrated modeling \cite{Grierson2018FST}. TRANSP uses a 2D axisymmetric magnetic equilibrium along with a 1D transport grid. Fundamentally, codes such as TRANSP exist to solve generic transport equations such as 

\begin{align}
\pderiv{X}{t} + \div F = S.
\end{align}

Here, $X(\rho,t)$ is a time evolving profile depending on a radial coordinate $\rho$ such as ion temperature, electron density, \etc The right hand side $S(\rho,t)$ represents sources and sinks, which can be a combination of measurements (for instance, radiated power) and calculations from models within TRANSP (such as auxiliary heating from neutral beams or radio frequency waves). While TRANSP can be run in many different ways, the most salient distinction for this work is interpretive \vs predictive TRANSP simulations. For interpretive simulations, all of the kinetic profiles are provided as inputs based on experimental observations or heuristic assumptions. This allows TRANSP to subsequently solve for the relevant flux $\div F(\rho,t)$ that is consistent with the measured time evolution of the profiles and modeled sources and sinks, yielding transport coefficients for experimental interpretation and analysis. For predictive TRANSP simulations (sometimes abbreviated as PTRANSP), the fluxes are instead calculated by numerical transport models at each time step, allowing a subset of the kinetic profiles to be predicted over time via the implicit transport solver \ptsolver \cite{Yuan2011APS,Yuan2012APS,Yuan2013APS}. For a full description of the current implementation of \ptsolver, see Sec. 9 of \citeref{Pankin2025CPC}. In this work, the electron and ion temperature profiles were predicted simultaneously in the region $\rho = 0 - 0.7$. Closer to the edge in the examined NSTX plasmas, experimental data is considered less reliable, and therefore has larger uncertainties which would influence the predictions and make comparisons to the data less meaningful \cite{Avdeeva2023NF,Avdeeva2024PPCF}. The density and rotation profiles were treated as inputs to the transport models (\eg, determined entirely by experimental data), instead of being predicted alongside the temperature profiles. To evolve $\Te$, the following electron energy conservation equation is solved by \ptsolver: 

\begin{multline}
\pderiv{}{t}\left[\frac{3}{2}V^\prime n_e \Te\right] + \pderiv{}{\rho}\left[V^\prime\langle\abs{\grad\rho}^2\rangle n_e \left(\chi_e \pderiv{\Te}{\rho} - \Te v_e\right)\right] \\ 
- \overdot{\xi}\pderiv{}{\rho}\left[\rho V^\prime \frac{3}{2} n_e \Te\right] = S_e V^\prime.
\label{eq:pt_te}
\end{multline}

Here, the toroidal magnetic flux is defined as $\Phi = \pi\rho^2 B_0$, where $\rho$ is a dimensioned radial coordinate and $\xi = \rho/\rho_\text{sep} = \sqrt{\Phi/\Phi_\text{sep}}$ is a normalized flux surface label. $V^\prime$ and $\langle\abs{\grad\rho}^2\rangle$ are geometric factors related to the transformation to flux coordinates. \rev{The total electron energy source term, $S_e$, includes contributions from auxiliary heating, Ohmic heating, collisional electron-ion coupling, radiated power loss, and neutral ionization, as described in detail in Sec. 4 of \citeref{Pankin2025CPC}.} Lastly, $\chi_e$ is the electron thermal conductivity and $v_e$ is the electron convective velocity. The \rev{prescribed} time evolution of the magnetic equilibrium is included via the term that depends on $\overdot{\xi}$. An analogous equation is solved for the ion temperature. Note that while \eqref{eq:pt_te} uses the variable naming conventions found in the \ptsolver references \cite{Yuan2011APS,Yuan2012APS,Yuan2013APS,Pankin2025CPC} to avoid confusion with those documents, the rest of this paper adopts the more common choice for the symbol $\rho$ to represent the normalized flux surface label when discussing radial profiles. 

\ptsolver solves \eqref{eq:pt_te} via finite difference methods with a Newton iteration scheme \cite{Jardin2008JCP,Yuan2011APS,Yuan2012APS,Yuan2013APS}. At each iteration, the inputs are passed to the transport models, which compute new temperature profiles and transport coefficients based on those inputs, iterating until a specified error tolerance is satisfied. 
Predictive TRANSP simulations also use an adaptive time stepping scheme, such that if the \ptsolver error tolerance is not satisfied within a specified maximum number of iterations, the time step will be automatically reduced and the Newton iterations will be restarted. Hence, poor convergence of \ptsolver can lead to extremely long computation times in some cases where the adaptive time step becomes very small. \secref{sec:cputime} compares the computational expense of time-dependent predictive TRANSP simulations with MMM \vs TGLF. 

The transport models used in TRANSP are fundamentally local calculations, such that the transport coefficients calculated by TGLF and MMM only depend on the values of quantities and gradients at each spatial grid point. Numerically, \eqref{eq:pt_te} is solved from the prediction boundary radially inwards. Hence the resulting temperature profile depends on the experimental temperature provided at the boundary ($\rho = 0.7$), with errors accumulating with radial distance from this fixed boundary. \rev{The equilibria are reconstructed with the EFIT \cite{Lao1985NF,Lao2005FST} or LRDFIT \cite{Menard2006PRL} codes and treated as inputs to the simulations. Although TRANSP is capable of evolving the current profile by solving a poloidal field diffusion equation and solving for a new equilibrium at each time step \cite{Pankin2025CPC}, these features are not employed in the simulations presented here, in order to isolate the predictive power of the transport models alone. The influence of equilibrium reconstruction accuracy on single time slice transport analysis in NSTX(-U) is discussed in \citeref{Avdeeva2024PPCF}.}

For the predictive TRANSP simulations presented in this paper, the same input settings for \ptsolver, MMM, and TGLF were used for each respective simulation, with the only exceptions described in \secref{sec:tglf_settings}. Numerical parameters specific to \ptsolver were varied for a few specific discharges to probe the numerical sensitivity, but none of these changes had a significant effect on the predicted profiles. The varied settings included reducing the residual tolerance for the Newton iteration scheme by a factor of ten beyond its default value and adjusting the time stepping ``implicitness'' parameter \cite{Pankin2025CPC}. All simulations use the NCLASS neoclassical transport solver \cite{Houlberg1997POP}. Neither TGLF nor MMM includes anomalous energetic particle transport in the simulations presented in this work, though TGLF can be configured to include low-$n$ \Alfven eigenmode instabilities \cite{Sheng2017POP} which have been used within a simplified critical gradient model of fast ion transport \cite{Bass2020NF}. The inclusion of energetic-particle-driven instabilities and consequent transport is an area of active development for MMM \cite{Weiland2023POP,Rafiq2023APS}. Hence, fast ion confinement is assumed to be classical, and all simulations of a given discharge should have comparable fast ion density profiles, differing only by indirect changes in neutral beam deposition due to differing temperature profile predictions from different models. Neither TGLF nor MMM is presently capable of capturing transport due to low-$n$ MHD modes. 

\begin{figure}[tb]
\includegraphics[width = \columnwidth]{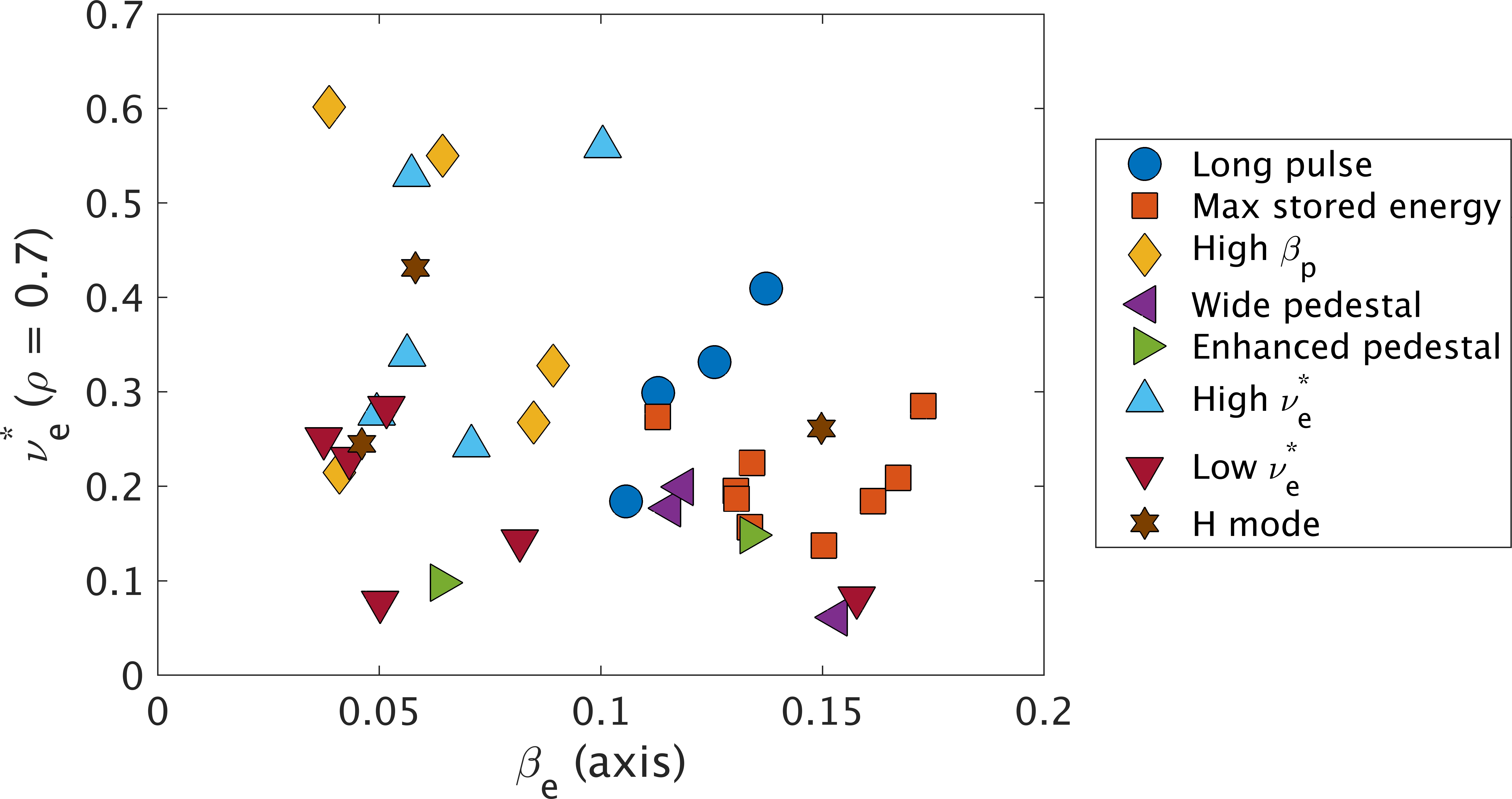}
\caption{Collection of well-analyzed NSTX discharges modeled in this study, classified by the property which made them noteworthy in previous NSTX transport studies (markers). The on-axis $\betae$ and normalized collisionality $\nustar$ near the plasma edge are shown for each discharge. The full list of discharges and associated references can be found in \tabref{tab:runids}.}
\label{fig:shots_beta_nu}
\end{figure}

\subsection{Selection of Discharges and Analysis Windows}
\label{sec:shots}

For each discharge, an analysis window was identified from experimental data by finding a period of relative MHD quiescence that is free of impulsive variations in the neutron rate and stored energy.  
 The chosen analysis time window for each discharge is given in \tabref{tab:runids}, with the median analysis window being 200 ms in duration. While kinetic profiles in NSTX often evolve on this timescale even during the current flat top -- especially the density -- an emphasis of this work is to evaluate the accuracy of \emph{time-dependent} transport predictions, which would be less meaningful if tested instead on a much shorter timescale with less plasma evolution. A detailed time slice analysis of transport in two relatively low $\beta$ NSTX discharges with the TGYRO flux matching code \cite{Candy2009POP} and electrostatic TGLF can be found in \citeref{Avdeeva2023NF}. 

In order to obtain results that are representative of a wide array of high performing NSTX scenarios, predictions were made for discharges of interest that had been previously analyzed thoroughly, and in most cases were high performing discharges in at least one aspect. The discharges are classified according to their descriptions in NSTX papers where they were analyzed, and shown in \figref{fig:shots_beta_nu} 
 \cite{Gerhardt2011NFat,Gerhardt2011NFcur,Ren2012POP,Guttenfelder2013NF,Ren2013NF,Gerhardt2014NF,Kaye2014POP,RuizRuiz2019PPCF,
 Ren2020NF,Battaglia2020POP,Clauser2022POP,Avdeeva2023NF,Dominski2024POP,Rafiq2024NF,Clauser2025POP}.
These classifications include: long pulse length, large stored energy, sustained high $\betap$, wide pedestal, enhanced pedestal, high normalized collisionality $\nustar$, low normalized collisionality $\nustar$, and a few generic well-analyzed NSTX H modes. All of the discharges analyzed in this work were beam-heated, 
\rev{injecting $2 - 6$ MW of power at $65 - 95$ keV.} 
These discharges did not make use of the high harmonic fast wave system. The chosen discharges cover a wide range of $\beta$ and collisionality in order to explore the performance of predictive TRANSP simulations with MMM and TGLF across various regimes in the NSTX parameter space. By comparison, a spherical tokamak reactor would likely have $\nustar$ about $1 - 2$ orders of magnitude lower than the typical $\nustar$ shown in \figref{fig:shots_beta_nu}, due to the order of magnitude increase in temperature anticipated for a reactor ($\nustar \propto n_e/\Te^2$). Although $\beta$ in such a reactor would be more sensitive to the specific design, generally speaking $\beta$ would be expected to be somewhat smaller for a high field spherical tokamak reactor than in NSTX, since $\beta \propto T/B^2$ and NSTX had a fairly low on-axis magnetic field $B < 0.5$ T.  

Since the chosen NSTX discharges were well-analyzed, and in most cases well-diagnosed, they also feature reliable experimental profiles and equilibrium reconstructions. Whenever possible, an interpretive TRANSP run previously prepared for experimental analysis is used as a starting point for the predictive runs. By starting with existing runs from several different experienced users, this also serves to reduce idiosyncracies in the performance of the transport predictions that could be unique to one particular user's style of fitting profiles, equilibria, \etc. Since many of these existing TRANSP runs were many years old, the input namelists typically needed to be modernized to be compatible with the current version of TRANSP. With these minor changes, the modernized interpretive runs reproduced the original simulations very closely, providing a useful starting point for the predictive simulations for comparison. The vast majority of the TRANSP simulations used a uniform spatial grid with 60 radial zones, with a few using 40 or 20 zones instead. 

In each predictive TRANSP run used in this paper, the simulation starts around $200 - 300$ ms before the analysis window. The simulation is initialized in interpretive mode, where the temperature profiles are initially taken from fitted experimental inputs. The first $100 - 200$ ms of the simulation are run this way, in order to allow time for the Monte Carlo neutral beam module NUBEAM to build up the fast ion density to its experimental value (since the beam ion injection begins at the simulation start time, even though the injection could have started earlier in the experiment) \cite{Grierson2018FST}. At this point, the predictive mode is enabled, evolving the temperature profiles with the transport models instead of experimental measurements. The next $50 - 100$ ms of the simulation are not included in the analysis window, as the predictive models typically require some time to converge to a steady solution from the initial experimental profiles it was given (see Fig. 5 of \citeref{Lestz2025pre2}). All of the analysis windows occur during the plasma current flat top. The start time of each TRANSP simulation and the time at which the predictive models are enabled can be found in the input files for each TRANSP run ID, which are stored in the NSTX MDSplus tree. 

\subsection{Accuracy Metrics}
\label{sec:metrics}

In order to evaluate how well the predicted temperature profiles agree with the experimental measurements, standard figures of merit are used \cite{ITER1999NF}, with notation borrowed from \citeref{Abbate2024POP}. For a given profile $X(\rho,t)$, let $\xtrue$ be the ``true'', measured experimental profile and $\xpred$ be the predicted profile from transport models. The additional subscript $j$ will be used to denote the index of the 1D spatial grid in $\rho$. Then the local error $\epsj$, the relative signed offset $f$, and the relative root mean square error (RMSE) $\sigma$ are given by

\begin{subequations}
\begin{align}
\label{eq:eps}
\epsilon_j &= \xtruej - \xpredj, \\
\label{eq:off}
f &= \ddfrac{\frac{1}{N}\sum_j^N \epsilon_j}{\sqrt{\frac{1}{N}\sum_j^N\xtruej^2}}, \\ 
\label{eq:rmse}
\sigma &= \sqrt{\ddfrac{\sum_j^N \epsilon_j^2}{\sum_j^N\xtruej^2}}.
\end{align}
\end{subequations}

In general, these quantities will be time-averaged over each analysis window. One shortcoming of these figures of merit is that $\abs{f}$ and $\sigma$ are at most 1 when $\xpred$ underpredicts $\xtrue$, whereas they are unbounded when they overpredict. This is important to keep in mind when assessing the significance of the error metrics, as a factor of two overestimate would give $\sigma = 1$, whereas the same factor of two underestimate gives $\sigma = 0.5$. Although not the main focus of this work, at times a linear (Pearson) correlation coefficient between two quantities from several different simulations will be discussed, which is defined in the usual way: 

\begin{align}
r_{xy} = \ddfrac{\sum_i^N{\left(x_i - \bar{x}\right)\left(y_i - \bar{y}\right)}}{\sqrt{\sum_i^N \left(x_i - \bar{x}\right)^2 \sum_i^N \left(y_i - \bar{y}\right)^2}}\text{, where } \bar{x} = \frac{1}{N}\sum_i^N x_i.
\label{eq:rxy}
\end{align}

It is important to note that the emphasis of this work is primarily a pragmatic assessment of the agreement between temperature profiles calculated by the reduced models and those observed in NSTX discharges. Determining how faithfully the reduced models theoretically reproduce the higher fidelity physics models which they are intended to represent (for instance, nonlinear gyrokinetics) is beyond the scope of this work. Hence, any references to accuracy, error, and related terms in this paper should be understood as shorthand for the level of agreement between a modeled profile with respect to an observed one, based on the above metrics. Additionally, measurement error in the underlying experimental data and possible systematic errors introduced when creating profile fits increase the level of uncertainty when making these comparisons, which is not quantified in this work. 

\begin{figure*}[tb] 
\subfloat[\label{fig:te_blow}]{\includegraphics[height = \thirdheight]{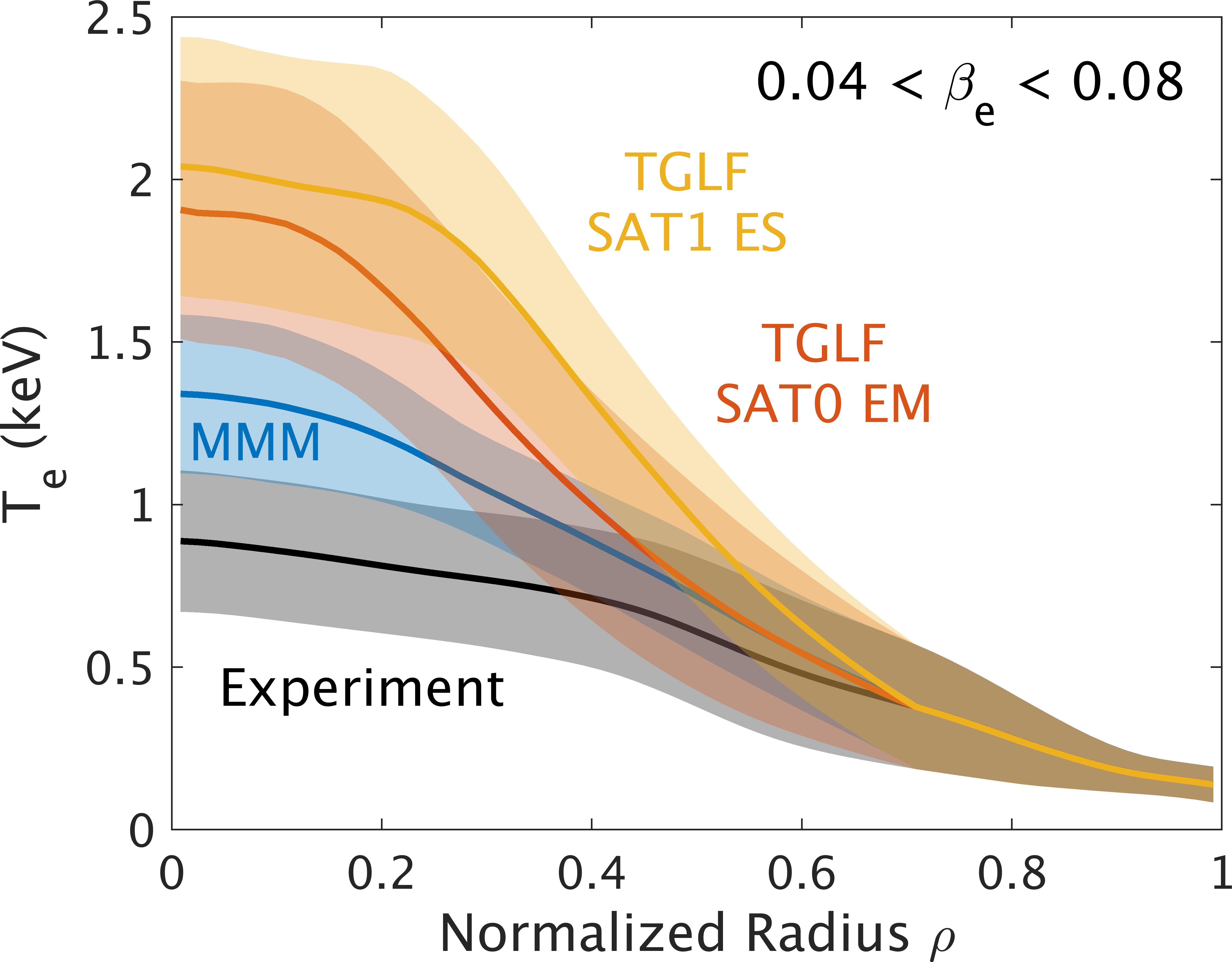}} \figsep
\subfloat[\label{fig:te_bmed}]{\includegraphics[height = \thirdheight]{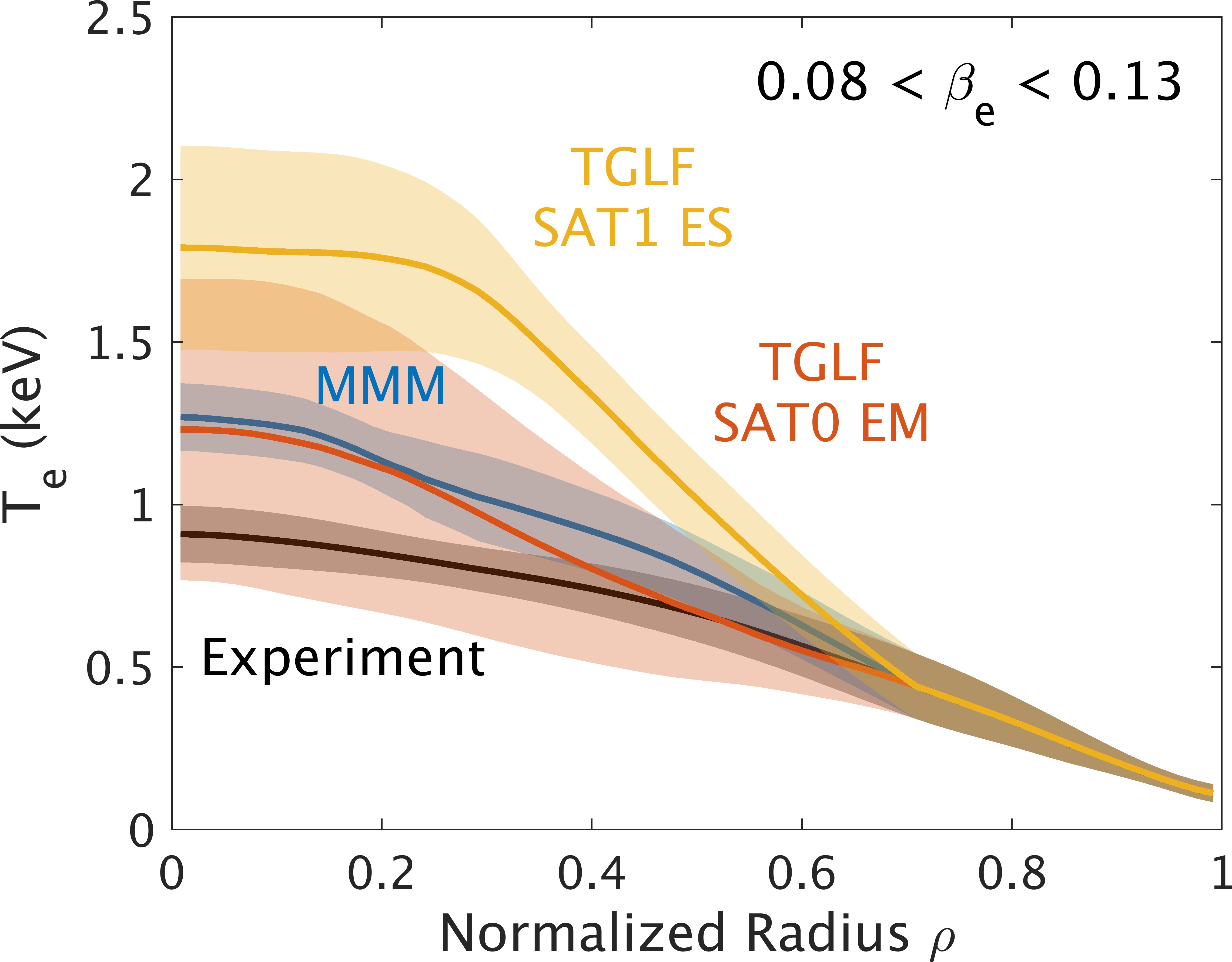}} \figsep
\subfloat[\label{fig:te_bhigh}]{\includegraphics[height = \thirdheight]{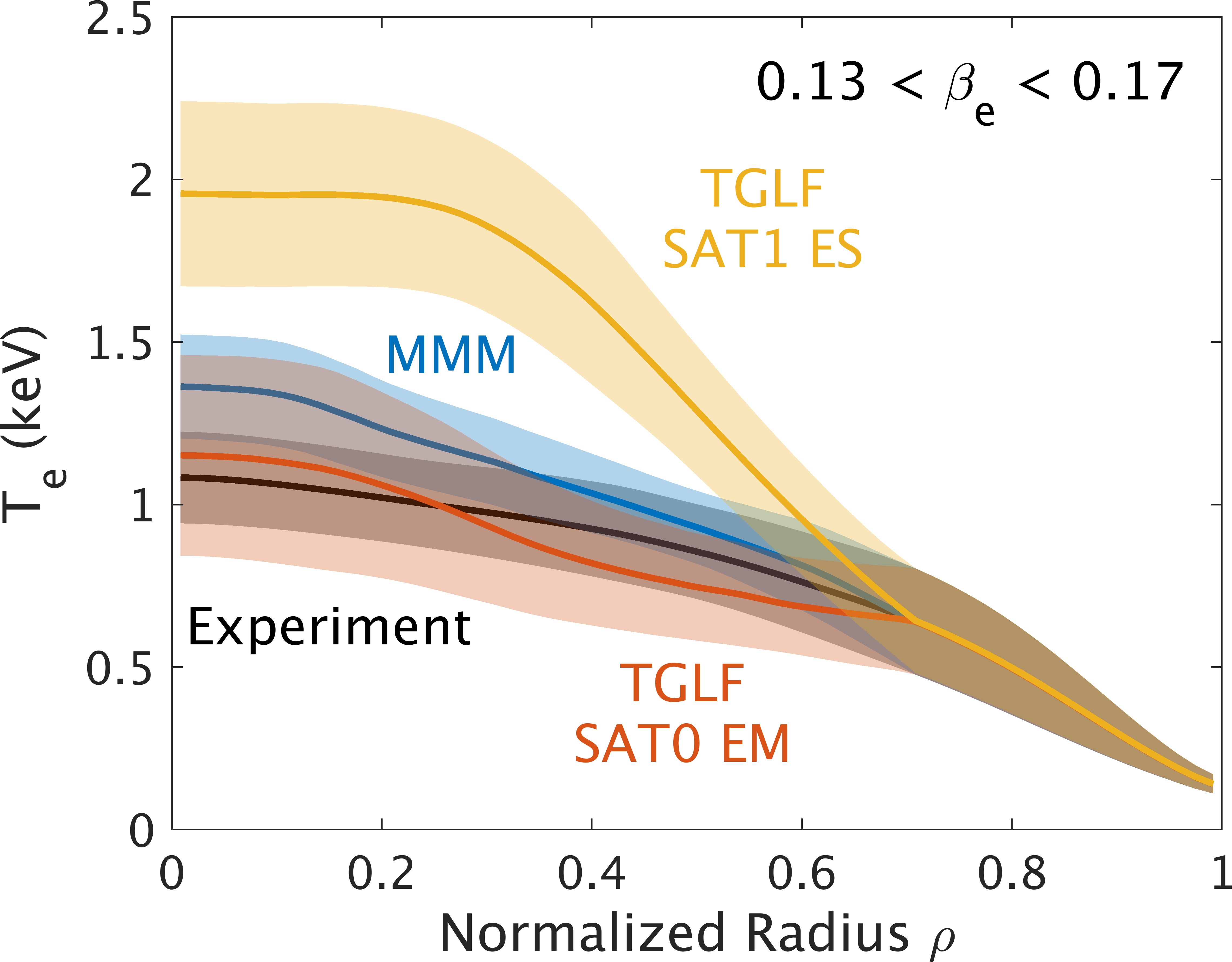}}
\caption{Electron temperature profiles averaged over all examined NSTX discharges for (a) relatively low $\betae$ on-axis, (b) intermediate $\betae$, and (c) relatively high $\betae$. Experimental profile fits are shown in black. Colors correspond to TRANSP simulations using MMM (blue), electromagnetic TGLF (orange), and electrostatic TGLF (gold). Thick curves show the average over all discharges in each range of $\betae$. Shaded regions show the standard deviation across the relevant discharges.}
\label{fig:meanprof_te}
\end{figure*}

\section{Temperature Profile Prediction Agreement With Experiment}
\label{sec:profpred}

\newcommand{\twotab}[2]{\begin{tabular}{c} #1 \\ #2 \end{tabular}}
\newcommand{\threetab}[3]{\begin{tabular}{c} #1 \\ #2 \\ #3 \end{tabular}}

\subsection{Electron Temperature Profile}
\label{sec:te}

First, consider the predictions for the electron temperature profile, shown in \figref{fig:meanprof_te} for each of the reduced turbulent transport models used within predictive TRANSP simulations in this work. The three panels divide all of the examined NSTX discharges into three groups, sorted by the experimental value of on-axis $\betae = 2\mu_0 P_e/B_T^2$, where $P_e = n_e T_e$ is the electron pressure. From low to high $\betae$, the three groups contain 15, 9, and 13 distinct NSTX discharges. In each panel, the thick black curve shows the experimental $\Te$ profile, averaged over all of the discharges in the specified range of $\betae$. The corresponding shaded region shows the standard deviation across the discharges in that group. The higher $\betae$ discharges tend to have somewhat higher experimental electron temperatures from the core all the way to the pedestal region. The other three curves and shaded regions correspond to time-dependent predictive TRANSP simulations with different turbulent transport models: blue for MMM, orange for electromagnetic TGLF with SAT0, and gold for electrostatic TGLF with SAT1. As in all simulations in this work, both the electron and ion temperature profiles are predicted simultaneously. The prediction boundary is at $\rho = 0.7$, outside of which the temperature profiles are set by experimental fits. These aggregate profiles give a sense of the character  of the predictions made by the different models. 

Namely, all of the turbulent transport models tend to overpredict the $\Te$ profile relative to the experimental observations. MMM is the most consistent, tending to predict $\Te$ a few hundred eV too high relative to the measurements in each of the $\betae$ groups. In contrast, the simulations using electrostatic TGLF exhibit a much more egregious $\Te$ overprediction. In those simulations, the calculated $\Te$ profiles are much too steep for $\rho \gtrsim 0.3$, overpredicting electron energy confinement over a large radial region, consequently lifting the core $\Te$ far above its observed value. Interestingly, the electrostatic TGLF $\Te$ predictions do not improve in the examined discharges with lowest $\beta$, where electromagnetic effects should be least important. However, it is worth emphasizing that even the low $\beta$ NSTX discharges modeled in this work would be considered high $\beta$ in present day conventional tokamaks. Conversely, the electromagnetic TGLF simulations show a strong sensitivity to $\beta$. In the lowest $\beta$ discharges shown in \figref{fig:te_blow}, the electromagnetic TGLF simulations predict $\Te$ to be nearly as high as in the TRANSP simulations using electrostatic TGLF, only lower due to a slightly shallower $\Te$ gradient near the prediction boundary. In the group of discharges with intermediate $\beta$ in \figref{fig:te_bmed}, the on-axis electron temperatures from electromagnetic TGLF simulations have dropped to a similar level as those predicted by MMM for the same discharges, albeit with significantly greater variance as indicated by the much larger shaded region. Notably, the edge $\Te$ gradient has dropped further, now aligning reasonably well with the experimental gradient in that region. Lastly, in the highest $\beta$ discharges shown in \figref{fig:te_bhigh}, the electromagnetic TGLF $\Te$ profile predictions have dropped well below MMM at all radii. Although the on-axis $\Te$ predicted by TGLF superficially matches that in the experiments, the shape of the predicted profile is qualitatively quite different, with a pronounced flat region near the prediction boundary and an overly steep gradient at mid-radius to compensate.

\begin{figure*}[tb]
\subfloat[\label{fig:te_err_hist}]{\includegraphics[height = \thirdheight]{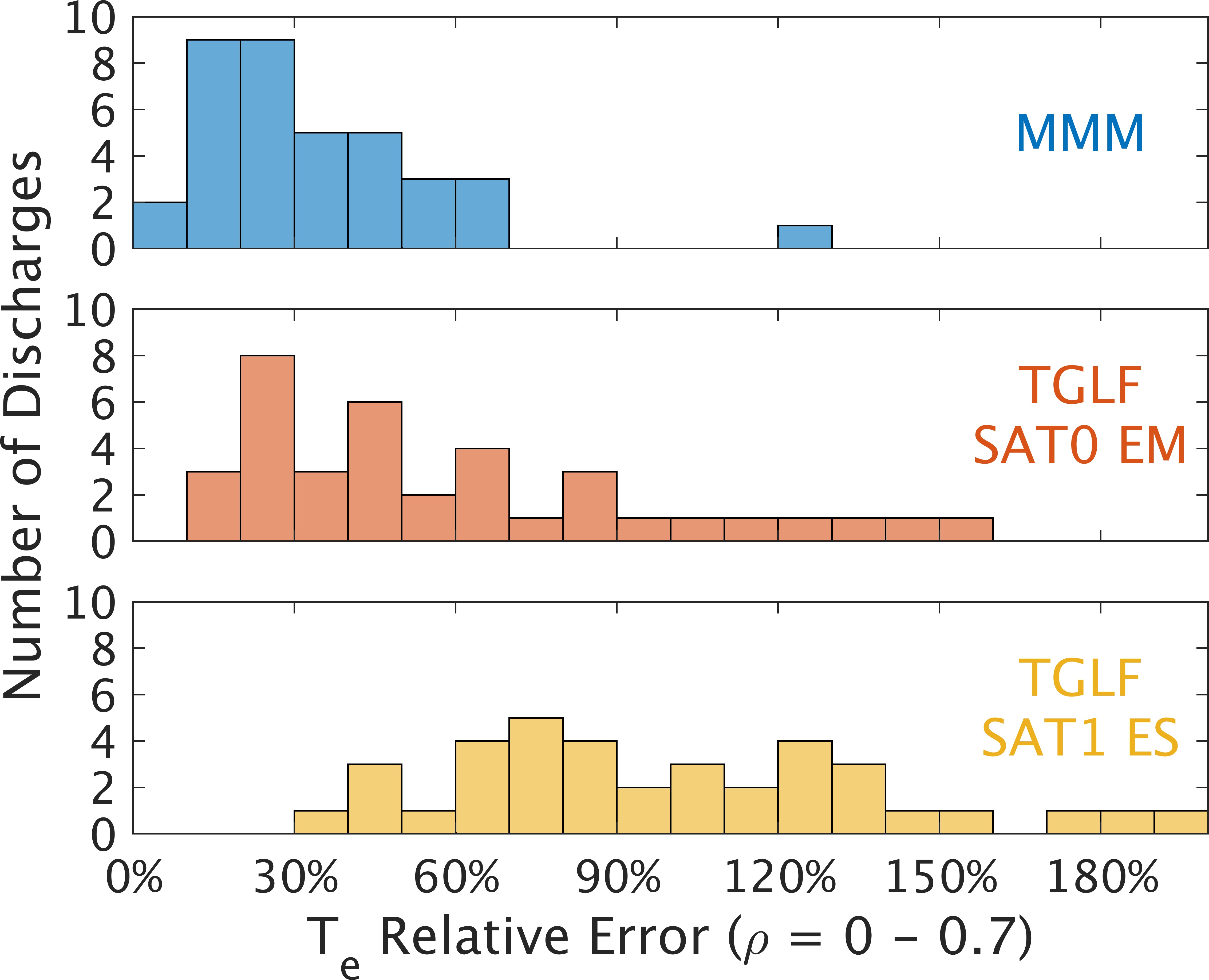}} \figsep 
\subfloat[\label{fig:te0_expred}]{\includegraphics[height = \thirdheight]{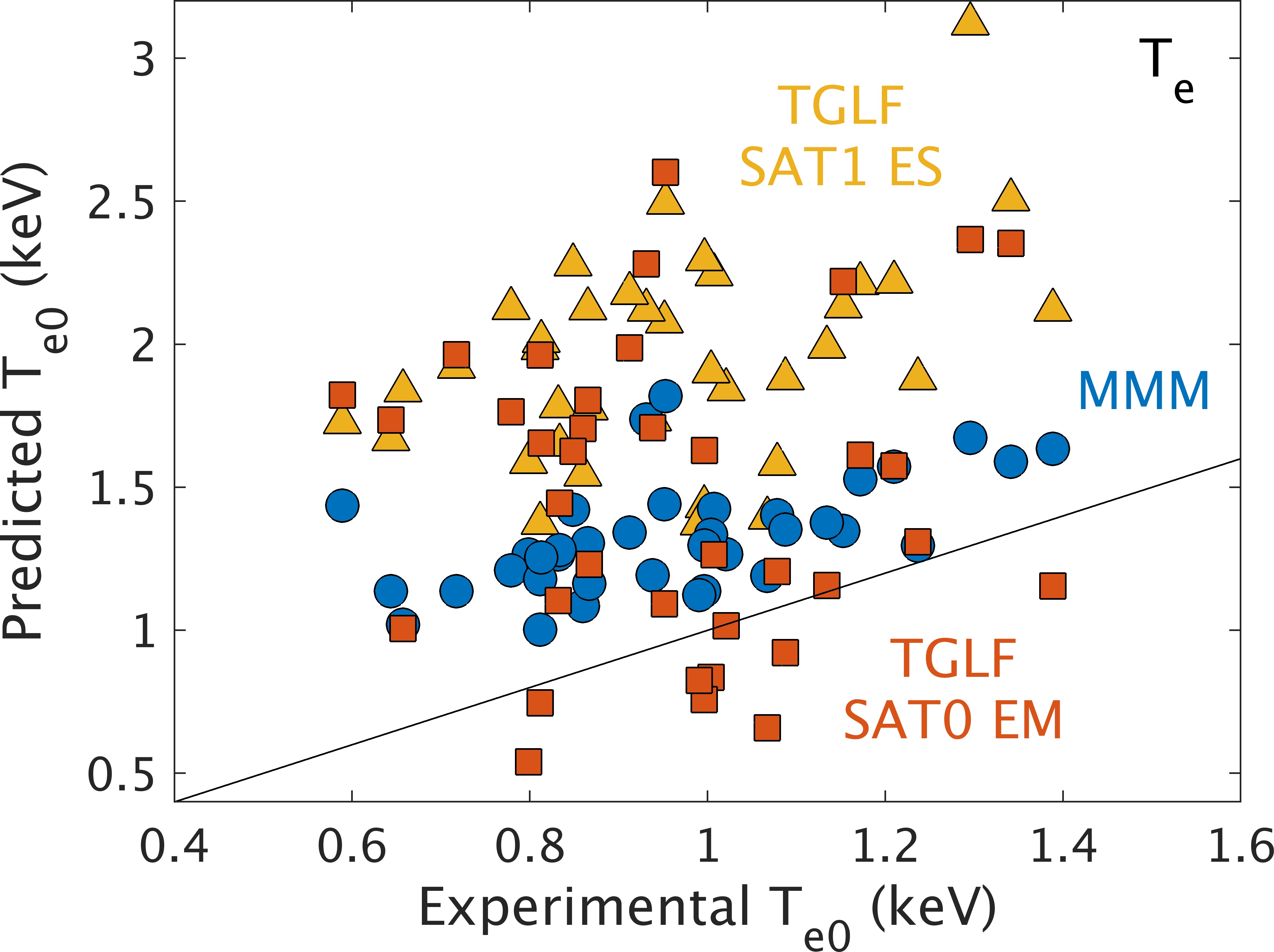}} \figsep
\subfloat[\label{fig:betat_teoff}]{\includegraphics[height = \thirdheight]{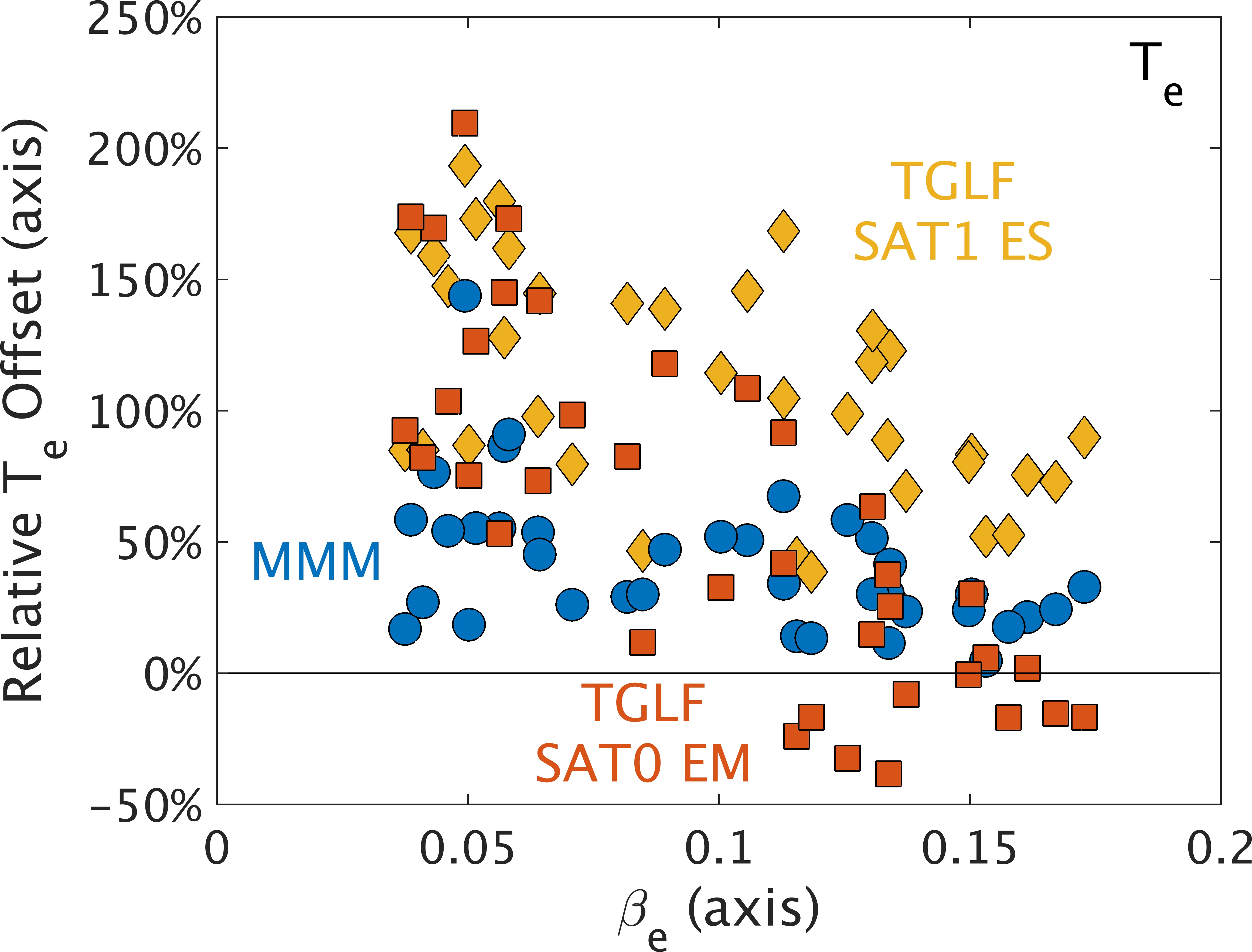}} 
 
\caption{(a) Histogram of relative error (RMSE over $\rho = 0 - 0.7$) between predicted \vs observed electron temperature profiles for all discharges. (b) Comparison of experimentally measured and predicted on-axis electron temperature. (c) Relative on-axis $\Te$ offset errors as a function of the experimental on-axis $\betae$. In (b) and (c), the solid line is for reference only, indicating zero offset between experimental and predicted $\Teo$. In all plots, blue shows TRANSP simulations using MMM, orange corresponds to electromagnetic TGLF, and gold to electrostatic TGLF.}
\label{fig:teax}
\end{figure*}

\definecolor{matlab_bluec}{RGB}{0,114,189}
\definecolor{matlab_orangec}{RGB}{217,83,25}
\definecolor{matlab_goldc}{RGB}{237,177,32}
\newcommand{\mmmtext}[1]{\textcolor{matlab_bluec}{#1}}
\newcommand{\temtext}[1]{\textcolor{matlab_orangec}{#1}}
\newcommand{\testext}[1]{\textcolor{matlab_goldc}{#1}}
\begin{table}\centering
\begin{tabular}{ccccc}
\hline\hline
Transport Model & $\Te$ Error & $\Ti$ Error & \threetab{Stored}{Energy}{Error} & \threetab{Energy}{Confinement}{Time Error} \\ 
\hline
MMM & $28 \pm 13\%$ & $27 \pm 7\%$ & $18 \pm 8\%$ & $15 \pm 10\%$ \\ 
TGLF SAT0 EM  & $46 \pm 30\%$ & $25 \pm 5\%$ & $17 \pm 14\%$ & $23 \pm 15\%$ \\ 
TGLF SAT1 ES & $93 \pm 27\%$ & $47 \pm 12\%$ & $59 \pm 15\%$ & $67 \pm 22\%$ \\
\hline\hline
\end{tabular}
\caption{Statistics for temperature profile predictions and stored energy (integrated from $\rho = 0 - 0.7$) for all simulated NSTX discharges. Reported as median root mean square error $\pm$ half interquartile range.}
\label{tab:allerrs}
\end{table}

To quantitatively compare the performance of different transport models in predicting $\Te$ across many discharges, the root mean square error is computed between the experimental profile and the predicted one for each simulation, with the sum in \eqref{eq:rmse} performed over the prediction region $\rho = 0 - 0.7$. When aggregating statistics for all of the TRANSP simulations, it is more representative to use the median and half the interquartile range (IQR, defined as the spread between first and third quartiles of data) instead of the mean and standard deviation. Characterizing the typical value and spread in this way prevents outliers in the error distributions from skewing the statistics towards towards large values. These statistics are tabulated in \tabref{tab:allerrs} and shown as histograms in \figref{fig:te_err_hist}, with the same color conventions as in \figref{fig:meanprof_te}. Overall, MMM predicts $\Te$ profiles that are in better agreement with experiment than either of the TGLF models, with a median RMSE of $28 \pm 13\%$ half IQR for MMM compared to $46 \pm 30\%$ for electromagnetic TGLF and $93 \pm 27\%$ for electrostatic TGLF. The larger variation in the accuracy of TGLF predictions can be attributed to the much longer tail of the TGLF histogram in \figref{fig:te_err_hist}, indicating a much larger fraction of discharges that are very poorly predicted than in MMM. Note that even when restricting to NSTX discharges with on-axis $\betae > 0.1$ (about half of the studied discharges) in order to exclude the most egregious $\Te$ overpredictions from electromagnetic TGLF, TRANSP simulations with MMM are still in better agreement with experiment than those with electromagnetic TGLF, with relative errors of $22 \pm 11\%$ \vs $34 \pm 9\%$, respectively. A detailed investigation of the variation in the $\Te$ profile prediction error for different plasma conditions in TRANSP simulations with MMM is the focus of \citeref{Lestz2025pre2}. 

Beyond the average prediction accuracy across all discharges, it is relevant to consider how closely the predictions track the experimental observations on a discharge to discharge basis. To this end, the experimental \vs predicted on-axis electron temperatures $\Teo$ are plotted in \figref{fig:te0_expred} for all three turbulent transport models. Whereas TRANSP simulations with MMM or electrostatic TGLF overpredict $\Teo$ in every discharge that was modeled in this work, electromagnetic TGLF underpredicts the on-axis electron temperature in 24\% of the cases, largely skewed towards discharges with high $\beta$, as in the discussion of \figref{fig:te_bhigh}. Moreover, MMM's prediction of $\Teo$ is highly correlated with the experimental value (linear correlation coefficient $r = 0.55$), indicating a systematic and relatively consistent overprediction of electron energy confinement. In contrast, electromagnetic TGLF's calculation of $\Teo$ is uncorrelated with the measured $\Teo$ ($r = 0.04$), suggesting more erratic predictions overall. However, the correlation for electromagnetic TGLF simulations becomes comparable to that for MMM when restricting to only the relatively high on-axis $\betae > 0.1$ discharges ($r = 0.53$). Surprisingly, even the predicted $\Teo$ from electrostatic TGLF simulations is similarly correlated with the experimental values ($r = 0.42$), despite the fact that this model predicts excessively high electron temperatures in nearly every case and would not be expected to capture high $\beta$, electromagnetic turbulence by construction. 

\begin{figure*}[tb] 
\subfloat[\label{fig:ti_blow}]{\includegraphics[height = \thirdheight]{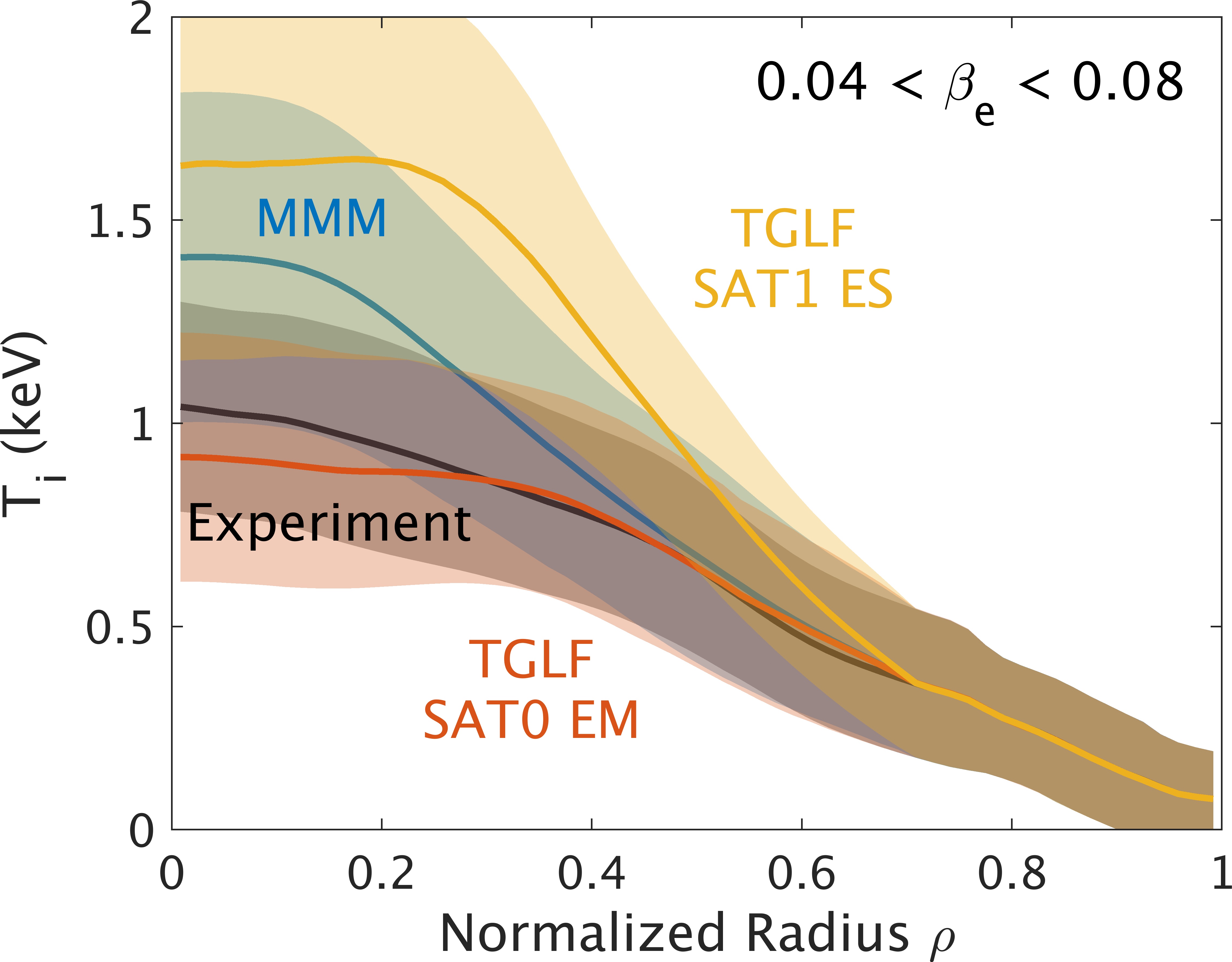}} \figsep
\subfloat[\label{fig:ti_bmed}]{\includegraphics[height = \thirdheight]{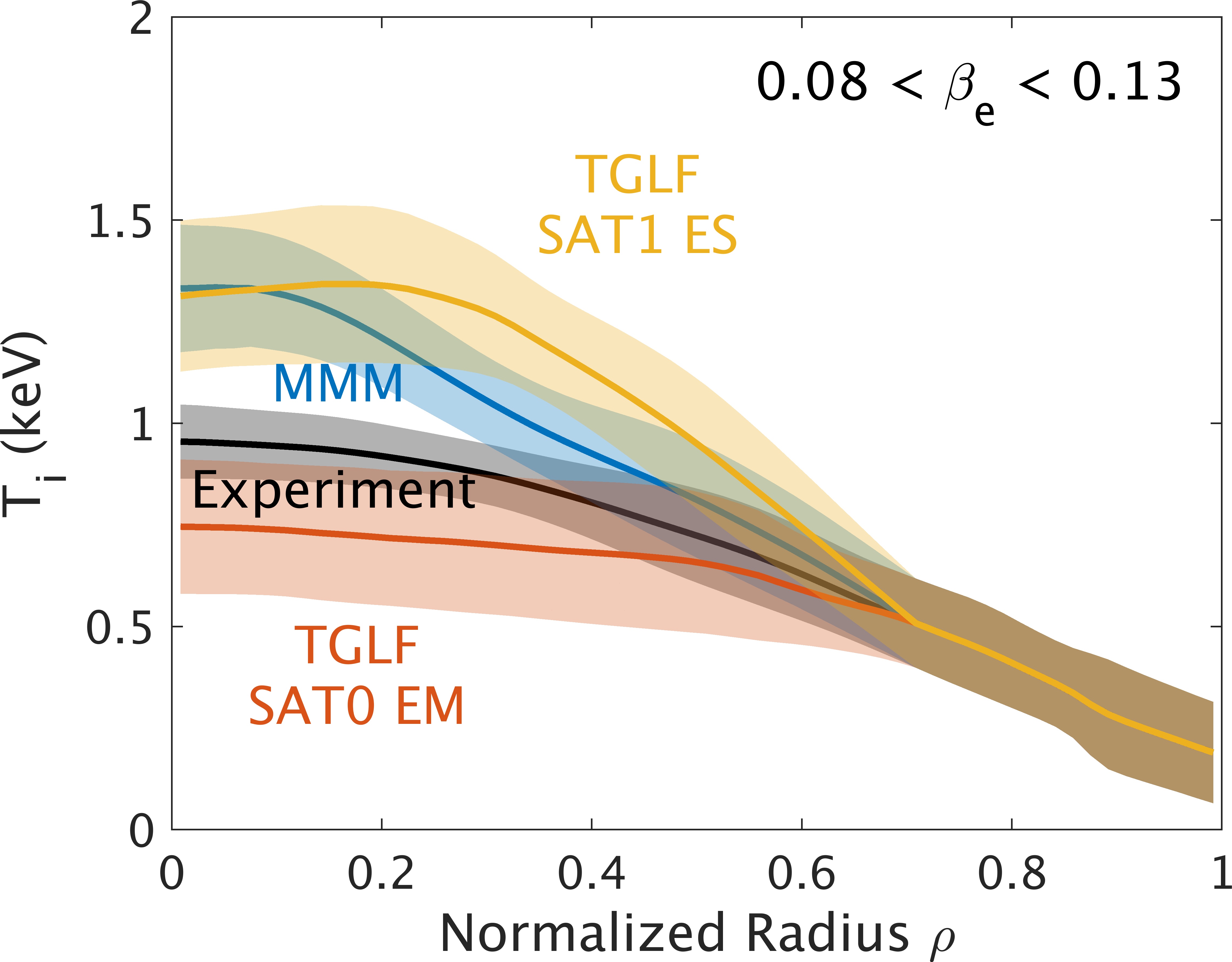}} \figsep
\subfloat[\label{fig:ti_bhigh}]{\includegraphics[height = \thirdheight]{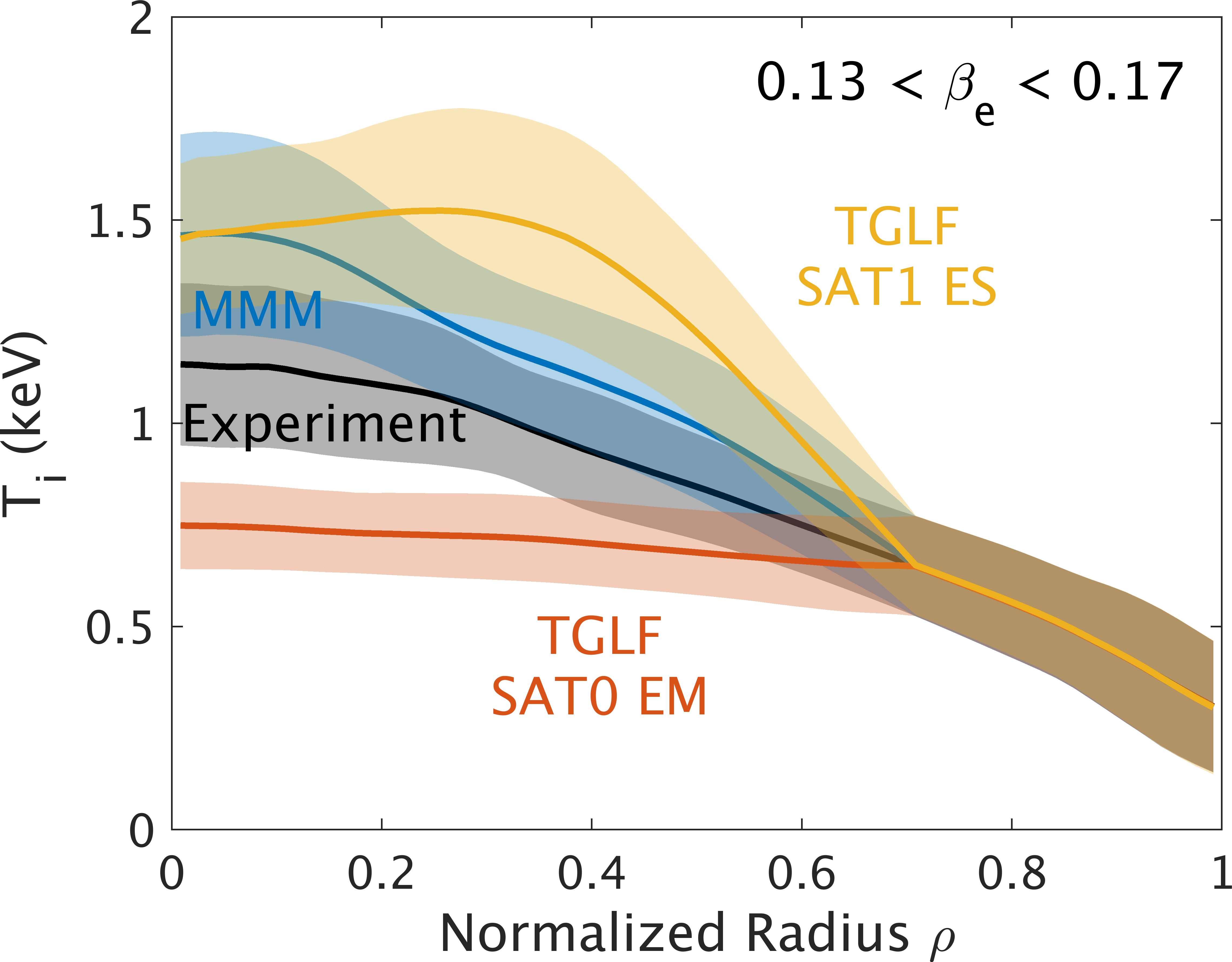}}
\caption{Ion temperature profiles averaged over all examined NSTX discharges for (a) relatively low $\betae$ on-axis, (b) intermediate $\betae$, and (c) relatively high $\betae$. Experimental profile fits are shown in black. Colors correspond to TRANSP simulations using MMM (blue), electromagnetic TGLF (orange), and electrostatic TGLF (gold). Thick curves show the average over all discharges in each range of $\betae$. Shaded regions show the standard deviation across the relevant discharges.}
\label{fig:meanprof_ti}
\end{figure*}

To further investigate the prediction sensitivity to $\beta$, \figref{fig:betat_teoff} plots the signed offset error (\eqref{eq:off}, relative to the experimental measurement) of the predicted on-axis $\Te$ for each of the models against on-axis $\betae$. For MMM and both types of TGLF simulations, the on-axis $\Te$ offset is decreased as $\betae$ is increased. This effect is most prominent for electromagnetic TGLF simulations, which exhibit the steepest slope in \figref{fig:betat_teoff} and also the highest linear correlation coefficient ($r = -0.80$). Strong correlation coefficients are also found with total $\beta$ instead of $\betae$, and also when evaluating $\beta$ at the prediction boundary or taking its volume average instead of using its on-axis value, supporting the robustness of this trend. By comparison, the $\Teo$ offsets from electrostatic TGLF simulations still have a moderately steep dependence on $\betae$ with a strong correlation ($r = -0.55$) and MMM has a more shallow dependence with correlation of comparable strength $(r = -0.49$). Evidently, all three models are underpredicting electron energy transport in low $\beta$ plasmas compared to experiment to a relatively higher degree than in higher $\beta$ plasmas. 

Note that part of this effect can be explained by improved electron energy confinement with higher $\beta$ in the examined NSTX discharges. While there is a correlation between the on-axis $\Te$ offset and $\betae$ for both MMM and electrostatic TGLF, the actual on-axis electron temperatures predicted by these models is uncorrelated with $\betae$ -- the shrinking offset is mostly due to an increase in the experimental $\Teo$ (which \emph{is} correlated with on-axis $\betae$, with $r = 0.46$). In particular, this helps explain the apparent improved predictions with electrostatic TGLF with increased $\betae$, where it would be expected to be less applicable by definition due to lacking the electromagnetic physics that become more relevant at higher $\beta$. On the other hand, the on-axis electron temperatures predicted by electromagnetic TGLF simulations decrease with increasing $\betae$ (as in \figref{fig:meanprof_te}), which compounds the effect of the experimental $\Te$ increasing with $\betae$ to create the very strong dependence of the relative $\Teo$ offset on $\betae$ demonstrated in \figref{fig:betat_teoff}.  

\rev{The influence of collisionality on the $\Te$ profile predictions and level of agreement with the experimental profiles was also explored but found to  generally be weaker than the sensitivity to $\betae$ for each of the different reduced transport models. When grouping the discharges into ranges of relatively low, moderate, and high normalized collisionality $\nustar$ evaluated at the prediction boundary $\rho = 0.7$ and averaging over the profiles analogously to what was done in \figref{fig:meanprof_te}, there are indeed qualitative trends in the predictions. Namely, the TRANSP simulations with electromagnetic TGLF tend to have the greatest overprediction of $\Te$ at relatively high $\nustar$ and become more similar to MMM for the discharges in the low and moderate $\nustar$ groups, though still with significantly greater variation than the MMM predictions. However, this trend is in line with what would be implied by the $\betae$ trend from \figref{fig:meanprof_te}, since $\Te \propto \betae \propto 1/\sqrt{\nustar}$. Furthermore, the correlation coefficients between $\betae$ and $\Te$ RMSE are consistently stronger than those between $\nustar$ and $\Te$. Whereas the correlations with $\betae$ are robust and relatively insensitive to precisely how $\beta$ is evaluated (on axis, near the boundary, averaged over some spatial region, \etc), the weaker ones with collisionality are also less consistent when evaluating $\nustar$ at different plasma locations, which could be in part due to $\nustar$ often exhibiting a steep profile near the prediction boundary. Broadly, the temperature profile predictions appear to depend more robustly on $\betae$ than $\nustar$ across all of the transport models explored in this study. \citeref{Lestz2025pre2} further explores the relative influence of $\betae$ and $\nustar$ for MMM by modeling a second database of NSTX discharges that is constructed to analyze these quantities independently, reaching a similar conclusion.} 

Overall, TRANSP simulations using MMM reproduce experimental NSTX $\Te$ profiles both more closely and consistently than either of the TGLF models tested. Electromagnetic TGLF SAT0 is moderately less accurate, but is highly sensitive to $\beta$. It predicts on-axis $\Te$ far too high for the low $\beta$ discharges while predicting on-axis $\Te$ in line with the experiment for high $\beta$ discharges, albeit with unusually flat gradients near the pedestal. Electrostatic TGLF SAT1 overpredicts $\Te$ by around a factor of two for most discharges, substantially worse than MMM or electromagnetic TGLF, underscoring the importance of electromagnetic turbulence for electron energy transport in NSTX.  

\subsection{Ion Temperature Profile and $\teti$ Ratio}
\label{sec:ti}

\begin{figure*}[tb]
\subfloat[\label{fig:ti_err_hist}]{\includegraphics[height = \thirdheight]{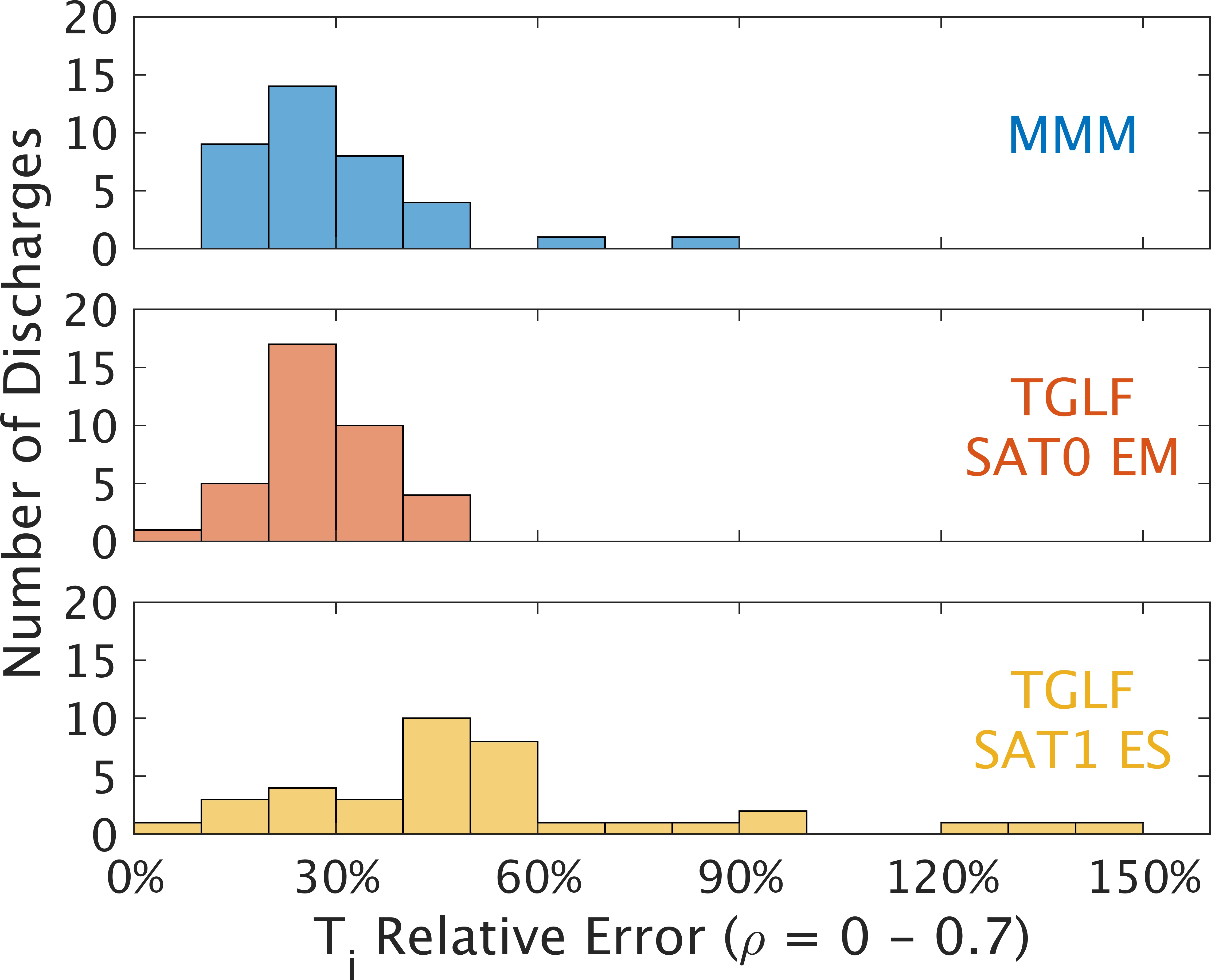}} \figsep
\subfloat[\label{fig:ti0_expred}]{\includegraphics[height = \thirdheight]{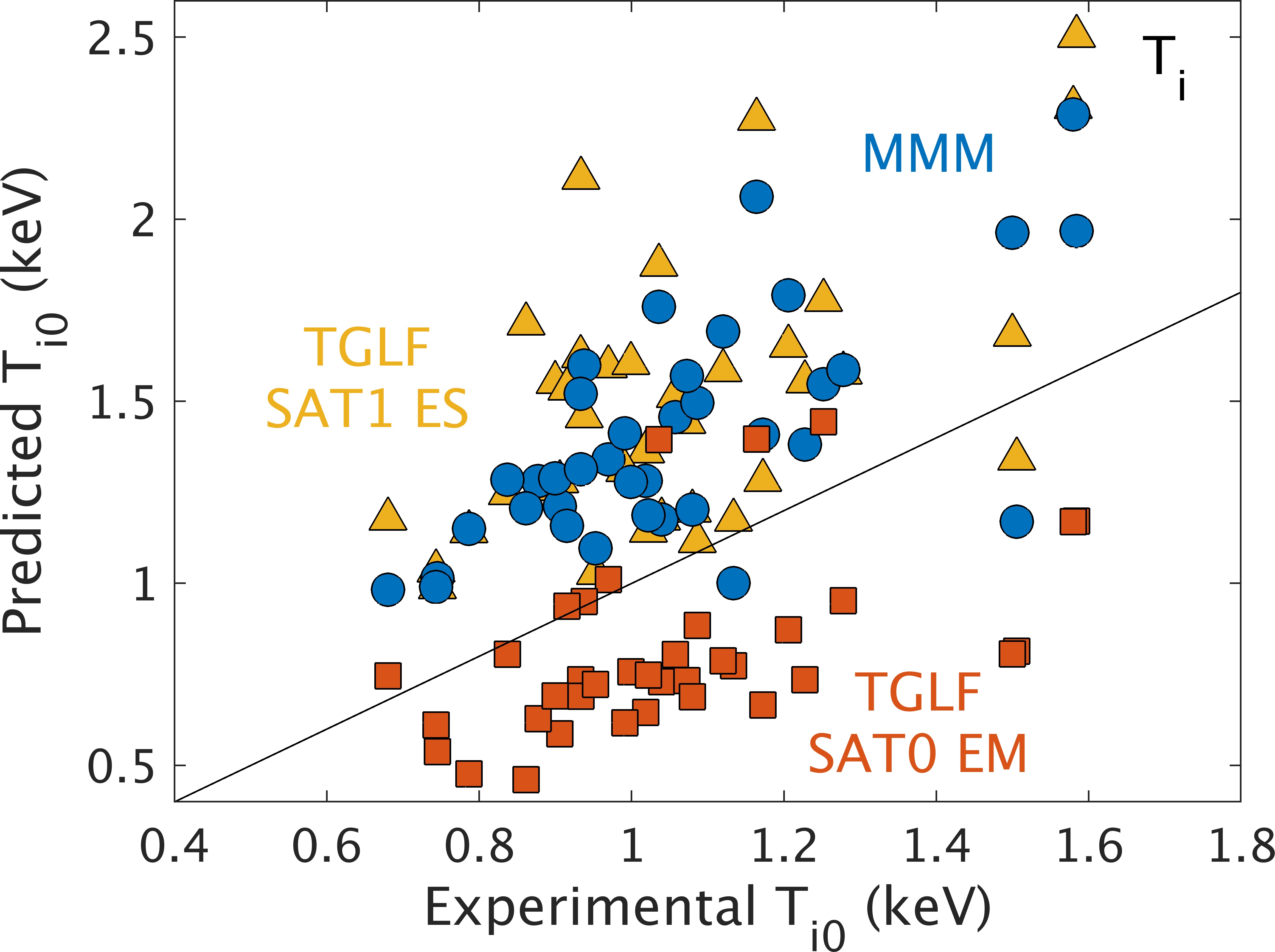}} \figsep
\subfloat[\label{fig:betat_tioff}]{\includegraphics[height = \thirdheight]{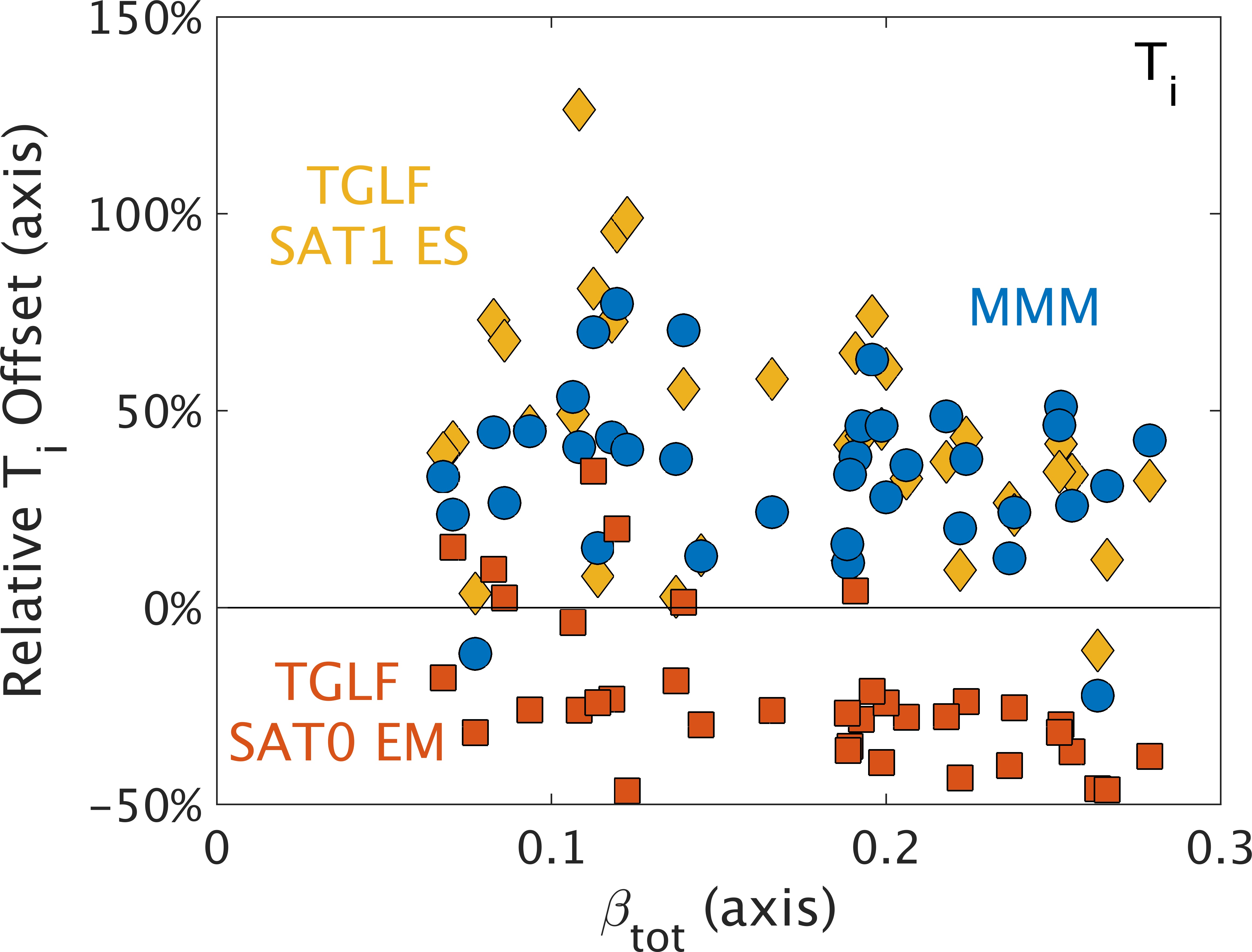}}
\caption{(a) Histogram of relative error (RMSE over $\rho = 0 - 0.7$) between predicted \vs observed ion temperature profiles for all discharges. (b) Comparison of experimentally measured and predicted on-axis ion temperature. (c) Relative on-axis $\Ti$ offset errors as a function of the experimental on-axis total $\beta$. In (b) and (c), the solid line is for reference only, indicating zero offset between experimental and predicted $\Tio$. In all plots, blue shows TRANSP simulations using MMM, orange corresponds to electromagnetic TGLF, and gold to electrostatic TGLF.}
\label{fig:tiax}
\end{figure*}

It is often assumed that ion transport in spherical tokamaks such as NSTX is dominated by neoclassical transport due to the large $\exb$ shear \cite{Kinsey2007POP,Roach2009PPCF,Kaye2021PPCF}. A comparison between the neoclassical and turbulent ion energy transport predicted by TRANSP simulations using MMM is made in \citeref{Lestz2025pre2} for the same set of discharges as in this work, finding that neoclassical transport generally dominates, except near the pedestal where there can be a small turbulent component. 
\rev{Moreover, the same trend is found in \secref{sec:tglf_nn} using TGLF-based surrogate models within a time slice flux matching transport solver, outside of TRANSP.} 
Since the same neoclassical model, NCLASS, is used for the MMM and TGLF simulations with TRANSP, one might expect that differences in the predicted turbulent ion transport between the models may be trivial in comparison to the neoclassical contribution. Despite finding similar root mean square errors in the $\Ti$ profile predictions from MMM and electromagnetic TGLF, these errors have very different parametric dependencies in the examined discharges. 

The ion temperature profile predictions are presented in \figref{fig:meanprof_ti} in the same fashion as in \figref{fig:meanprof_te} -- divided into the same three ranges of on-axis $\betae$ and then averaged over the modeled discharges in each group. As for the electron temperature profile predictions, both MMM and electrostatic TGLF simulations have a strong tendency to overpredict $\Ti$ across all values of $\betae$. Also qualitatively similar to the $\Te$ profiles, the MMM predictions of $\Ti$ are quite consistent across the three groups of discharges and the electrostatic TGLF $\Ti$ profiles have very steep gradients from the prediction boundary inwards to a mid-radius region. Unlike the other models, the TRANSP simulations with electromagnetic TGLF tend to underpredict $\Ti$, with this underprediction becoming progressively worse at higher $\betae$. In the relatively low $\beta$ discharges shown in \figref{fig:ti_blow}, the $\Ti$ profiles predicted by electromagnetic TGLF are in excellent agreement with the observed profiles for $\rho = 0.3 - 0.7$, with an underprediction in the core due to profile flattening in that region. For the intermediate $\beta$ discharges in \figref{fig:ti_bmed}, the underprediction becomes more substantial as the predicted $\Ti$ gradients flatten over a much wider region. Lastly, in the highest $\beta$ discharges in \figref{fig:ti_bhigh}, the $\Ti$ profile predicted by electromagnetic TGLF is almost entirely flat, yielding the largest underprediction. 

When comparing the level of experimental agreement of each model in predicting $\Ti$ for the entire set of examined discharges, the performance of MMM and electromagnetic TGLF is quite similar. The average MMM error for the $\Ti$ profile is $27 \pm 7\%$, compared to $25 \pm 5\%$ for electromagnetic TGLF, where both quantities are reported as median $\pm$ half interquartile range, as before. Electrostatic TGLF has larger errors of $47 \pm 12\%$, with about twice the degree of disagreement with experiment as electromagnetic TGLF. The full distribution of errors for each model is shown in \figref{fig:ti_err_hist}. Overall, MMM predicts both $\Te$ and $\Ti$ with about the same level of agreement with experiment, whereas both TGLF models predict $\Ti$ with almost a factor of two better agreement with experiment than their predictions for $\Te$, and with significantly less variance. The qualitative trends highlighted in \figref{fig:meanprof_ti} for the profiles predicted by the different models are quantified in \figref{fig:ti0_expred} by comparing the experimental \vs predicted on-axis ion temperatures. MMM tends to moderately overpredict $\Tio$ relative to its experimental value, similar in magnitude to its overprediction of $\Teo$. In contrast, electromagnetic TGLF underpredicts $\Tio$ in the vast majority of cases. 

The degree to which $\Ti$ is underpredicted in the electromagnetic TGLF simulations is most strongly influenced by total $\beta$, as illustrated in \figref{fig:betat_tioff}. While the overall spread in offsets is much smaller for the on-axis predicted $\Ti$ than $\Te$, the on-axis $\Ti$ offset is still strongly correlated with $\beta$, with a linear correlation coefficient of $r = -0.58$. Hence, electromagnetic TGLF underpredicts $\Ti$ more severely at higher $\beta$. Similar strength correlations are found with $\betae$ or total $\beta$ evaluated at other locations or with different spatial averaging. In contrast, there is no correlation of the on-axis $\Ti$ offset calculated by MMM with $\beta$, unlike for the $\Te$ predictions where MMM had a qualitatively similar but weaker dependence on $\betae$ as TGLF. The $\Ti$ offset in electrostatic TGLF simulations is also reasonably correlated with $\beta$ ($r = -0.39$), though \figref{fig:meanprof_ti} shows that this is at least in part due to higher $\Ti$ in the highest $\beta$ discharges, since the electrostatic TGLF $\Ti$ profiles have a tendency to broaden with increased $\beta$, but the predicted $\Tio$ does not drop. 
\rev{Similar to what was found when analyzing the $\Te$ profile predictions, the predicted $\Ti$ profiles are more sensitive to $\beta$ than to normalized collisionality.}

\begin{figure*}[tb]
\subfloat[\label{fig:teti_ax}]{\includegraphics[height = \thirdheight]{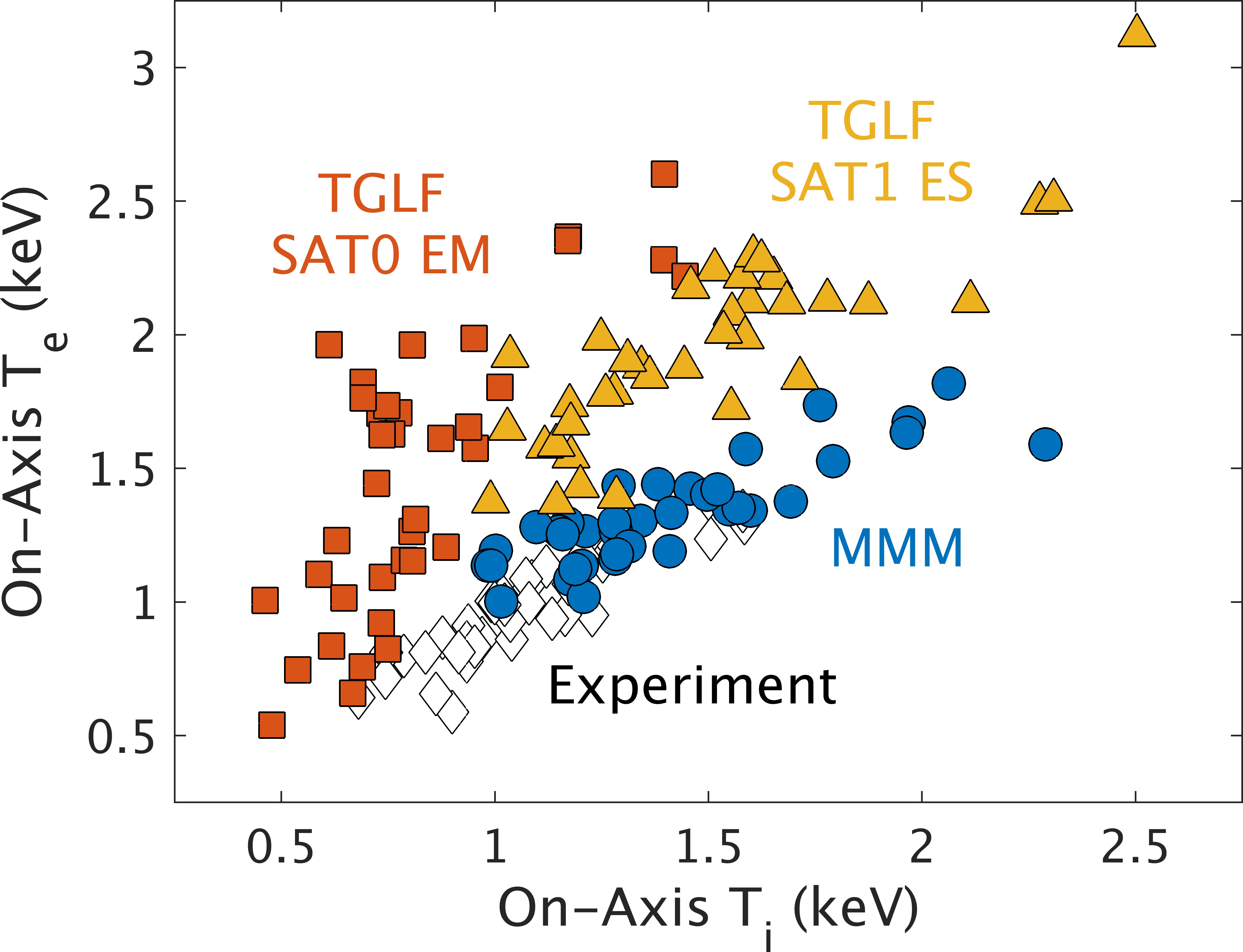}} \figsep
\subfloat[\label{fig:tepeak_expred}]{\includegraphics[height = \thirdheight]{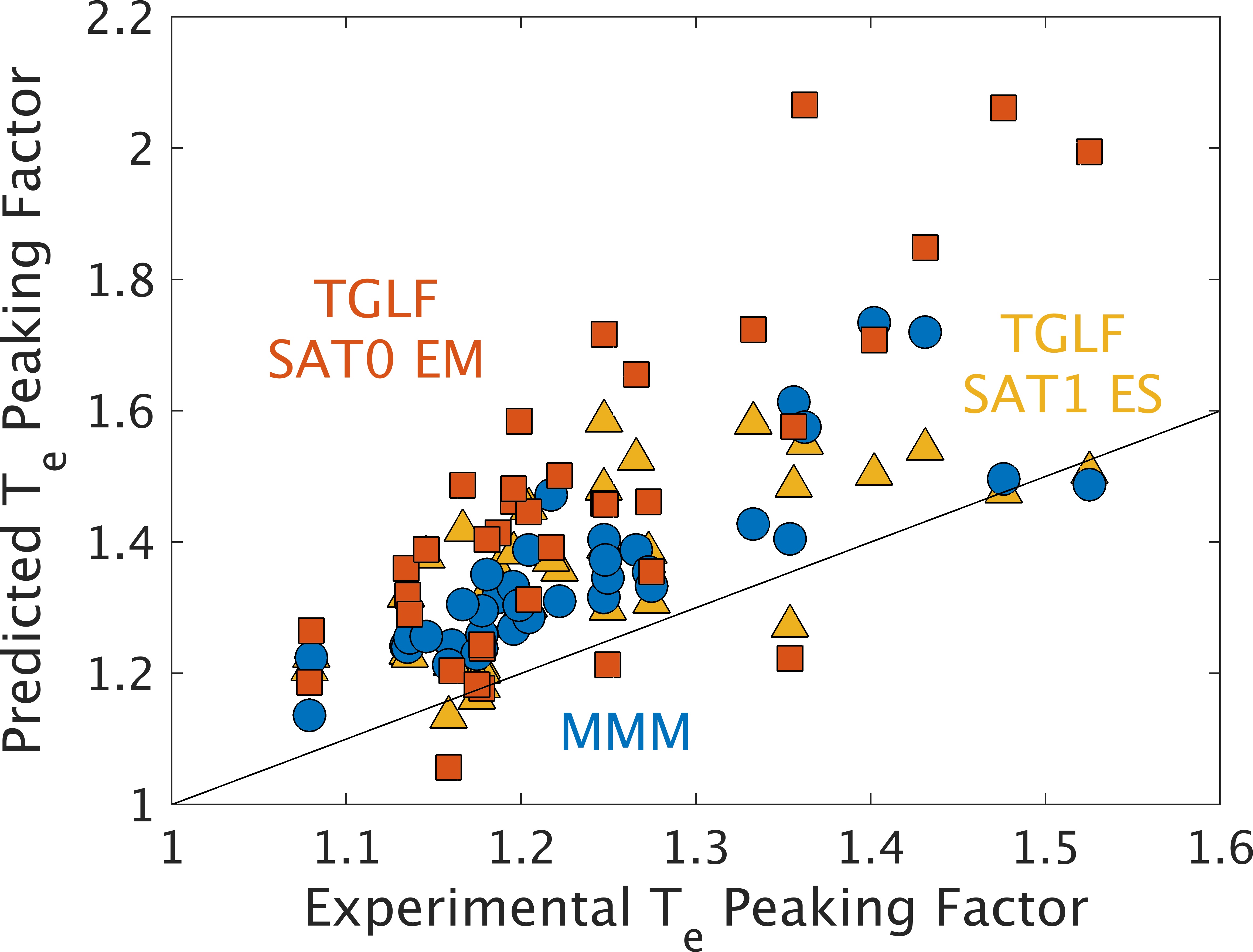}} \figsep
\subfloat[\label{fig:tipeak_expred}]{\includegraphics[height = \thirdheight]{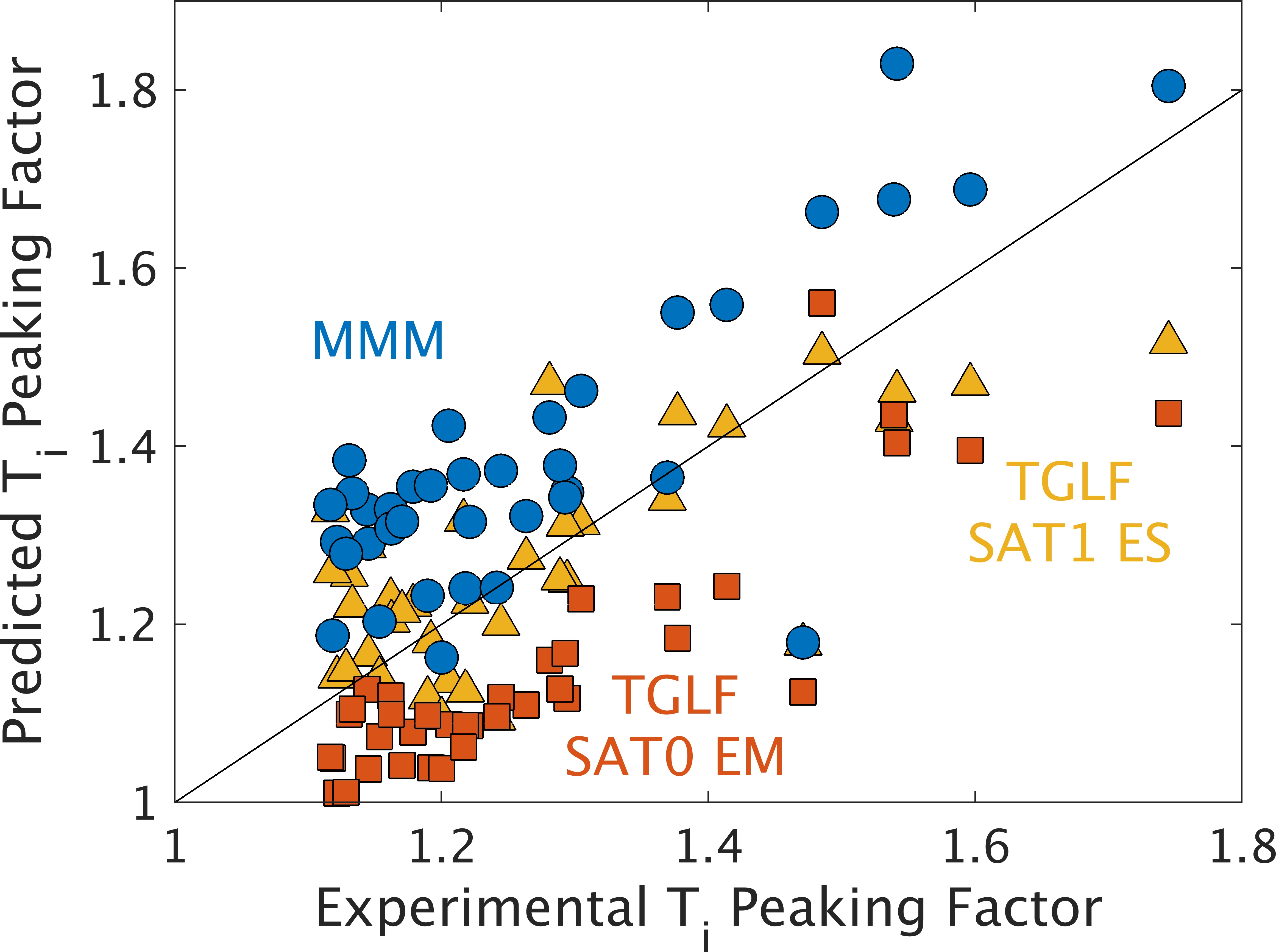}}
\caption{(a) On-axis $\Te$ \vs on-axis $\Ti$ for all examined discharges. Open diamonds shows the experimental values and filled markers correspond to simulations with different transport models. (b) and (c): comparison of experimentally measured and predicted temperature profile peaking factors for (b) $\Te$ and (c) $\Ti$. The solid lines are for reference only, indicating zero offset between experimental and predicted values. In all plots, blue points are TRANSP runs using MMM, orange use electromagnetic TGLF, and gold use electrostatic TGLF.}
\label{fig:peak_scatter}
\end{figure*}

Recalling that electromagnetic TGLF overpredicts the $\Te$ profile except in some of the high $\beta$ discharges, it turns out that electromagnetic TGLF is significantly overestimating the $\teti$ ratio in general. \figref{fig:teti_ax} compares the on-axis values of $\Te$ and $\Ti$ for all discharges for the experimental values (open diamonds) and TRANSP simulations with each transport model (filled markers). The blue points for MMM fall on nearly the same line as the experimental points, albeit shifted to higher temperatures due to overestimating the confinement. Quantitatively, the average ratio for MMM is $\teti = 0.96$, very close to the experimental average of 0.92, reproducing a similar balance between ion and electron transport channels as in the experiment. In contrast, the simulations using electromagnetic TGLF had an average $\teti = 1.81$, double the experimental value, as the combined consequence of both underpredicting $\Ti$ and overpredicting $\Te$. The electrostatic TGLF simulations fall in between with an average $\teti = 1.33$, due to overpredicting $\Te$ to a greater degree than $\Ti$. 

In addition to the difference in predicted $\teti$ ratios, comparison of \figref{fig:meanprof_te} and \figref{fig:meanprof_ti} illustrate that the TGLF predicted $\Te$ profiles tend to be characteristically more peaked than the MMM predictions while the $\Ti$ profiles from TGLF have more flattening. To quantify how peaked a profile is, a temperature profile peaking factor is defined as the ratio of the on-axis temperature to the radially averaged temperature, up to the prediction boundary at $\rho = 0.7$. Assuming non-hollow profiles, a peaking factor of 1 corresponds to a perfectly flat profile, while the maximum peaking factor for a monotonic profile is $T_0/T_\text{edge}$. A comparison of the predicted \vs experimentally measured peaking factor for the $\Te$ and $\Ti$ profiles are shown in \figref{fig:tepeak_expred} and \figref{fig:tipeak_expred}, demonstrating a clear tendency for the electromagnetic TGLF simulations to predict overly flat $\Ti$ profiles, while MMM predicts overly peaked ones with about the same absolute difference from the experimental peaking factor. The $\Ti$ profile peaking factors from both models are correlated with the experimental fits for each discharge. The $\Ti$ profile peaking factors from electrostatic TGLF simulations are less consistent, in part due to non-monotonic profiles predicted at high $\beta$. In contrast, MMM and TGLF both predict $\Te$ profiles that are too peaked relative to the experiment, as shown in \figref{fig:tepeak_expred}, with the electromagnetic TGLF profiles being the most substantially overly peaked. 

Since it is of practical importance to understand the reliability of both the ion and electron temperature predictions when using reduced models such as MMM or TGLF, it is worth emphasizing that the $\Te$ and $\Ti$ profile predictions are strongly coupled in TRANSP. Specifically, there is nontrivial ion-electron energy coupling in the examined NSTX discharges, proportional to $\Ti - \Te$ and resulting from collisions. Hence, even though the ion energy transport is dominantly neoclassical such that one might not expect the choice of turbulent transport model to have a significant effect, the predicted $\Ti$ profiles can nonetheless vary due to differing turbulent electron transport between the models. The coupling between $\Te$ and $\Ti$ profile predictions is examined in greater detail in \citeref{Lestz2025pre2} by comparing the resulting $\Te$ and $\Ti$ profiles from MMM simulations that predict both $\Te$ and $\Ti$ simultaneously (as in this work) \vs individually. 

\subsection{Stored Energy and Confinement}
\label{sec:storedenergy}

A related quantity for assessing how well each transport model captures the experimental confinement is the plasma's total stored energy. Since the temperature profiles are only being predicted for $\rho = 0 - 0.7$, the stored energy examined here is likewise only volume integrated up to $\rho = 0.7$, in order to exclude the large contribution at larger radius due to large plasma volume, which is not predicted in these simulations. As in the individual profile predictions, electrostatic TGLF has much worse agreement with the experiment for the stored energy than the other models. For both MMM and electromagnetic TGLF, the predicted stored energy more closely matches the experiment than either the $\Te$ or $\Ti$ profiles do individually -- with median error of $18 \pm 8\%$ for MMM and $17 \pm 14\%$ for TGLF. This may be due to at least two reasons. First, the beam ion density is not being predicted by MMM, but is rather using the same Monte Carlo calculation as in the fully interpretive TRANSP runs \cite{Pankin2004CPC}. While there may be slight differences in the calculated beam density due to the different resulting temperature profiles, there is very low neutral beam shine-through in these high performing NSTX plasmas, such that the total stored energy in the beam power is comparable for the experimental and predicted temperature profiles. Since the fast ion stored energy makes a significant contribution to the total plasma stored energy in NSTX (routinely 10 - 60\% \cite{Fredrickson2014NF}, on average 20\% for the discharges examined here), this component being similar for the interpretive and predictive TRANSP runs inherently reduces the error in the prediction. Second, whereas the $\Te$ and $\Ti$ profiles tend to have their largest error on the axis since the predicted errors accumulate when solving \eqref{eq:pt_te} inwards from the boundary that is set to the experimental temperature at $\rho = 0.7$, the region near the axis only makes a small contribution to the stored energy due to the small plasma volume there. Conversely, the largest contribution comes from the largest radius included in the integration, where $\Te$ and $\Ti$ are still close to their experimental values due to being so close to the fixed boundary condition. Hence it is unsurprising that the stored energy has better experimental agreement than either of the temperature profiles. 

\begin{figure}[tb]
\subfloat[\label{fig:utotl_expred}]{\includegraphics[height = \thirdheight]{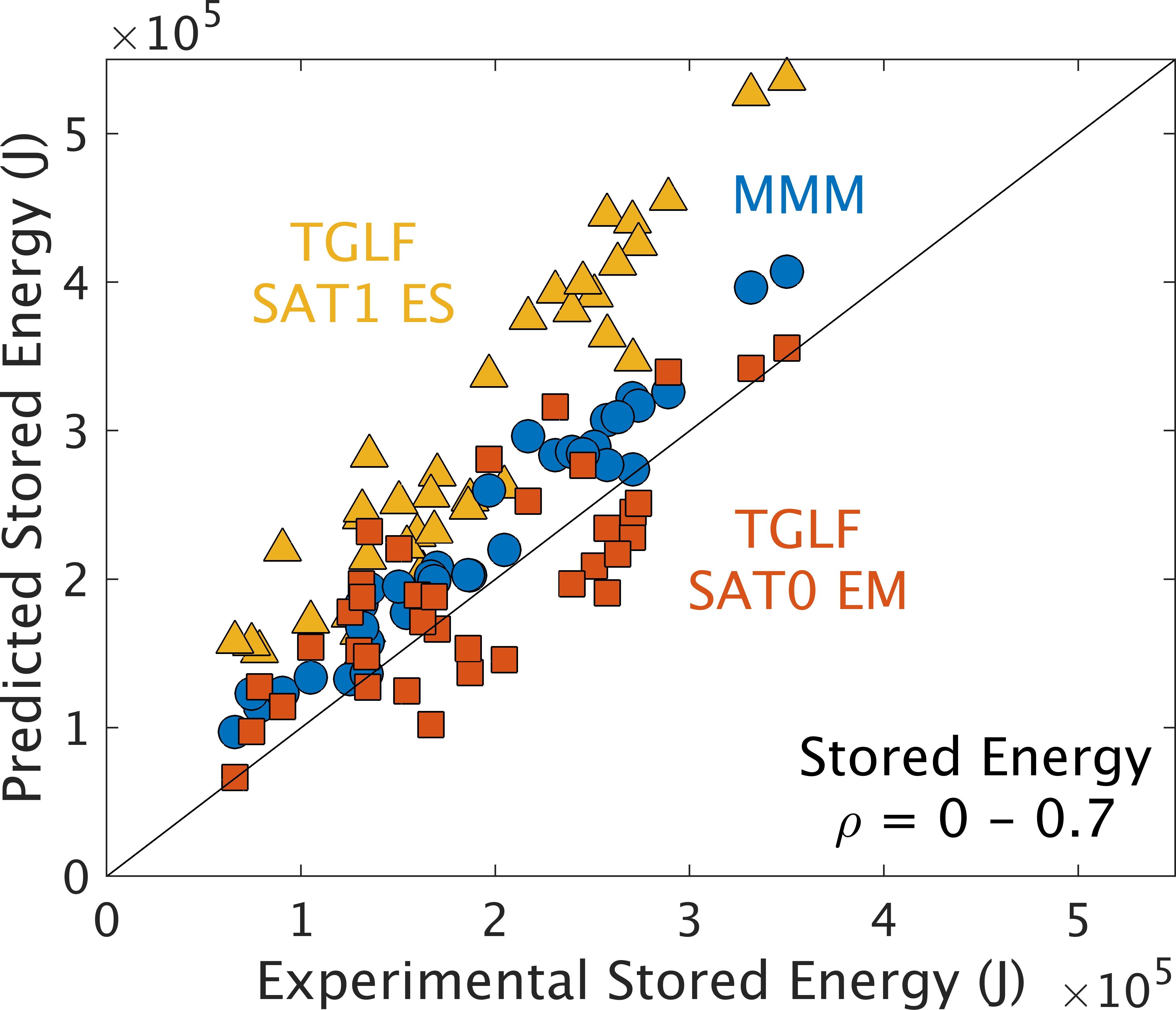}} \\
\subfloat[\label{fig:tauea_expred}]{\includegraphics[height = \thirdheight]{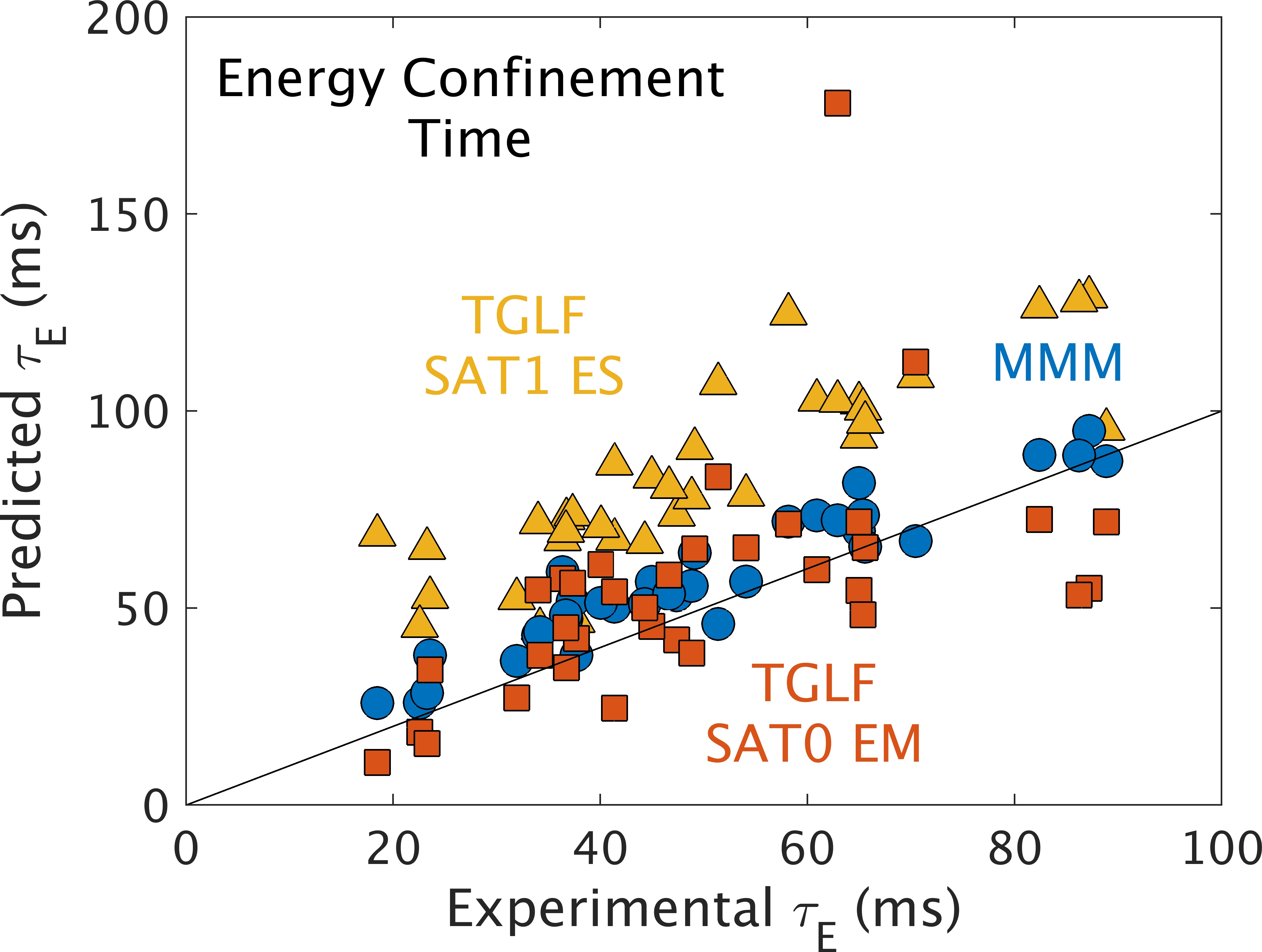}}
\caption{Comparison of experimentally measured and predicted quantities characterizing energy confinement in high performing NSTX discharges: (a) stored energy, integrated from $\rho = 0 - 0.7$ (prediction boundary) and (b) energy confinement time. In both plots, blue points are TRANSP runs using MMM, orange points correspond to electromagnetic TGLF, and gold points correspond to electrostatic TGLF. The solid lines are for reference only, indicating zero offset between experimental and predicted values.}
\label{fig:mmm_tglf_scatter}
\end{figure}

As shown in \figref{fig:mmm_tglf_scatter}, all three models have a high correlation between the experimentally measured \vs predicted stored energy and energy confinement time when comparing across all examined discharges. Not shown are two outlier discharges where the energy confinement time calculated by TRANSP was unphysically large, even in the interpretive simulations without any turbulent transport model. Predictive TRANSP simulations with MMM and electrostatic TGLF overpredict the stored energy in the vast majority of examined discharges, while the electromagnetic TGLF simulations are split about evenly between calculating too high or low stored energy, due to competing tendencies to overpredict $\Te$ and underpredict $\Ti$ relative to experimental values. Quantitatively, the global energy confinement time calculated by MMM 
is $15\% \pm 10\%$ higher than the experimental value, when taking the median and half interquartile range across all investigated discharges, compared to a median RMSE of $23\% \pm 15\%$ for electromagnetic TGLF.    

\subsection{Influence of Microtearing Modes}
\label{sec:mtm}

\begin{figure}[tb]
\includegraphics[height = \thirdheight]{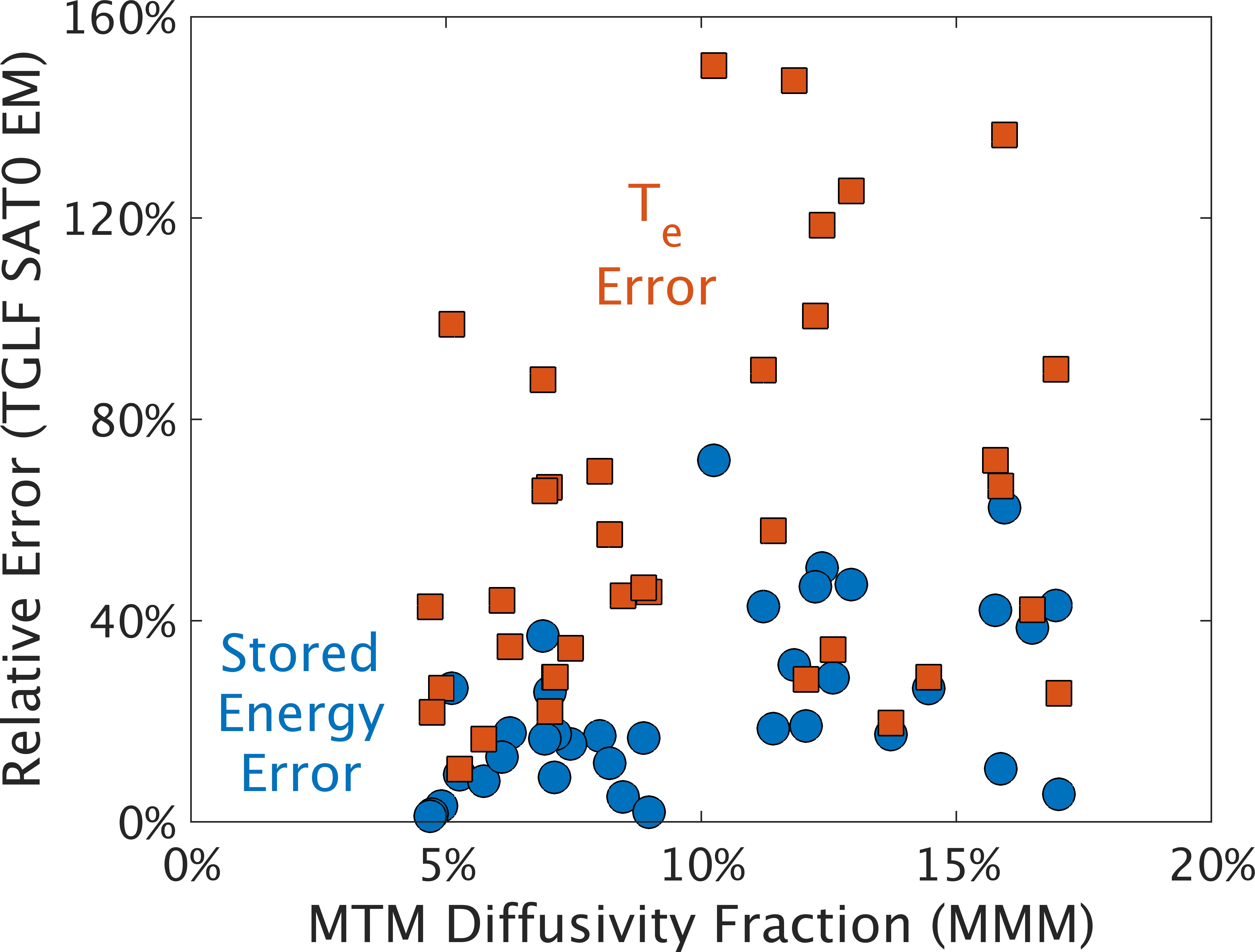}
\caption{Dependence of relative errors in electromagnetic TGLF predictions for stored energy (blue) and electron temperature (orange) on the fraction of electron energy diffusivity attributed to MTMs by MMM. Both the TGLF errors and MMM diffusivity fraction are averaged over the prediction region $\rho = 0 - 0.7$.}
\label{fig:mtm_tglf}
\end{figure}

\rev{
One important difference between MMM and TGLF is that the TGLF model can not accurately capture MTMs, which dominantly drive electron transport. As discussed in \citeref{Lestz2025pre2}, MMM predicts that MTMs can drive substantial transport near the plasma edge in the high performing NSTX discharges that were simulated, especially at high $\betap$. Hence, it is natural to consider whether the error in the TGLF $\Te$ predictions are due to the missing MTM physics. To investigate this, \figref{fig:mtm_tglf} plots the RMS error in both the electromagnetic TGLF-predicted stored energy and electron temperature profiles against the fraction of electron energy diffusivity attributed to MTMs by MMM, spatially averaged over the prediction region $\rho = 0 - 0.7$. The correlation is stronger for the stored energy ($r = 0.51$) than $\Te$ ($r = 0.32$), though both are clearly weaker than the $\beta$ dependence previously discussed in \secref{sec:te}. Moreover, these correlations only emerge when considering the MTM diffusivity averaged over all predicted radii, despite the fact that MMM predicts that the MTM contribution to the electron heat diffusivity is mostly at large radius. Note that while \figref{fig:mtm_tglf} indicates that MMM predicts that the MTM contribution is at most 20\% of the total diffusivity when averaging over $\rho = 0 - 0.7$, several of the modeled shots have an MTM diffusivity fraction exceeding 50\% in the region $\rho = 0.6 - 0.7$.
The errors in TGLF's electron temperature predictions are uncorrelated with the MTM diffusivity calculated by MMM in this edge region. Thus, while $\beta$ effects have a strong influence on TGLF's overprediction of the on-axis electron temperature, the lack of MTMs in the TGLF model appears to play a much more minor role.
}

\begin{figure}[tb] 
\subfloat[\label{fig:133964_xke}]{\includegraphics[height = \thirdheight]{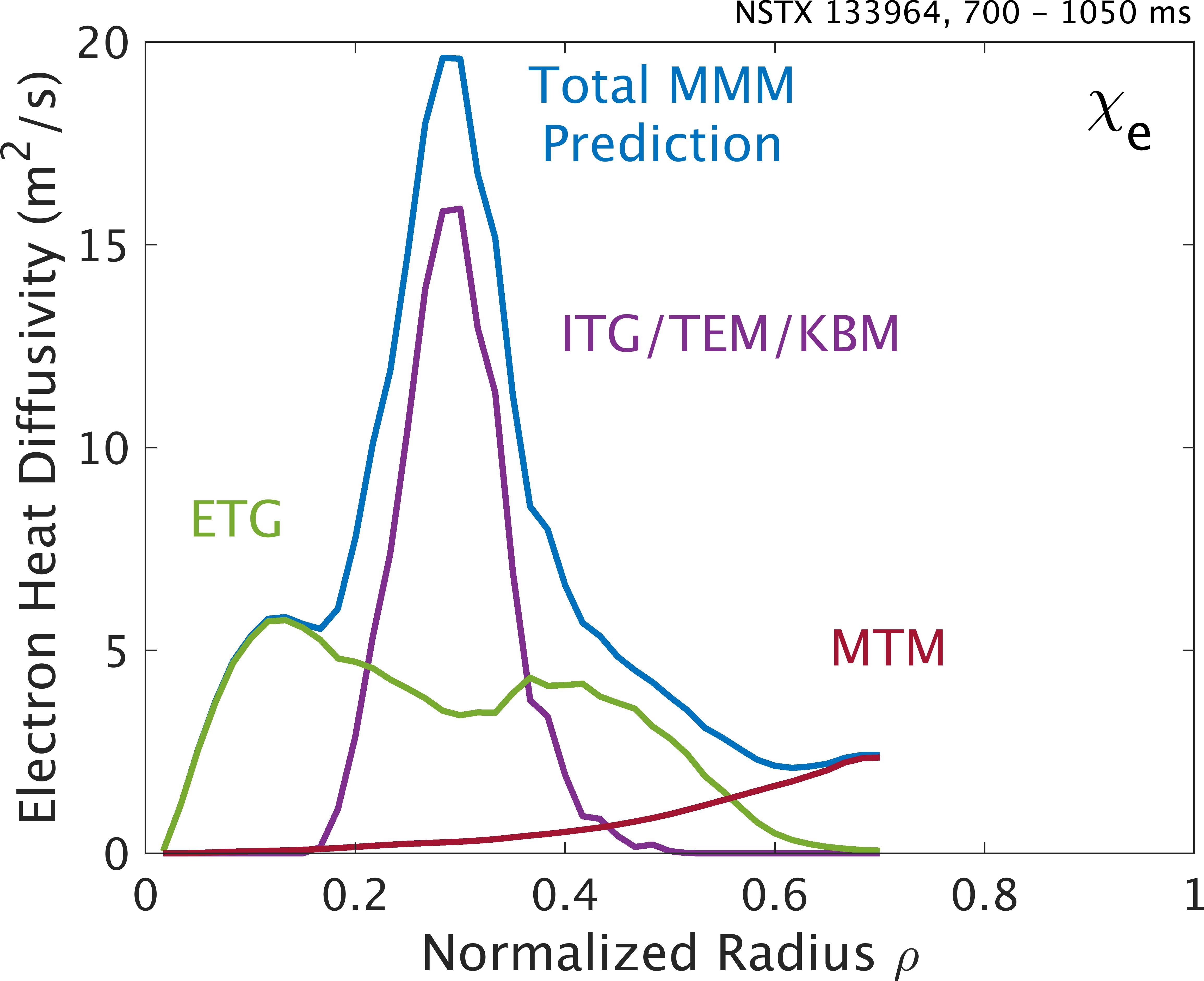}} \\
\subfloat[\label{fig:133964_sub_te}]{\includegraphics[height = \thirdheight]{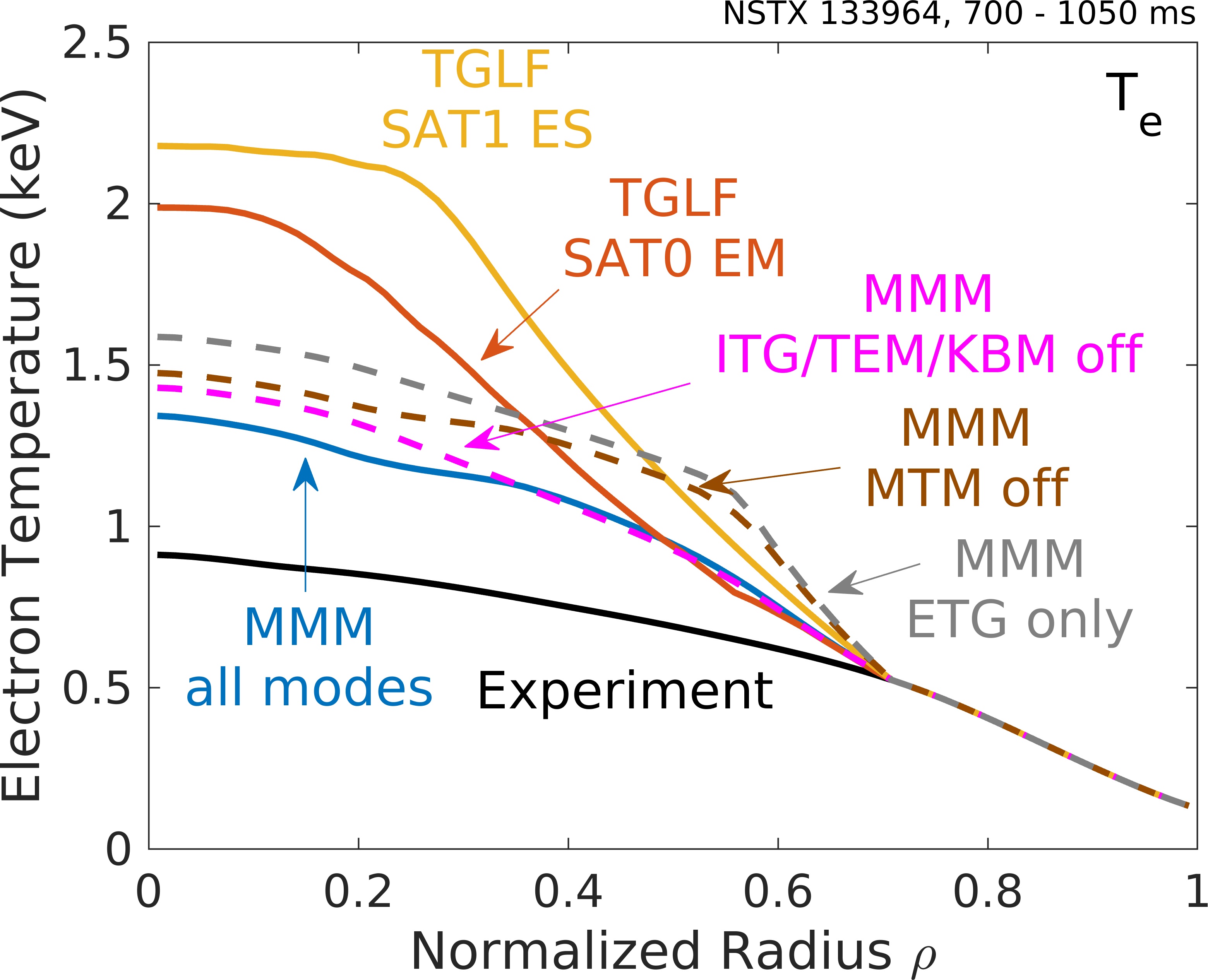}} 
\caption{(a) Electron heat diffusivity profile predicted by each submodel of MMM. Green shows the ETG contribution, purple shows the contribution from the Weiland model (ITG/TEM/KBM), and red the MTMs. Blue indicates the total diffusivity, the sum of each contribution. (b) Electron temperature profiles predicted by TRANSP simulations using MMM (solid blue), electromagnetic TGLF (solid orange), and electrostatic TGLF (solid gold). Dashed curves show separate TRANSP simulations where different instability models were disabled in MMM. Magenta shows a simulation where only the Weiland model is disabled. Brown shows a simulation where only the MTM model is disabled. Gray shows a simulation where both the Weiland and MTM models are disabled, leaving only the electromagnetic ETG model. Black shows the experimental $\Te$ profile.} 
\label{fig:133964sube}
\end{figure}

\rev{
As a concrete example, consider NSTX discharge 133964. This was a high $\betap$, high performing discharge that had the lowest flat-top average loop voltage of any NSTX discharge and achieved a sustained $\betan > 4.5$ \cite{Gerhardt2011NFcur}, where $\betan = \beta a B_T/I_p$ and $I_p$ is the plasma current. The chosen analysis window was $700 - 1050$ ms, towards the end of the current flat top. As shown in \figref{fig:133964_xke} for discharge 133964, MMM predicts that ETGs dominate the electron heat transport near the axis, for $\rho < 0.2$. Then from about $\rho = 0.2 - 0.4$, the heat transport becomes dominated by the diffusivity predicted by the Weiland model, which includes ITG/TEM/KBM, and other modes. At even larger $\rho$, MTMs begin to become important, and eventually dominant for the heat transport at $\rho > 0.6$. The relative importance of each instability with respect to radial location in this discharge is representative of the typical case when considering all discharges, which is discussed in further detail in \citeref{Lestz2025pre2}. Moreover, \citeref{McClenaghan2025PPCF} recently studied this same NSTX discharge in detail with electromagnetic TGLF and linear gyrokinetics in a time slice flux matching analysis. There it was found that KBMs were not predicted to be unstable near the pedestal unless $\betae$ was scaled up beyond its experimentally measured value for this shot, consistent with the MMM predictions that the transport induced by KBMs is restricted to $\rho = 0.2 - 0.4$ for discharge 133964. 
}

\rev{
The $\Te$ profile predictions for this discharge are shown in \figref{fig:133964_sub_te}. The solid curves show TRANSP simulations simulations using MMM (blue), electromagnetic TGLF (orange), and electrostatic TGLF (gold), representing one of the 9 NSTX discharges that are averaged over in the temperature profiles shown in \figref{fig:te_bmed}, since this discharge has $\betae = 0.089$ on axis. Overall, the $\Te$ profile predictions for this discharge are fairly representative of the averages shown in \figref{fig:te_bmed}, with TGLF overpredicting $\Te$ to a greater degree than MMM does. Additional TRANSP simulations were performed with MMM that disabled different combinations of submodels for instabilities to isolate their effect on the temperature profile predictions. These $\Te$ profile predictions are shown as dashed curves in \figref{fig:133964_sub_te}. Disabling the Weiland model (dashed magenta) -- equivalent to artificially stabilizing all ITG, TEM, and KBM transport -- barely affects the $\Te$ relative to the MMM simulation where all models are enabled (solid blue). Although \figref{fig:133964_xke} demonstrates that there is significant transport predicted to be caused by these instabilities near $\rho = 0.3$, this contribution to the diffusivity is relatively narrow in space. Hence, while the ITG/TEM/KBM turbulence is predicted to flatten the $\Te$ gradient in that region, it has a relatively small effect on the overall $\Te$ profile. 
}

\rev{
When disabling the MTM model (dashed brown), the predicted electron temperature profile in the absence of MTMs is much too steep relative to the fitted experimental data in the edge region where MTMs dominate the transport, indirectly lifting up the predicted core $\Te$. However, since the overly steep gradient in MMM simulations that artificially suppress MTMs is restricted to $\rho > 0.5$, the $\Te$ profile predicted by MMM without MTMs is only modestly higher than MMM with MTMs. When both the Weiland and MTM model are disabled in MTM, leaving only ETG transport, the predicted $\Te$ profile raises slightly further due to removing transport generated by another group of instabilities. Hence, this set of simulations demonstrates that the absence of MTM-induced transport in TGLF is not large enough to explain the more substantial overprediction of $\Te$ by TGLF than MMM, based on how much the core $\Te$ is raised in the MMM simulations where MTM transport is suppressed. Moreover, the $\Te$ profile predicted by TGLF is significantly steeper than that predicted by any of the MMM simulations in the mid-radius region of $\rho \approx 0.2 - 0.5$, indicating that TGLF is underpredicting the electron transport in this region, far deeper in the core than the MTMs are expected to be relevant in this discharge. 
}

\begin{table*}\centering
\begin{tabular}{cccc}
\hline\hline
Transport Model & \twotab{CPU Hours per}{Simulated Second} & \twotab{CPU Seconds per}{Newton Iteration} & \twotab{Newton Iterations per}{Simulated Second} \\ 
\hline
MMM           & $0.7  \pm 2.9$  & $1.5 \pm 5.4$  & $2,814  \pm 970$   \\ 
TGLF SAT0 EM  & $3,201 \pm 8,339$ & $710 \pm 7,716$ & $11,758 \pm 7,961$  \\ 
TGLF SAT1 ES  & $5,791 \pm 5,764$ & $908 \pm 2,028$ & $14,700 \pm 11,453$ \\
\hline\hline
\end{tabular}
\caption{Statistics for computational cost for all simulated NSTX discharges. Reported as median $\pm$ half interquartile range of each quantity.}
\label{tab:allcpu}
\end{table*}

\section{Computational Cost}
\label{sec:cputime}

\begin{figure}[tb]
\includegraphics[height = \thirdheight]{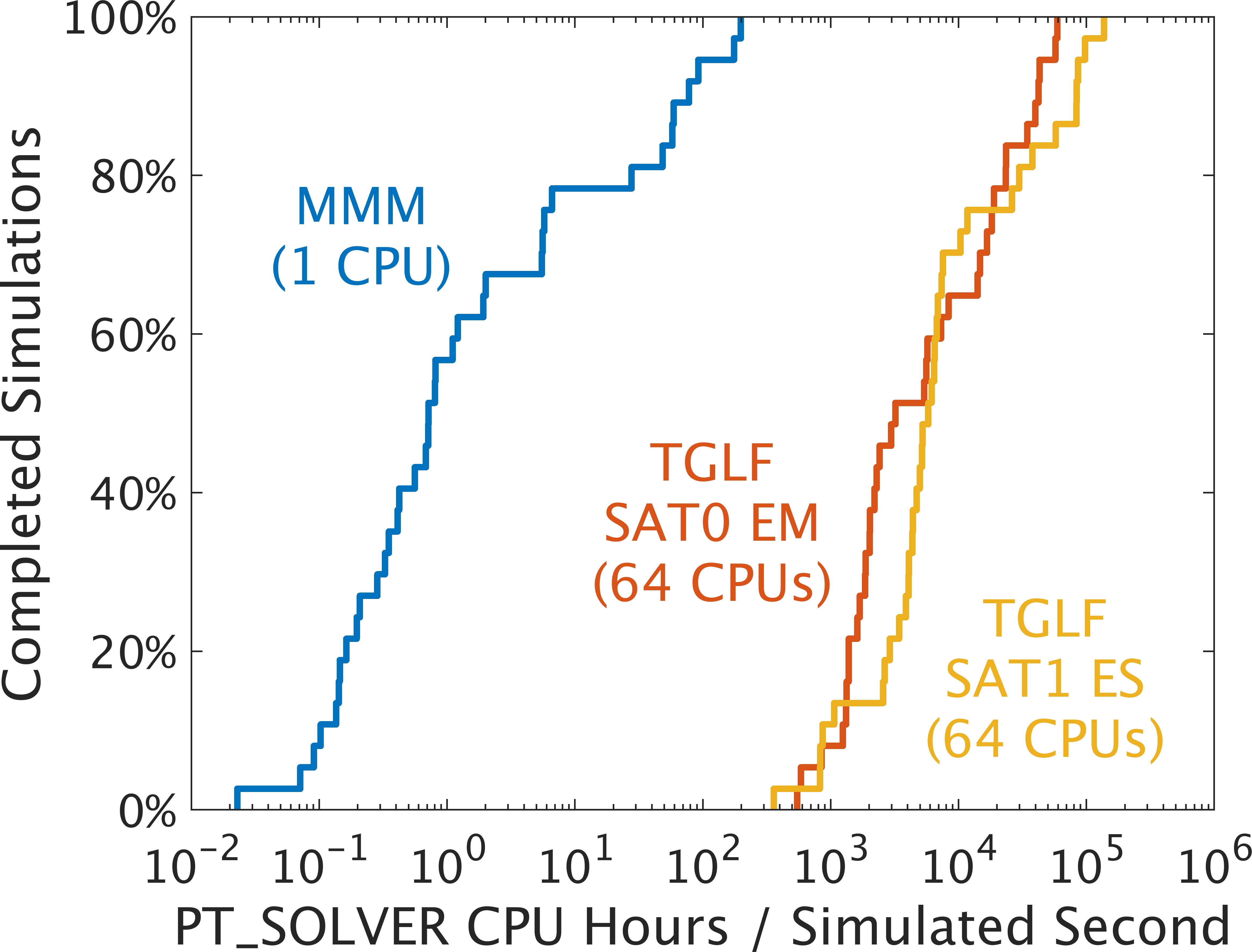}
\caption{Comparison of computational cost of different transport models within \ptsolver, quantified as CPU hours spent on the \ptsolver portion of the TRANSP simulation, normalized to a full second of simulated plasma time. The curves show the fraction of the simulations for all of the shots that have completed within a certain amount of time. Blue: MMM, orange: electromagnetic TGLF, gold: electrostatic TGLF.} 
\label{fig:cputime}
\end{figure}

To contextualize the performance of MMM and TGLF, it is important to also consider the relative differences in computation time. In general, the predictive TRANSP simulations that used MMM required orders of magnitude less CPU time than when using TGLF for the same discharge, though with significant variance for both models. \figref{fig:cputime} shows the number of CPU hours spent in the \ptsolver portion of the TRANSP simulations for each transport model, normalized to one second of simulated plasma time to make a fair comparison across discharges which had analysis windows of different durations. A cumulative probability distribution is used to represent the fraction of shots with simulations that have completed in less than a specified computational cost. Note the logarithmic axis that is necessary to capture the multiple orders of magnitude range of computation times, even for a specific transport model. This quantity also accounts for the fact that MMM was run in serial on a single CPU while the simulations with TGLF used 64 CPUs in parallel. Hence the wall clock time for the TGLF simulations would be a factor of 64 smaller than the CPU time shown in \figref{fig:cputime}. The reported CPUs times are for TRANSP simulations performed on the PPPL cluster. 

Statistics for the CPU hours per simulated second are listed in \tabref{tab:allcpu}, with values reported as median $\pm$ half interquartile range. Notably, the interquartile ranges are comparable to or greater than the medians, reflecting the long tail of the distribution of computational costs. Compared to MMM, TGLF is orders of magnitude more costly. The typical CPU time for TGLF to simulate one second of a discharge within \ptsolver is thousands of hours, whereas the vast majority of simulations with MMM required less than 10 hours, often significantly less. It is worth repeating that TGLF was run in parallel and usually simulated around 200 ms of a discharge, such that the wall clock time of a typical simulation with TGLF was on the order of $50 - 100$ hours in total on the PPPL cluster, not thousands like it would be if run in serial for a full second.

In order to distinguish between the inherent computational cost of the transport model itself \vs the \ptsolver Newton iteration scheme, \tabref{tab:allcpu} also includes statistics for the CPU time spent on a single Newton iteration and the number of Newton iterations used per simulated second. 
TGLF's greater computational expense relative to MMM is both due to the inherent cost of the TGLF model and slower convergence of \ptsolver when using TGLF. On average, TGLF is several hundreds times slower than MMM for a single iteration and needed significantly more Newton iterations to converge in \ptsolver. This was despite using slightly  relaxed \ptsolver residual tolerances for some TGLF runs in order to allow them to complete at all. Electromagnetic and electrostatic TGLF simulations had similar computational cost, with the electrostatic ones being modestly more expensive. While the significant computational expense of TGLF was tolerable for the scope of this study, it did limit the extent to which different TGLF physics models could be tested with predictive TRANSP, and would likely become prohibitive when modeling long pulse scenarios. This computational cost could be ameliorated by the many orders of magnitude speedup provided by the TGLF surrogate models that will be discussed briefly in \secref{sec:tglf_nn}, suggesting their integration into TRANSP as a fruitful avenue for future code enhancement. 

\section{Sensitivity to TGLF Settings}
\label{sec:tglf_settings}  

Given the substantially different temperature profile predictions found when using the electrostatic SAT1 \vs electromagnetic SAT0 TGLF models in TRANSP, it is natural to wonder whether the different saturation rule or inclusion of electromagnetic effects is the key difference. While a supervised learning approach has recently been undertaken to study the sensitivities of TGLF accuracy in time slice flux matching analysis for DIII-D and MAST-U \cite{Neiser2024APS}, a careful study of the relative performance of different combinations of TGLF input settings in time-dependent TRANSP simulations is computationally infeasible. Instead, this section varies pertinent TGLF settings on a much smaller subset of discharges than was used to test electromagnetic TGLF SAT0 and electrostatic TGLF SAT1 in the earlier sections of this paper. While most of the TGLF runs use the same \ptsolver settings, some of the simulations needed to relax numerical tolerances in order to allow the Newton iteration scheme to converge in a non-prohibitive amount of time. Independent of the predictive TRANSP simulations, this section briefly discusses neural net surrogate models for TGLF that have been trained on NSTX plasmas for the first time, allowing a broad, systematic survey of different TGLF settings that would otherwise be computationally infeasible. These TGLF calculations were performed entirely outside of TRANSP with a time slice flux matching transport solver, which takes a different approach from the solver within \ptsolver that was described in \secref{sec:transp}. 

\subsection{Electromagnetic Effects, Spectral Shift Model, and Sensitivity to Saturation Rules}
\label{sec:tglf_other}

Five NSTX discharges are chosen to explore several different combinations of TGLF settings and further probe the differences seen between electrostatic TGLF SAT1 and electromagnetic TGLF SAT0 in the previous sections. The specific TGLF input settings used for each simulation can be found in the MDSplus archive of the input files for the TRANSP runs listed in \tabref{tab:tglf_runs}. The additional TRANSP simulations presented in this section are performed with electrostatic SAT0 and electromagnetic SAT1, SAT2, and SAT3 saturation rules. Although higher numbers x in the SATx label indicate saturation rules that were developed more recently, and thus should benefit from more comprehensive physics models and nonlinear simulations to fit free parameters, the older saturation rules are by no means deprecated. A variety of preferences exist among TGLF users regarding which saturation rule is best-suited to which regimes even for the better-studied conventional tokamaks. 

\begin{figure*}[tb] 
\subfloat[\label{fig:133959_tglf_te}]{\includegraphics[height = \thirdheight]{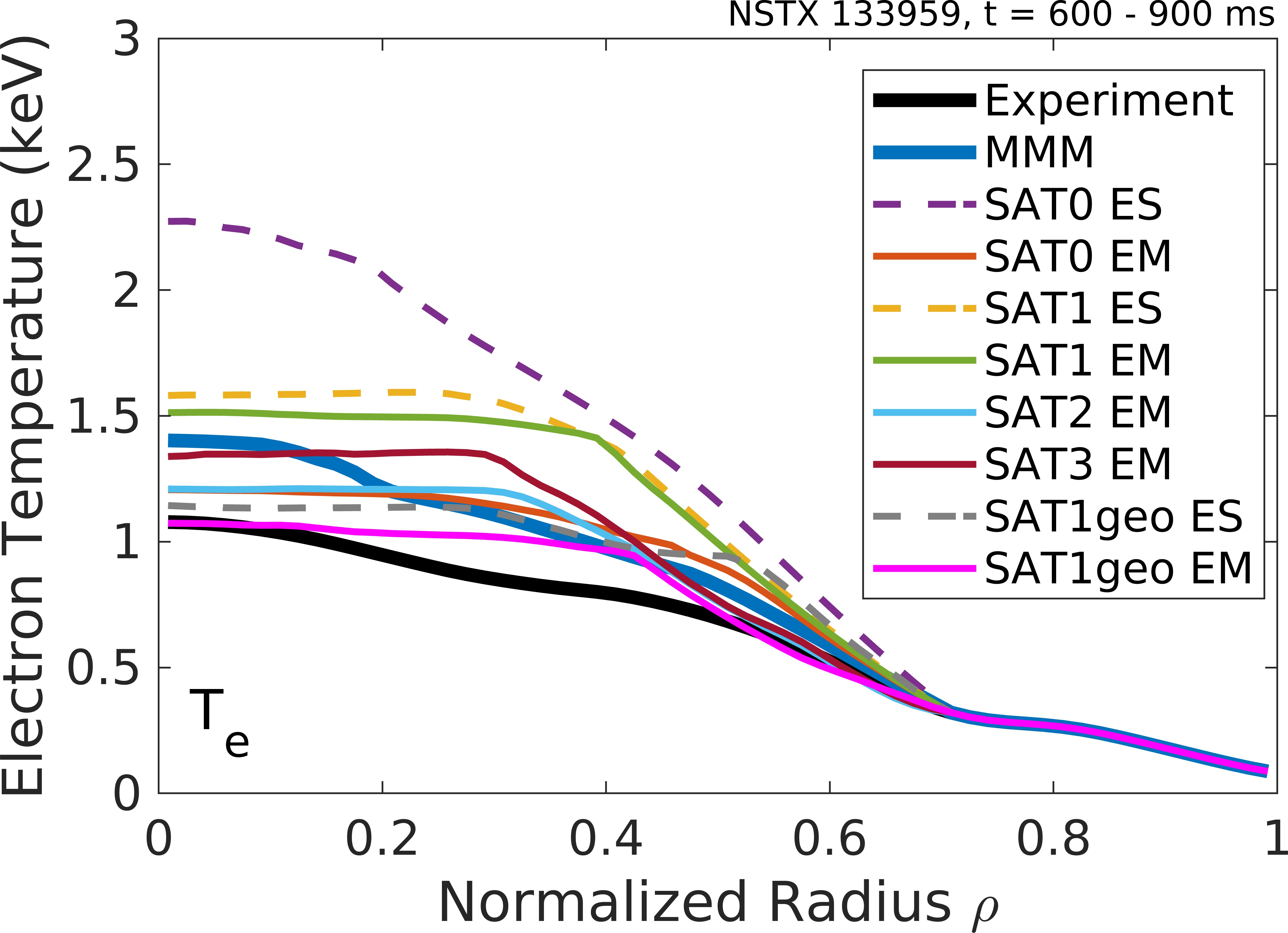}} \figsep
\subfloat[\label{fig:133959_tglf_ti}]{\includegraphics[height = \thirdheight]{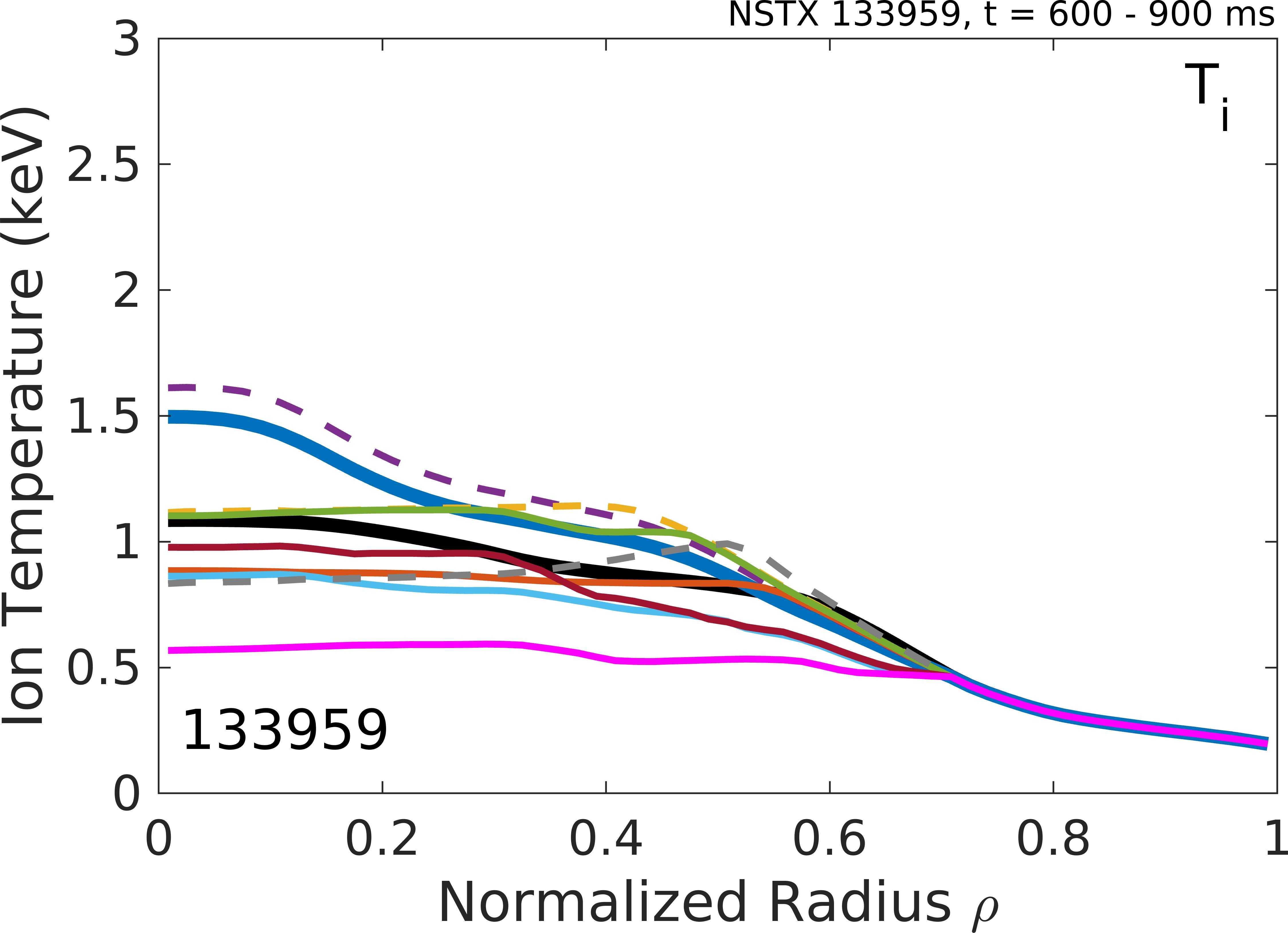}} \\
\subfloat[\label{fig:133964_tglf_te}]{\includegraphics[height = \thirdheight]{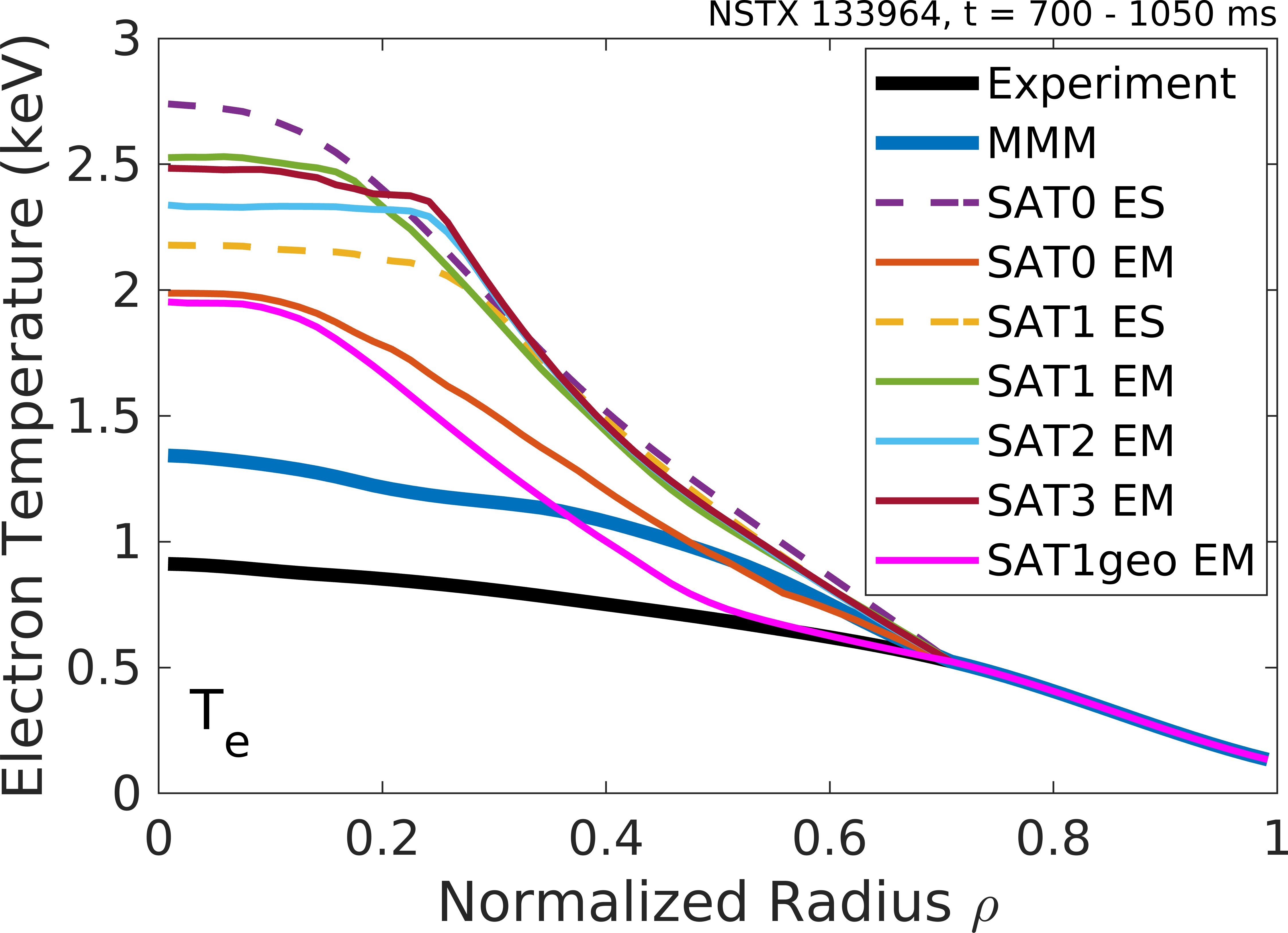}} \figsep
\subfloat[\label{fig:133964_tglf_ti}]{\includegraphics[height = \thirdheight]{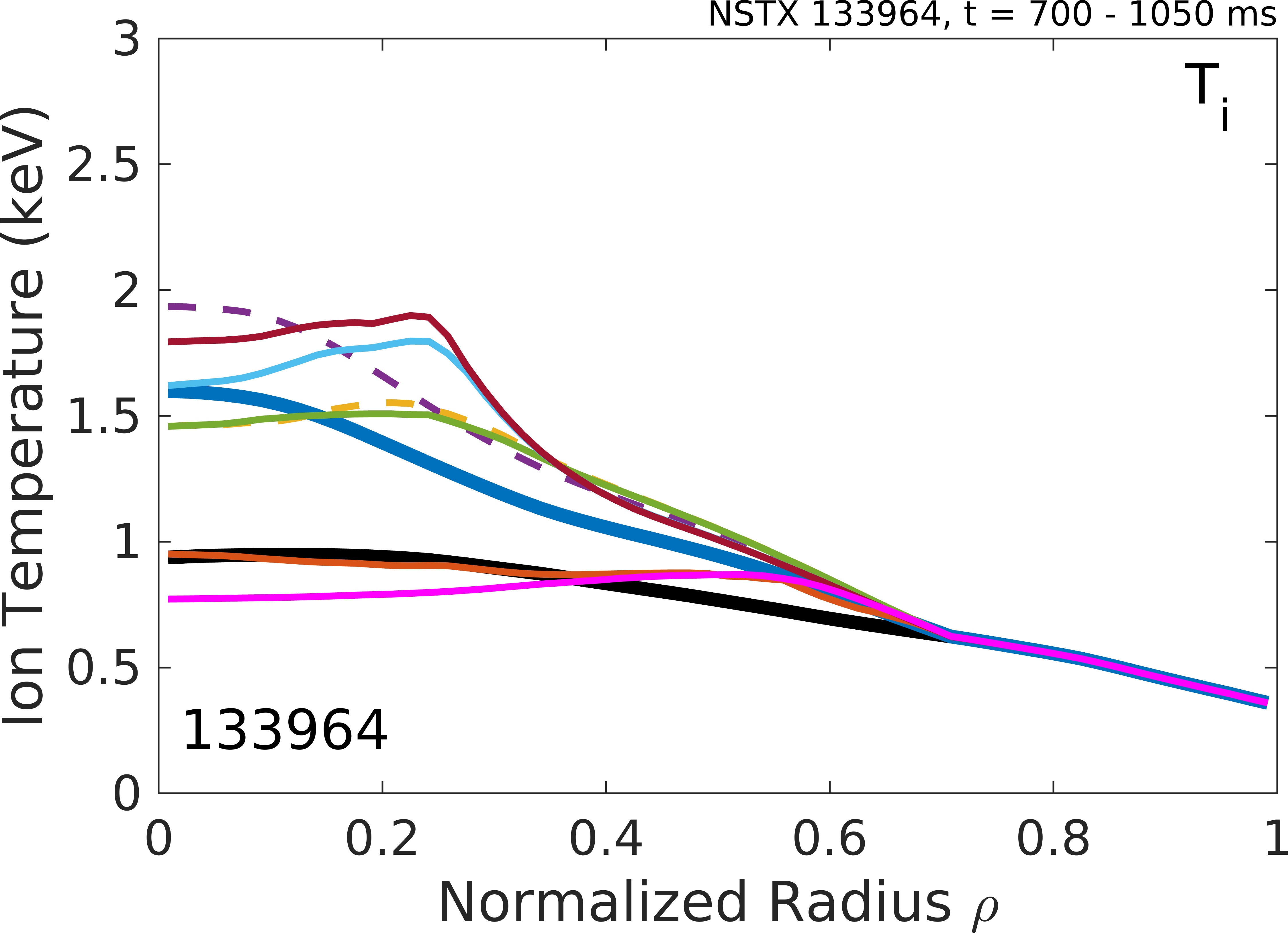}}
\caption{Electron and ion temperature profiles predicted by TGLF for discharge 133959 (top row) and 133964 (bottom row) for a variety of model settings. Thick black and dark blue curves show the experimental profile fit and MMM prediction for reference, respectively. All other curves are from separate TGLF simulations, where SATx labels the saturation rule, ES stands for electrostatic, EM for electromagnetic, and SAT1geo for runs using SAT1 with the recalibrated spectral shift model from SAT2. Dashed curves indicate electrostatic TGLF simulations and solid curves are used for electromagnetic.}
\label{fig:profiles_tglf}
\end{figure*}

A brief description of the TGLF saturation rules SAT0 and SAT1 was given in \secref{sec:intro}, but SAT2 and SAT3 have not yet been discussed. SAT2 extends the physics from SAT1 by incorporating more robust geometric effects and also recalibrates the spectral shift model that was first introduced in SAT1 \cite{Staebler2021NF,Staebler2021PPCF}. SAT3, the most recent saturation rule released to date, uses a different description of trapped electron mode saturation than for ion temperature gradient modes in order to incorporate the isotope scaling reversal observed in nonlinear gyrokinetic simulations, which was not captured by previous TGLF saturation rules \cite{Dudding2022NF,Dudding2022thesis}. Additionally, the effect of using the recalibrated spectral shift model from SAT2 \cite{Staebler2021NF} with the SAT1 saturation rule is explored in this section to isolate the effect of the change in the spectral shift model from the other differences between the SAT1 and SAT2 saturation rules.

A variety of behavior is found in response to the different TGLF model settings, though there are some tendencies that can be commented on. The $\Te$ and $\Ti$ profiles predicted by TGLF with these different model settings are shown for two example NSTX discharges in \figref{fig:profiles_tglf}, 133959 and 133964, with the experimental profiles and MMM predictions also shown for reference. Both are similar high $\betap$ discharges with large non-inductive current fractions \cite{Gerhardt2011NFcur}, where 133964 is the same example discharge that was analyzed in \secref{sec:mtm} and recently investigated with TGLF and linear gyrokinetic simulations within a flux matching framework in \citeref{McClenaghan2025PPCF}. The same set of simulations were performed for three other discharges that are listed in \tabref{tab:tglf_runs} but not shown in the figure. 

In the five examined discharges, including electromagnetic effects in TGLF often reduces the predicted $\Te$ and $\Ti$, but it is also not uncommon for it to have negligible effect or even increase the predicted temperatures. In \figref{fig:profiles_tglf}, the electrostatic TGLF simulations are indicated with dashed curves. For instance, simulations that used electrostatic SAT0 \vs electromagnetic SAT0 correspond to the dashed purple and solid orange curves, respectively. In all four of the panels, electromagnetic effects reduce the predicted temperature by $0.5 - 1$ keV. However, when making the same comparison instead with SAT1 (dashed gold for electrostatic and solid light green for electromagnetic), there is barely any change in $\Ti$ for either discharge. In comparison, discharge 133959 has a small decrease in core $\Te$ for the electromagnetic SAT1 simulation while discharge 133964 has a small increase. Hence, the substantial differences in profiles predictions from electromagnetic SAT0 \vs electrostatic SAT1 TGLF presented in \secref{sec:profpred} can not be conclusively attributed to either electromagnetic effects or the two different saturation rules based on the results in this subsection, and instead should be the subject of future work.  

Focusing on electromagnetic TGLF simulations, the solid orange, light green, cyan, and maroon curves in \figref{fig:profiles_tglf} show simulations that use SAT0, SAT1, SAT2, and SAT3, respectively. Across all five discharges, there were no consistent trends regarding which saturation rule more accurately reproduced the experimental profile, or even which one tended to predict the highest or lowest temperatures. In contrast, there is a published example where a large overprediction of the $\Te$ profile in an L mode NSTX plasma by SAT0 was greatly improved upon when using SAT1 instead (see discussion of Figure 2 in \citeref{Kaye2019NF}, which used partially electromagnetic TGLF by keeping $\dbperp$ fluctuations but suppressing $\dbpar$). Typically for the simulations presented here, varying the saturation rule had less of an effect on the predicted profiles than whether electromagnetic effects were included or which version of the spectral shift model was used. 

Lastly, employing the recalibrated spectral shift model from SAT2 in TGLF SAT1 simulations consistently had a pronounced effect in lowering both the predicted $\Te$ and $\Ti$ profiles. This combination of settings is commonly referred to as SAT1geo. For discharge 133959, the dashed gray and solid magenta curves use the recalibrated spectral shift model with SAT1 for electrostatic and electromagnetic simulations, respectively. These can in turn be compared to the dashed gold and solid light green curves, respectively, which use the same settings except using the original spectral shift model. For discharge 133964, only the electromagnetic case is available, as the electrostatic simulation with the recalibrated spectral shift model was unable to complete successfully due to a numerical issue. For both discharges shown, and also characteristic of those not shown, TGLF simulations with the recalibrated spectral shift model predict significantly lower temperatures relative to equivalent simulations where the original spectral shift model was used, for both electrostatic and electromagnetic simulations. Since TGLF SAT1 with the original spectral shift model tends to overpredict $\Te$, this temperature reduction often improves the agreement between the predicted $\Te$ profile and the experiment in those simulations, but it can also lead to a substantial underprediction in cases where the calculated profile with the original spectral shift model was not much higher than the experiment, as is the case for the $\Ti$ profile for discharge 133959. 

Although omitted from \figref{fig:profiles_tglf}, electromagnetic simulations were also performed that included $\dbperp$ fluctuations but set $\dbpar = 0$. When $\dbpar$ fluctuations are suppressed in this way, the $\Te$ profile is mildly to moderated more overpredicted than in the fully electromagnetic TGLF simulations. This is consistent with but generally less substantial than the trends found in \citeref{McClenaghan2025PPCF} which analyzed the influence of electromagnetic effects and especially the role of $\dbpar$ fluctuations in a subset of NSTX discharges presented in this paper. In that work, calculations with the TGYRO time slice flux matching transport solver \cite{Candy2009POP} using TGLF found that including electromagnetic effects substantially increased the linear growth rates of turbulent instabilities and quasilinear heat fluxes compared to electrostatic simulations. In discharge 133964, inclusion of $\dbpar$ fluctuations led to a transition of the dominant instability from MTM to KBM in linear gyrokinetic CGYRO simulations, which was associated with improved agreement in the temperature profile predictions when the linear CGYRO spectra and growth rates were used in place of those from TGLF when performing time slice flux matching with the QLGYRO solver \cite{Patel2021thesis,McClenaghan2025PPCF}. 

To summarize, the most consistent trend found from varying the TGLF settings on several discharges was that using the recalibrated spectral shift model usually decreases predicted temperatures by a moderate to substantial amount. Electromagnetic effects usually reduce the predicted temperatures as well, though this effect is more consistent when using SAT0 than SAT1, as the influence in SAT1 simulations tends to be less significant and can even lead to higher predicted temperatures. Given the tendency for TGLF to overpredict $\Te$ in NSTX, the inclusion of either electromagnetic effects or the recalibrated spectral shift model usually results in predicted $\Te$ profiles that more closely match the experiment, but they can also lead to an excess of transport and consequent underprediction of $\Te$. Varying the saturation rule for electromagnetic simulations can lead to some variation, but the consequences on the predicted profiles are idiosyncratic and not possible to generalize from this limited investigation. Unfortunately, the significant differences discussed in \secref{sec:profpred} between electromagnetic SAT0 and electrostatic SAT1 TGLF simulations performed on the large database of NSTX discharges could not be conclusively attributed to electromagnetic effects or different saturation rules. Electrostatic SAT0 simulations with the original spectral shift model stand out in these five discharges as having the largest disagreement with experiment due to strongly overpredicting both $\Te$ and $\Ti$ in all five cases. Despite the common assumption that ion transport in NSTX is dominated by neoclassical transport, varying the TGLF settings leads to an enormous variation in the predicted $\Ti$ profile as an indirect effect from sensitivities of the $\Te$ profile on the TGLF settings. No specific combination of TGLF settings was found which consistently performed better than MMM across these five discharges, and it is worth noting the wide range of responses to different TGLF settings illustrated in \figref{fig:profiles_tglf} even for two similar NSTX discharges. Given its fairly consistent effect in this small set of NSTX discharges, predictive TRANSP simulations with TGLF that use the recalibrated spectral shift model should be explored further in the future. 

\newcommand{\nnheight}{4cm}
\newcommand{\wtheight}{\nnheight}
\begin{figure*}[tb]
\subfloat[\label{fig:nnbars}]{\includegraphics[height = \nnheight]{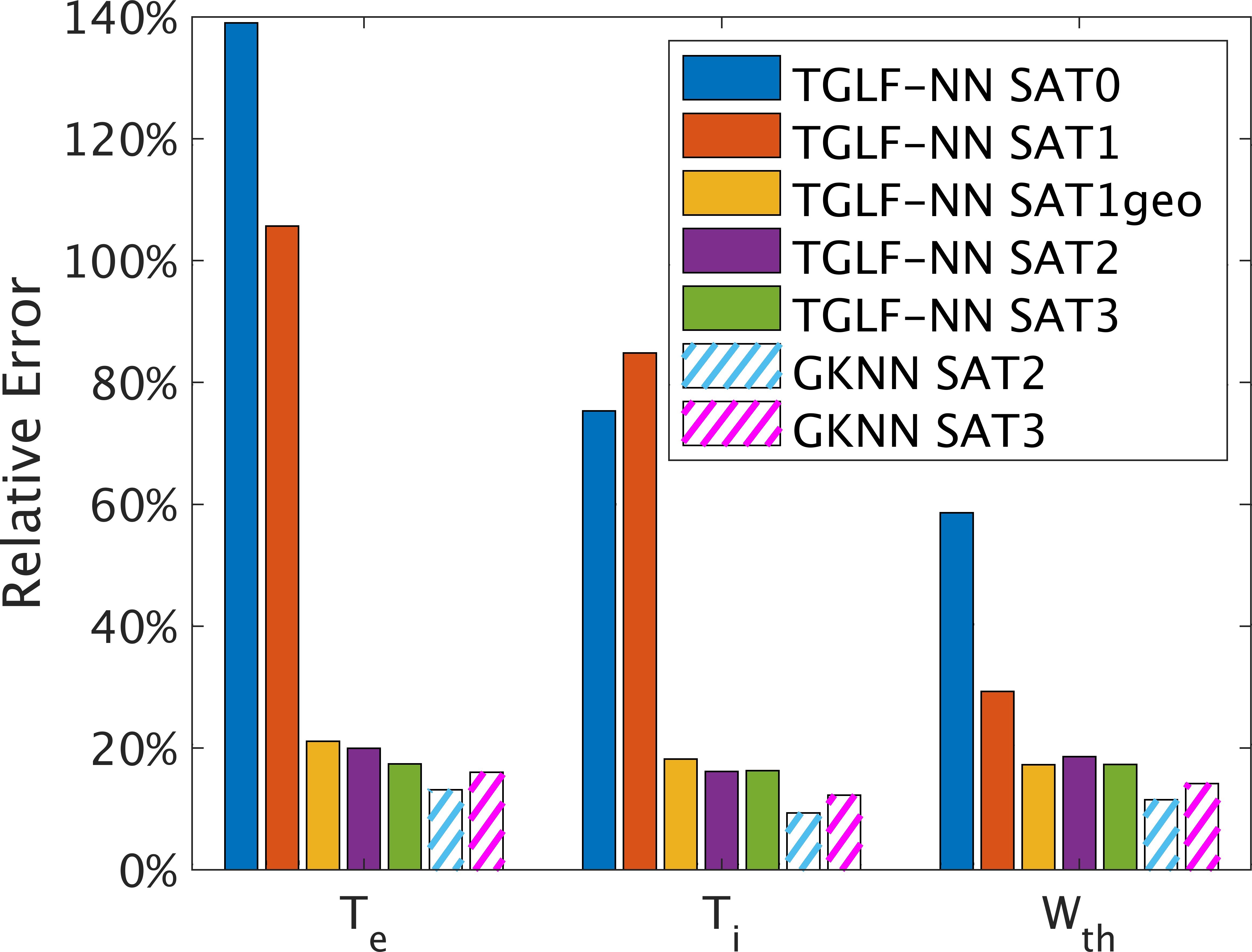}} \figsep
\subfloat[\label{fig:nnprofte}]{\includegraphics[height = \nnheight]{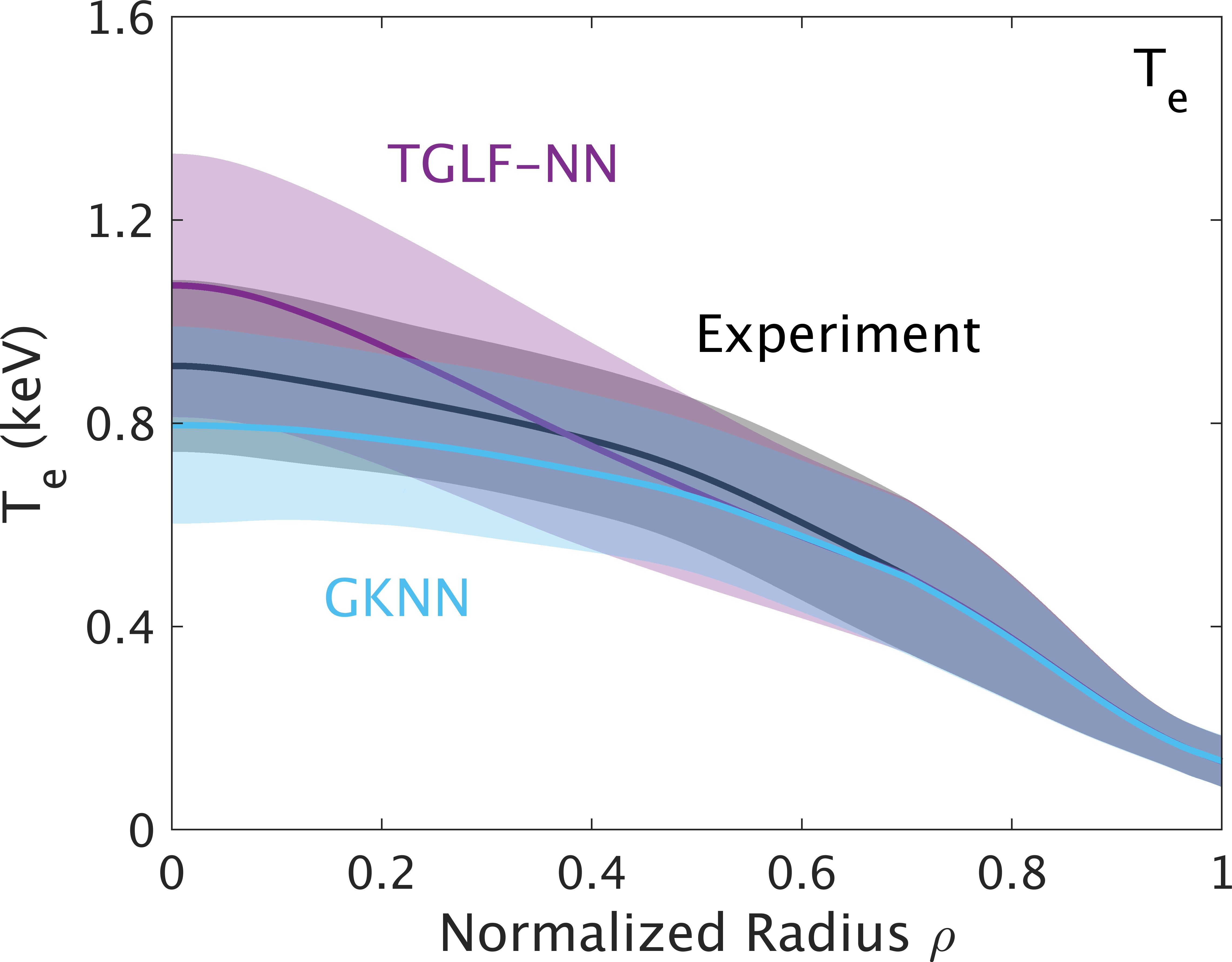}} \figsep
\subfloat[\label{fig:nnprofti}]{\includegraphics[height = \nnheight]{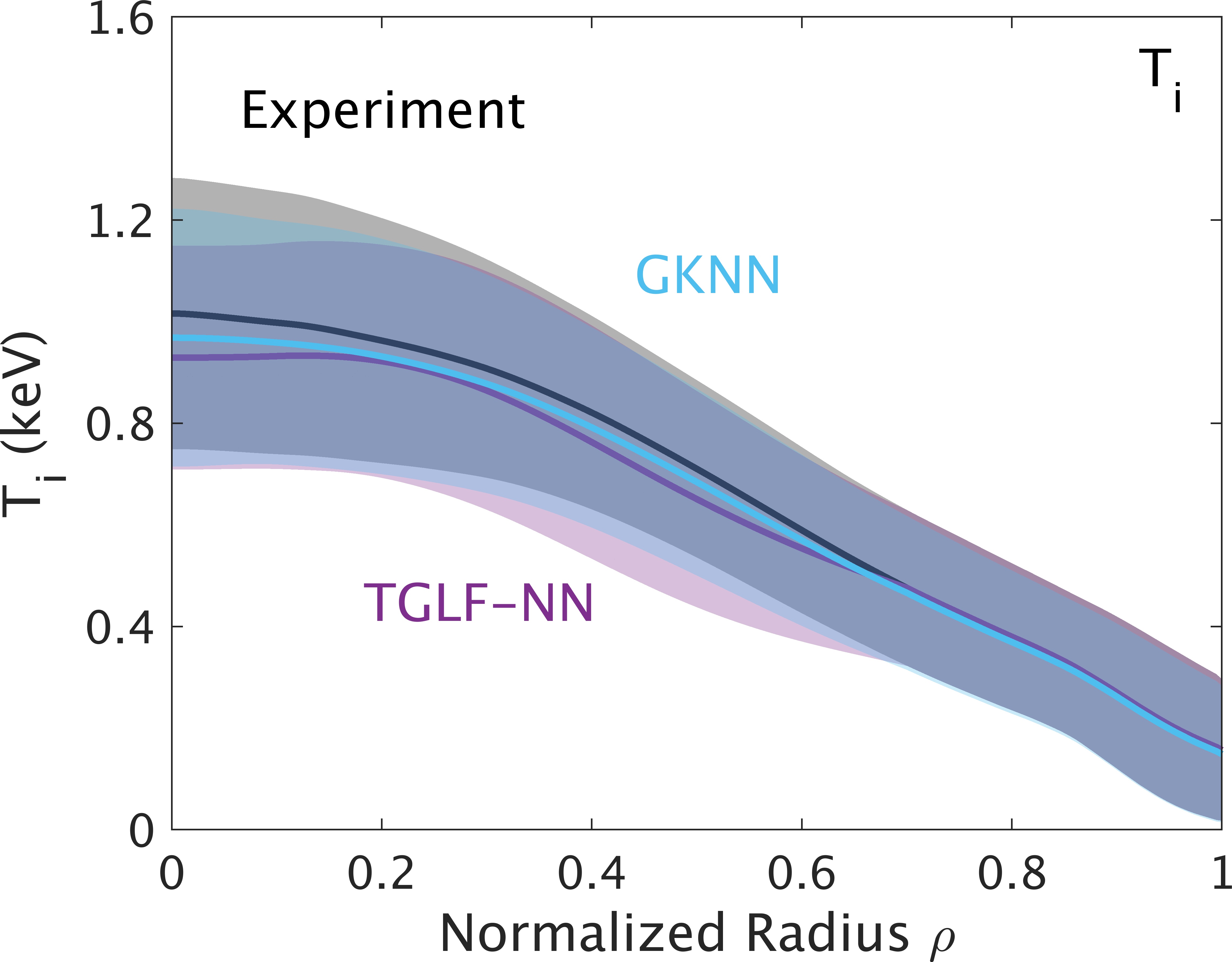}} \\
\subfloat[\label{fig:nnwth}]{\includegraphics[height = \wtheight]{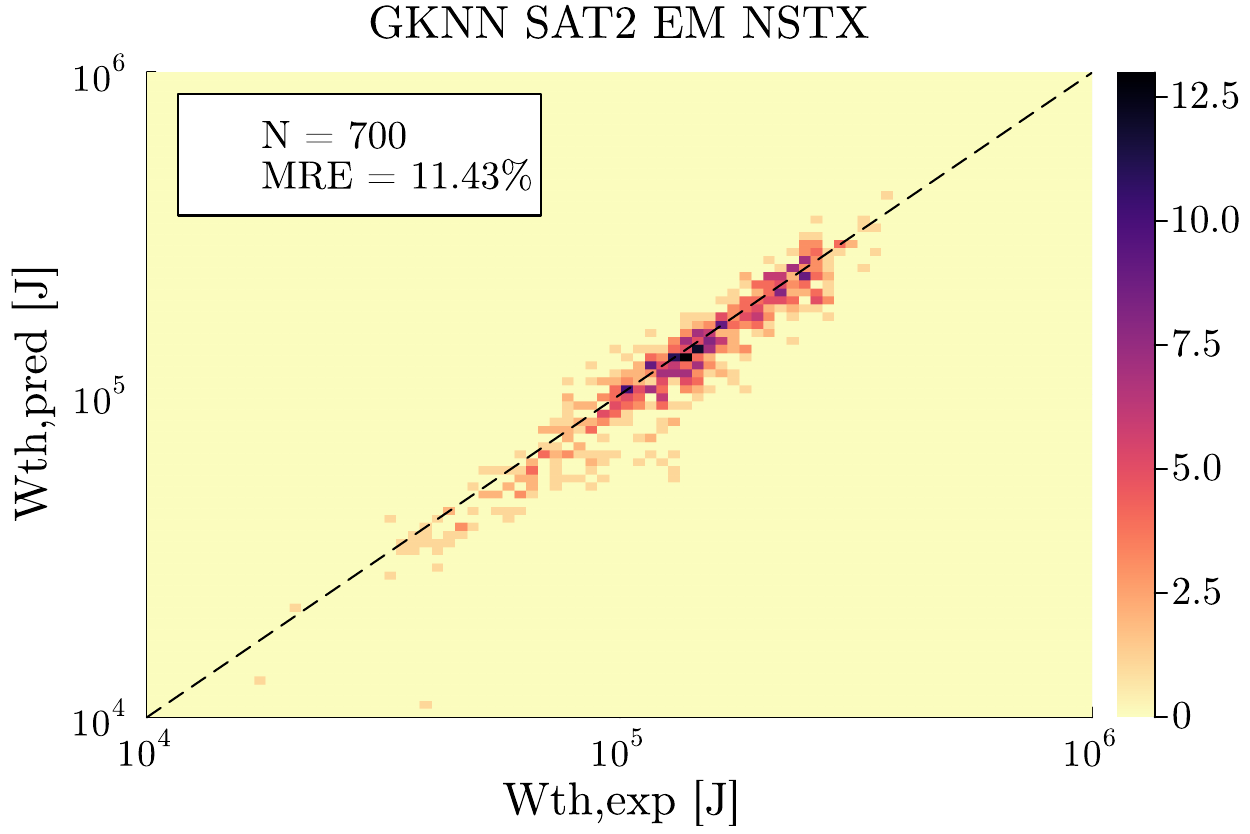}} \figsep
\subfloat[\label{fig:nnprofne}]{\includegraphics[height = \nnheight]{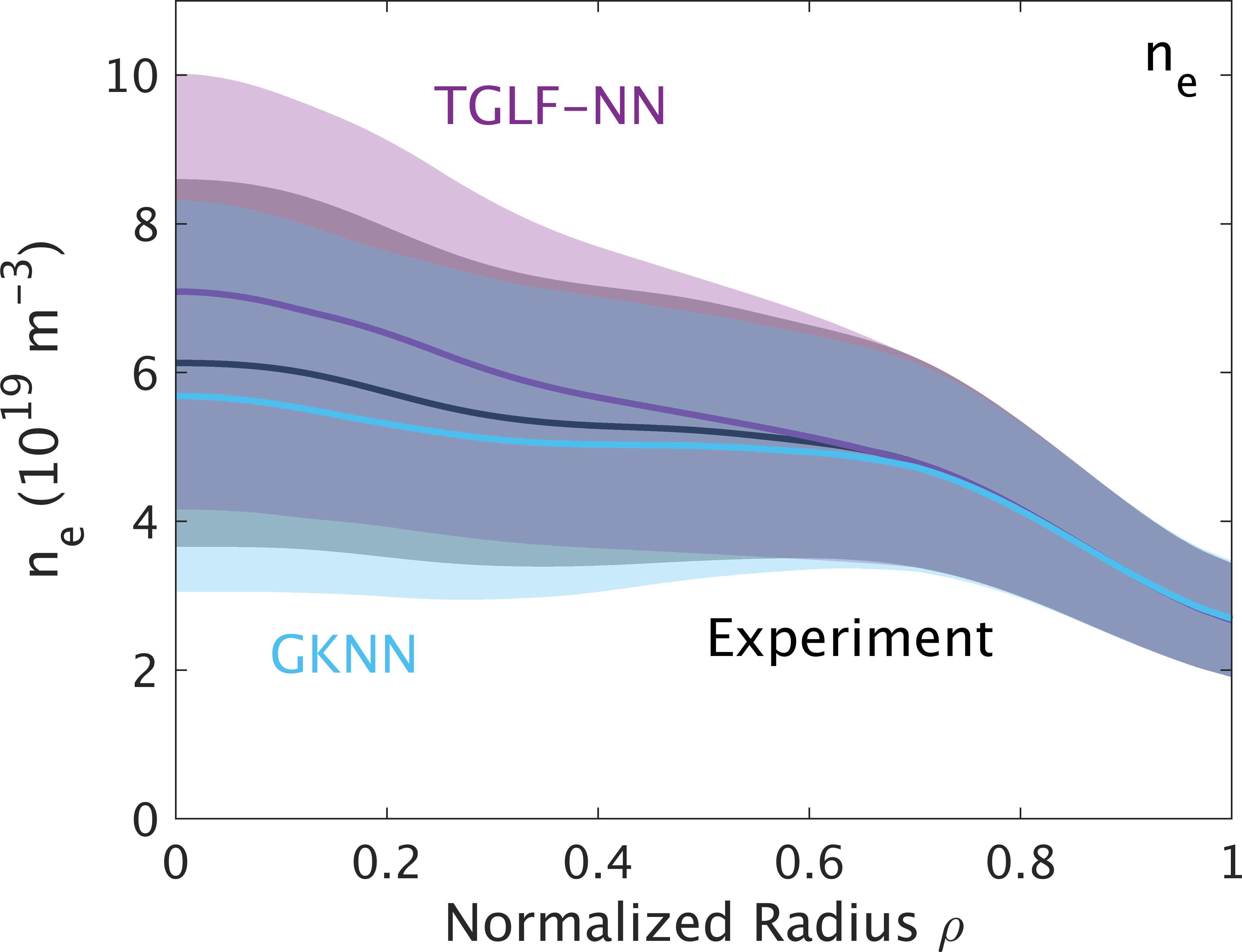}} \figsep
\subfloat[\label{fig:nnprofom}]{\includegraphics[height = \nnheight]{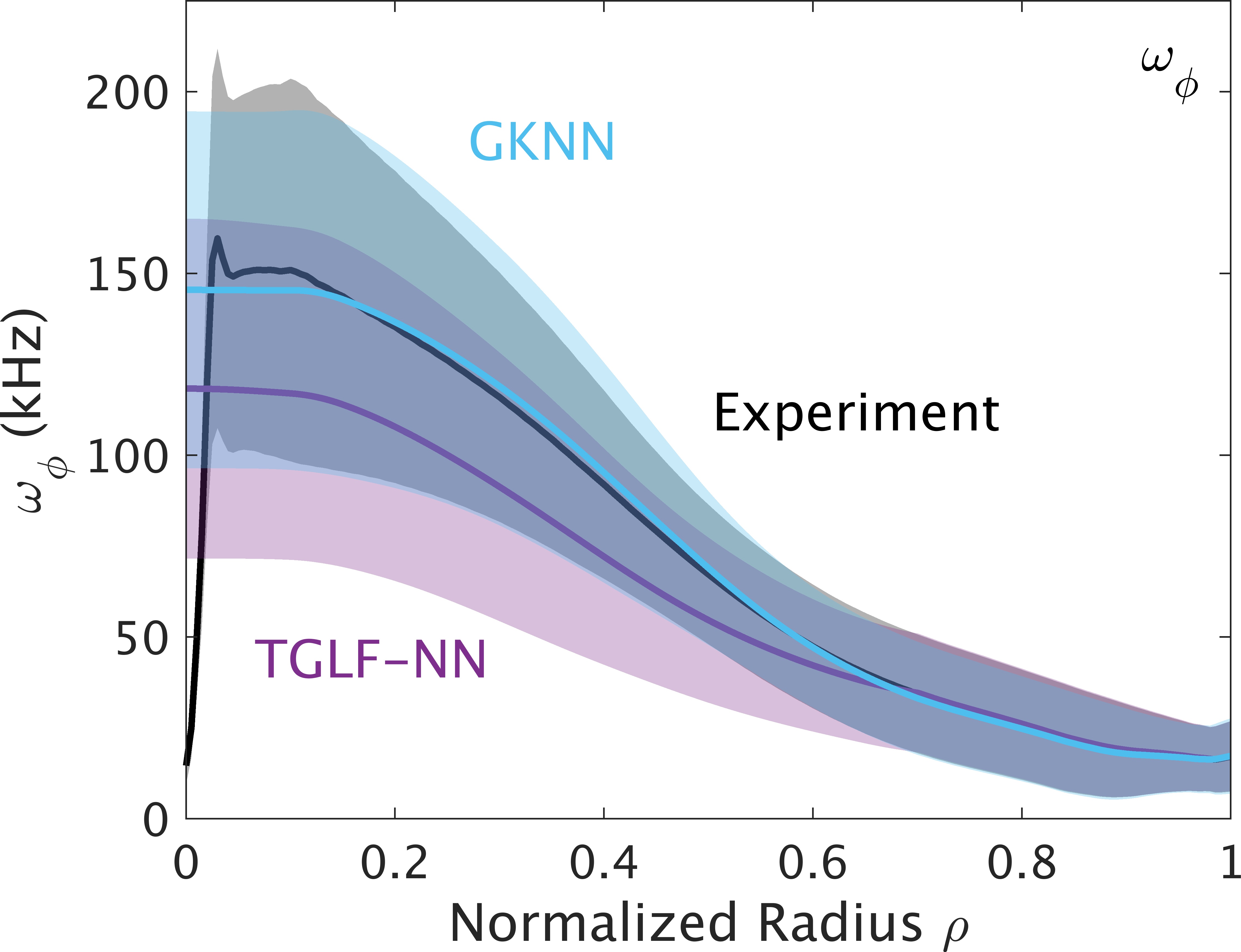}}
\caption{Calculations of kinetic plasma profiles and thermal stored energy with neural net surrogate models based on electromagnetic TGLF (TGLF-NN) and QLGYRO (GKNN). (a) RMSE of different surrogate models for $\Te$, $\Ti$, and the thermal stored energy for $\rho = 0 - 0.7$. Solid bars indicate TGLF-NN models with different saturation rules. Hatched bars indicate GKNN models. (b,c,e,f) TGLF-NN SAT2 (purple) and GKNN SAT2 (light blue) predictions of kinetic profiles, in comparison to experimental data (black). Thick curves and shaded regions indicate the average profile and standard deviation across all time slices. (d) GKNN prediction accuracy of the thermal stored energy. Colorscale indicates number of samples in each logarithmically sized bin.}
\label{fig:tglf_nn}
\end{figure*}

\subsection{Neural Net Optimization of TGLF Settings}
\label{sec:tglf_nn}

Given the computational cost of predictive TRANSP simulations that use TGLF, as discussed in \secref{sec:cputime}, it is impractical to systematically vary each of the several potentially relevant TGLF settings in order to determine which are best suited to capturing the transport physics in a representative database of NSTX plasmas. An alternative approach to optimizing the TGLF settings is to use machine learning to train a series of neural net surrogate models of TGLF \cite{Meneghini2017NF,Neiser2022APS}, each with different settings, and use an iterative, time slice flux matching transport solver to make profile predictions. Once the prediction errors relative to the experimental profile fits are computed, one can efficiently evaluate the performance of TGLF with different settings over a very large database. In addition to identifying the optimal model settings, performant surrogate models for TGLF that are orders of magnitude faster than running TGLF directly have made possible more sophisticated between-shot transport analysis \cite{Smith2024IEEE} and rapid optimization for reactor design studies \cite{Meneghini2024arxiv,Slendebroek2025arxiv}. This exercise was recently performed for both DIII-D and MAST-U, providing new insights into the accuracy of the underlying physics in the various TGLF models \cite{Neiser2024APS}.

A new set of TGLF surrogate models has now been trained on the same set of high performing NSTX discharges used for the predictive TRANSP simulations in this study. 
\rev{TGLF was run with different physics settings on approximately 1000 profiles taken from time slices in these discharges, both within and outside of the identified quiescent analysis windows, performing calculations on a radial grid spanning $\rho = 0.1 - 0.9$ in steps of $\Delta\rho = 0.1$. The experimental profiles were also systematically perturbed to ensure a robust data set for the training of TGLF surrogate models. In total, this training set encompassed five million TGLF calculations.} 
Along with the TGLF surrogate models, the flux matching solver within the FUSE framework \cite{Meneghini2024arxiv,Neiser2024APS} uses the NEO drift-kinetic code to calculate neoclassical transport \cite{Belli2008PPCF,Belli2011PPCF}. The training of the surrogate models and the evaluation of their accuracy were both performed entirely outside of TRANSP. Beyond the neural nets that were trained with TGLF as the turbulent transport model (TGLF-NN), transfer learning methods were used to develop more sophisticated, multi-fidelity surrogate models (GKNN), by training the residual model on the ratio between TGLF-NN heat fluxes and those predicted by QLGYRO \cite{Patel2021thesis,Neiser2023APS,McClenaghan2025PPCF}, a flux matching code that combines linear eigenmode spectra calculated by the electromagnetic gyrokinetic code CGYRO with quasilinear fluxes calculated by TGLF. 
\rev{Due to the greater computational expense of CGYRO relative to TGLF, GKNN was trained on a smaller data set of approximately 5000 QLGYRO simulations (each with a separate CGYRO simulation for a range of 21 $k_y$ values) representing 1000 NSTX plasmas that was combined with the existing training set of 5000 MAST-U simulations (2000 plasmas), and 20,000 DIII-D simulations (7500 plasmas) \cite{Neiser2024APS}.}
Combining TGLF and QLGYRO in this way is expected to provide more accurate linear growth rates of ETGs and TEMs than TGLF while also incorporating MTM physics that is present in CGYRO but absent in TGLF. 
\rev{Once the training of the surrogate models is complete, a massive speedup is obtained in the time slice flux matching transport solver, since each iteration of the solver has a trivial computational cost when employing the surrogate models instead of the underlying transport model. Additionally, the smooth fluxes that are predicted by the neural nets require fewer iterations for the flux matching solver to converge on a solution than when using the full transport model. Specifically, the flux matcher finds a solution 600 times faster when using TGLF-NN instead of TGLF and one million times faster when using GKNN instead of QLGYRO.}

Unlike the predictive TRANSP simulations where only $\Te$ and $\Ti$ were evolved in the simulations, the time slice flux matching solver iterates $\Te$, $\Ti$, electron density, and toroidal rotation profiles simultaneously, using a radial grid spanning $\rho = 0.1 - 0.7$ in steps of $\Delta\rho = 0.05$. 
\rev{To evaluate the performance of the TGLF-NN and GKNN models that were trained, the transport solver is run on time slices taken at every 20 ms in the identified analysis windows, corresponding to over 700 unique cases where the solver converged.} 
The root mean square errors for $\Te$, $\Ti$, and stored energy with respect to experimental measurements are shown in \figref{fig:nnbars}. All of these models are fully electromagnetic, with SAT1geo denoting a model trained on TGLF that used SAT1 with the recalibrated spectral shift model from SAT2. The largest difference in agreement is between TGLF-NN SAT0 and SAT1 in comparison to the rest of the models. This discrepancy results from an overestimation of $\exb$ shear which was improved in later saturation models that used the recalibrated spectral shift model. As discussed in \secref{sec:tglf_other}, using the recalibrated spectral shift model with SAT1 also had a pronounced effect on the temperature profile predictions in time-dependent TRANSP simulations. Among the other trained TGLF-NN models, the agreement with experiment is not substantially affected by the use of different saturation rules. However, the GKNN models, represented by hatched instead of solid bars in \figref{fig:nnbars}, do find consistently better agreement with observations than the corresponding TGLF-NN models that use the same saturation rules (\eg, compare solid purple to hatched light blue for SAT2 and solid green to hatched magenta for SAT3). The best performing model overall was GKNN using the SAT2 saturation rule from TGLF. When comparing GKNN and TGLF-NN with SAT2, GKNN has RMSE for $\Te$ and $\Ti$ of $13\%$ and $9\%$, respectively, compared to $20\%$ and $16\%$ for TGLF-NN, reflecting the higher physics fidelity from incorporating results from linear gyrokinetics into the training of GKNN. 

\begin{figure}[tb]
\includegraphics[height = \nnheight]{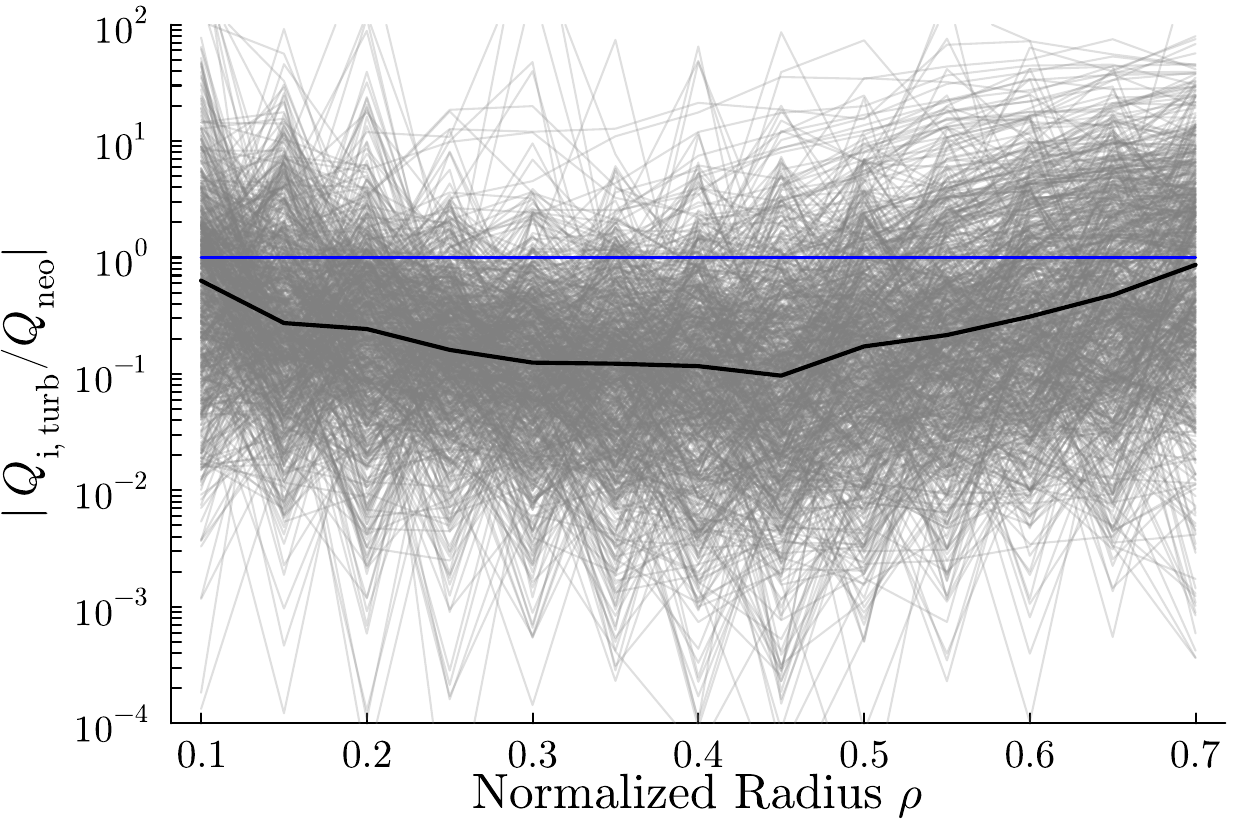}
\caption{Ratio of turbulent ion heat flux, calculated by GKNN SAT2, to neoclassical ion heat flux, calculated by NEO. Each gray curve corresponds to a solution for a single time slice. The thick black curve shows the median ratio. The horizontal blue line indicates equal turbulent and neoclassical heat fluxes.}
\label{fig:gknn_neo}
\end{figure}

Results from GKNN SAT2 and TGLF-NN SAT2 are shown in the remaining panels of \figref{fig:tglf_nn} that have yet to be discussed. A strong linear correlation is found between the predicted and experimental stored energy across the examined time slices, as shown in \figref{fig:nnwth}. The differences in the predicted profiles between TGLF-NN and GKNN are relatively small. As demonstrated in \figref{fig:nnprofti}, both the TGLF-NN and GKNN models shown have a tendency to underpredict $\Ti$, similar to electromagnetic TGLF SAT0 in the TRANSP simulations, though to a much lesser degree. Whereas TGLF-NN usually overpredicts $\Te$, also similar to the TRANSP simulations, GKNN is more likely to underpredict $\Te$ while more closely capturing the experimental profile shape. Although the electron density and rotation profiles were not predicted in the TRANSP simulations, these are shown in \figref{fig:nnprofne} and \figref{fig:nnprofom} for completeness, since these profiles were also predicted simultaneously with $\Te$ and $\Ti$ by the surrogate models within the FUSE flux matcher. 

\rev{Moreover, the calculations performed with the FUSE time slice flux matching transport solver can be used to compare the turbulent ion heat flux that is predicted by the surrogate model GKNN SAT2 to the neoclassical ion heat flux predicted by NEO. The ratio of these fluxes is shown in \figref{fig:gknn_neo}, where each gray curve corresponds to one of the 700 unique time slices described previously in this section and the thick black curve is their median. These results demonstrate that the ion heat transport is dominantly neoclassical in the flux matching predictions performed for the examined NSTX discharges, in line with previous NSTX analysis and consistent with the predictions made by MMM in TRANSP simulations \cite{Lestz2025pre2}. Unfortunately, the TRANSP code does not presently output the variables needed to compare the profiles of the predicted turbulent and neoclassical heat fluxes when using TGLF. Hence, the results shown in \figref{fig:gknn_neo} do not directly reflect the predictions made by TGLF within the time-dependent TRANSP simulations. However, they can still be considered representative of the ion transport regime that is predicted by TGLF.}

The success of the TGLF surrogate models in reproducing the NSTX temperature profiles suggests that the greater disagreement with experiment found with TGLF within predictive TRANSP simulations may be in part due to differences between the solvers and time-dependent \vs individual time slice calculations. \rev{To reiterate, the results in this section used a time slice flux matching solver outside of TRANSP, whereas the TRANSP simulations presented in the rest of this work uses \ptsolver to iteratively solve \eqref{eq:pt_te} to evolve the profiles in time, which does not feed back on a flux matching target.} 
A more detailed discussion of the results of the newly trained surrogate models for NSTX will be the focus of a future publication. While this section only discussed neural nets using TGLF, it is worth noting that machine-learning-based surrogate models for MMM have also been developed for DIII-D \cite{Morosohk2021NF,Morosohk2024NF} and the anticipated NSTX-U operating space \cite{Leard2024IEEE}, \rev{as well as data-driven neural net models \cite{Chung2025NF}. The implementation of these models in TRANSP could further speed up time-dependent simulations for scenario development and design studies \cite{Leard2025FED}.}

\section{Summary of TGLF \vs MMM Agreement With Experiment}
\label{sec:tglfmmm_summary}

In this work, time-dependent predictive TRANSP simulations were performed on a large set of well-analyzed NSTX discharges in order to compare the MMM and TGLF reduced turbulent transport models for the simultaneous prediction of electron and ion temperature profiles over hundreds of milliseconds. The main trends will be summarized here. One of the main conclusions of this study is that when coupled to TRANSP via \ptsolver, MMM provides more robust predictions than TGLF, especially when considering its much reduced computational cost. The reasonable agreement with experiment found with MMM along with its low computational cost makes it suitable for investigating full pulse, non-inductive scenario development on NSTX-U. 

For the set of NSTX discharges examined in this work, MMM predicts the $\Te$ profiles moderately better than electromagnetic TGLF, with typical errors of $15 - 40\%$ and $15 - 75\%$ relative to the experiment, respectively, based on the median and half interquartile range. MMM overpredicts $\Te$ in almost every case, while electromagnetic TGLF typically also overpredicts $\Te$ in general, but underpredicts $\Te$ in discharges with relatively high $\beta$. Electrostatic TGLF is found to overpredict $\Te$ more substantially, by around a factor of two on average. The electromagnetic TGLF predictions exhibit a very strong dependence on $\beta$, predicting much higher $\Te$ at lower $\beta$, in contrast to experimental observations. The level of experimental agreement of MMM predictions also has some dependence on $\beta$, but the effect is less pronounced than for electromagnetic TGLF. Beyond RMSE, the MMM calculations of $\Te$ are better correlated with the experimental values, featuring similar trends, than those from electromagnetic TGLF. Moreover, the electromagnetic TGLF simulations exhibit a larger variance in how well they predict $\Te$ than simulations using MMM. 
\rev{There is only a tenuous connection between discharges with stronger MTM activity predicted by MMM and worse TGLF agreement with experiment, much less robust than the influence of $\beta$.}

In contrast to the $\Te$ profile, the predictions for the $\Ti$ profile, stored energy, and energy confinement time have comparable agreement with experiment for MMM and electromagnetic TGLF. The typical range of MMM RMSE relative to experiment of $20 - 35\%$ overlaps with the typical range for electromagnetic TGLF, $20 - 30\%$. Both models predict the total stored energy with characteristic accuracy of around $5 - 40\%$. Electrostatic TGLF simulations had larger disagreement with experimental $\Ti$ than simulations with electromagnetic TGLF or MMM, though less dramatically so than for $\Te$. While MMM overpredicts $\Ti$ relative to its experimental value, electromagnetic TGLF underpredicts $\Ti$, with the underprediction becoming more substantial at higher $\beta$. Consequently, the $\teti$ ratio inferred from electromagnetic TGLF's calculated profiles is approximately double the typical experimental value ($\Te \approx 0.9 \Ti$), whereas MMM closely reproduces the observed ratio. While MMM predicts both $\Te$ and $\Ti$ profiles to be more peaked than in the experiment, electromagnetic TGLF predicts $\Te$ to be even more overly peaked and calculates $\Ti$ profiles that are overly flat. 
 
When further varying the TGLF settings for a small subset of the examined NSTX discharges, switching the TGLF saturation rule for converting the linear spectrum into quasilinear fluxes generally had a less significant effect than the inclusion of electromagnetic fluctuations. Employing a recalibrated spectral shift model consistently had a pronounced effect of driving stronger transport and hence decreasing $\Te$ and $\Ti$ offsets. In general, no specific combination of TGLF settings was found to perform better than MMM on the small subset of discharges where several TGLF model settings were systematically varied. For this reason, the companion paper \citeref{Lestz2025pre2} to this work focuses on investigating the sensitivities of the MMM predictions in NSTX in greater depth.  

\section{Discussion of Experimental Agreement in the Context of Other Tokamaks and Transport Solvers}
\label{sec:discussion}

Since this work focused on understanding time-dependent simulations of NSTX that used TGLF or MMM as the turbulent transport model, it is worthwhile to consider how these results compare to other recent studies that examined these models using different transport solvers or on different tokamaks. Specifically, the discussion will emphasize work that 1) used time slice flux matching solvers instead of time-dependent simulations and/or 2) investigated these reduced transport models on conventional tokamaks instead of spherical tokamaks.

First, other recent work using TGLF has found better experimental agreement in spherical tokamaks using time slice flux matching transport solvers than the time-dependent TRANSP simulations with TGLF presented here. Specifically, the same electrostatic TGLF SAT1 settings were recently used to analyze two representative low $\betae$ NSTX discharges in \citeref{Avdeeva2023NF}. While there was some spread in the predicted temperature profiles due to variation in input parameters, equilibrium reconstruction, \etc, the overall TGLF predictions using the TGYRO flux matcher did not typically exhibit the very large overpredictions that were found in this work. Moreover, a recent large database time slice study found better experimental agreement for MAST-U H modes than presented here for NSTX using similar electromagnetic TGLF SAT0 settings in a flux matching solver within the FUSE framework \cite{Meneghini2024arxiv,Neiser2024APS}. 
\rev{When training surrogate models for TGLF on individual time slices from the same NSTX discharges used for time-dependent TRANSP simulations in this work, the experimental agreement of temperature profile predictions from the FUSE flux matching analysis was found to be relatively better for the TGLF saturation rules that included the recalibrated spectral shift model, as discussed in \secref{sec:tglf_nn}. Conversely, the TGLF surrogate models without the recalibrated spectral shift model (SAT0 and SAT1) predicted profiles within FUSE with greater disagreement with the measured profiles than when those same TGLF models were used within predictive TRANSP simulations.} 
Hence, one should not conclude from this work that TGLF is inadequate for analyzing transport in spherical tokamaks, but rather that further investigation is needed to determine the cause of the larger experimental disagreement of the TGLF predictions within predictive time-dependent TRANSP simulations and the underlying \ptsolver iteration scheme in comparison to time slice flux matching transport solvers. Once these differences were understood, neural net based surrogate models for TGLF, such as those introduced in \secref{sec:tglf_nn}, could be implemented in TRANSP to make full discharge integrated modeling with TGLF much less computationally expensive. To our knowledge, there has not been any published work examining profile predictions made by MMM within a time slice flux matching solver.

Beyond spherical tokamaks, in \citeref{Abbate2024POP}, a large database of discharges from the conventional tokamak DIII-D (218 discharges, semi-randomly selected) was used to validate time-dependent TRANSP predictions of $\Te$ and $\Ti$ using TGLF SAT2 with partial electromagnetic effects (contributions from $\dbperp$ but not $\dbpar$) and the prediction boundary set to $\rho = 0.8$. In that work, it was found that TRANSP simulations with TGLF predicted H mode temperature profiles with a characteristic RMSE of $15 - 30\%$ and L modes with $20 - 40\%$, predicting $\Ti$ around $5 - 10\%$ better than $\Te$ on average. Hence, TGLF was somewhat more successful in predicting $\Te$ in DIII-D than in the set of NSTX discharges in this work, which is not surprising given that the saturation rules for TGLF were developed based on gyrokinetic simulations of conventional aspect ratio tokamaks. This is also consistent with a recent finding that TGLF surrogate models reproduce experimental profiles slightly better for DIII-D than the MAST-U spherical tokamak \cite{Neiser2024APS}. In contrast, $\Ti$ was predicted by TGLF with similar accuracy for DIII-D and NSTX, which may be due to the competing effects of TGLF being better suited to conventional tokamaks and ion transport being dominantly neoclassical in NSTX, reducing the influence of errors in the turbulent transport model. In the predictive TRANSP simulations of DIII-D, TGLF had a tendency to overpredict $\Te$ and underpredict $\Ti$, similar to what was found here for NSTX with fully electromagnetic TGLF (see \figref{fig:teti_ax}). However, \citeref{Abbate2024POP} also made time-dependent predictions with TGLF with the ASTRA code \cite{Pereverzev2002report,Fable2013PPCF} instead of TRANSP, finding that the solver within ASTRA tends to underpredict both $\Te$ and $\Ti$, demonstrating variability between different time-dependent solvers. Lastly, the DIII-D simulations found that the stored energy was predicted more accurately by TGLF than either of the individual temperature profiles, just as was found for the NSTX simulations with both TGLF and MMM. 
\rev{In the conventional tokamak KSTAR, a validation of electrostatic TGLF SAT0 has been performed within predictive ASTRA simulations of 30 discharges \cite{Lee2021NF}. It was found that the ratio of the volume-averaged predicted $\Te$ to the observed profile in KSTAR was $0.92 \pm 0.09$ (mean $\pm$ standard deviation), with an analogous ratio of $1.02 \pm 0.14$ for $\Ti$. In that work, the prediction boundary was set at $\rho = 0.8$ and these ratios were volume averaged over the entire plasma (including $\rho > 0.8$ where the predictions do not extend to) \cite{Lee202507}. For comparison, these same quantities have been computed for the NSTX TRANSP simulations using electromagnetic TGLF SAT0, yielding ratios of $1.19 \pm 0.31$ for $\Te$ and $0.91 \pm 0.16$ for $\Ti$. Although a direct comparison of these quantities is imprecise due to the different prediction boundaries used in the two works, it appears that TGLF more reliably reproduces the KSTAR temperature profiles than those in NSTX, similar to the DIIII-D comparison discussed above, further underscoring its greater reliability at large aspect ratio.}

For MMM, temperature profile predictions have been previously performed with TRANSP and compared against experimental data from large aspect ratio tokamaks EAST, KSTAR, JET, and DIII-D, with the most extensive comparisons made for KSTAR and JET plasmas \cite{Rafiq2023plasma}. As context, the prediction boundary was set to $\rho = 0.8$ for KSTAR and $\rho = 0.9$ for JET, both larger than the $\rho = 0.7$ used here for NSTX. For KSTAR, the average $\Te$ RMSE was $5 - 12\%$ depending on the plasma scenario, with average $\Ti$ RMSE of $6 - 20\%$. For JET, the MMM simulations had average RMSE of $10 - 13\%$ for $\Te$ and $6 - 15\%$ for $\Ti$. Hence, TRANSP simulations with MMM presented here do not agree as well with observed NSTX temperature profiles as they do for either of these conventional tokamaks, though the difference is only moderate. In the JET discharges, there was a clear tendency for MMM to underpredict both $\Te$ and $\Ti$, in contrast to the NSTX simulations where MMM tended to overpredict both (see \figref{fig:teti_ax}). In the KSTAR discharges, the reported temperature profile offsets were more balanced. Predictive TRANSP simulations were also performed in \citeref{Kim2017NF} using TGLF for the same JET discharges that were studied with MMM in \citeref{Rafiq2023plasma}. Validation of TGLF against the JET discharges found a systematic underprediction of $\Te$, unlike the electromagnetic TGLF simulations of NSTX in this paper which only underpredicted $\Te$ at high $\beta$. However, the specific TGLF settings were not listed in that work, so it may not be a one-to-one comparison with the simulations performed here. It was reported that the average temperature profile prediction RMSE was comparable for TGLF and MMM for those JET discharges \cite{Rafiq2023plasma}. Moreover, \citeref{Rafiq2023plasma} quotes that in total, the TRANSP simulations with TGLF used 895 times the number of CPU hours that MMM used for the same set of JET discharges. A similar ratio was found when comparing the computational cost of the predictive TRANSP simulations of NSTX discharges in this work, where the simulations using electromagnetic TGLF SAT0 and electrostatic SAT1 required 1005 and 1738 times the CPU hours that the ones using MMM did, respectively. Based on these works, both TGLF and MMM appear to more accurately reproduce experimentally observed temperature profiles in conventional tokamaks than NSTX, underscoring the challenges of reduced transport modeling in high performing spherical tokamaks.  

\section{Suggested Future Work} 
\label{sec:future} 

There are several natural avenues for future work that would build upon this study. First, this work was motivated by a need to understand the reliability of existing reduced transport models in predicting kinetic profiles for plasmas in the unique high $\beta$ spherical tokamak regime. With the improved understanding of the models' capabilities and uncertainties detailed in this paper, these models can now be soberly applied to forecasting and scenario development for upcoming NSTX-U campaigns. In support of this activity, the investigation of the different models could be further refined. Namely, the study of the effect of different TGLF saturation rules, and especially the influence of the spectral shift model, was not comprehensive, and will be explored further by leveraging the surrogate models discussed in \secref{sec:tglf_nn}. The differences found in time-dependent TRANSP simulations using MMM and TGLF motivate a direct comparison of the two models in a time slice flux matching solver such as TGYRO, which would lay the groundwork to compare profile predictions across a hierarchy of physics models (TGYRO flux matching \cite{Candy2009POP} \vs QLGYRO flux matching \cite{Patel2021thesis,McClenaghan2025PPCF} \vs fully nonlinear CGYRO simulations \cite{Candy2016JCP}), all using the current best tools for consistent kinetic equilibrium reconstruction on NSTX(-U) \cite{Avdeeva2024PPCF}. In such a study, it would also be natural to compare the spectra of unstable modes and fluxes calculated by the different transport models. As a first step, only the temperature profiles were allowed to evolve in the TRANSP simulations performed for this work, leaving the density and rotation profiles fixed to their experimental fits. It would be of interest to extend this study to predict those profiles as well, to see how the prediction accuracy is affected by removing additional experimental constraints. 

Additionally, predictive TRANSP simulations require a prescribed boundary condition, set at $\rho = 0.7$ in this work, such that the core profiles directly depend on the value given for the boundary, which is usually supplied by experimental measurements or some assumptions about expected pedestal height. For fully integrated predictive modeling, it would be useful to instead either interface with a code that can make predictions of the pedestal or use a heuristic scaling to provide the edge temperature boundary condition \cite{Parisi2024NFa,Parisi2024NFb}. 
For instance, core-edge integrated predictive simulations using TRANSP with GLF23 \cite{Kinsey2005FST}, a predecessor of TGLF, were recently applied to the high field ST40 spherical tokamak, finding a strong sensitivity of the core plasma predictions to the boundary condition provided by the edge model of the scrape-off-layer \cite{Zhang2025POP}. 
Another component of integrated modeling that could be incorporated to make the simulations more comprehensive is anomalous fast-ion transport or macroscopic MHD instabilities. While the analysis windows were chosen in this work to be relatively quiescent, NSTX plasmas with strong fast ion transport could also be modeled by making use of additional reduced models that have been implemented in TRANSP to treat fast-ion transport. These include the kick model \cite{Podesta2014PPCF} and the resonance broadened quasilinear model (RBQ) \cite{Gorelenkov2019POP}, which are both capable of calculating fast-ion transport induced by \Alfven eigenmodes. The kick model has also been extended to include sawteeth \cite{Kim2018NF} and neoclassical tearing modes \cite{Bardoczi2019PPCF}. However, both the kick model and RBQ currently require using separate codes outside of TRANSP to calculate the instability spectrum and their phase-space-dependent interaction with fast ions, such that further development would be needed before this physics could be captured in a fully automated way within TRANSP. 

Moreover, whereas NSTX discharges studied in this work were exclusively beam-heated, most reactor designs rely heavily on radio frequency (RF) heating and current drive techniques, many of which are in use or being tested on existing tokamaks. Hence, it would be of interest to perform a similar test of predictive TRANSP simulations to determine if their accuracy is sensitive to the mixture of auxiliary ion \vs electron heating methods, since RF heating sources have much more narrow power deposition profiles than neutral beam heating, which can result in large localized diffusion coefficients. Given that qualitative differences were already observed in NSTX when using different mixtures of high harmonic fast wave heating and neutral beam injection (NBI), including the highest observed $\Teo$ occurring with HHFW \cite{Taylor2010POP}, differences in internal transport barriers \cite{Yuh2009POP}, and suppression of \Alfven eigenmodes \cite{Fredrickson2014NF}, such a study could be performed across both NSTX and NSTX-U. This investigation would be timely given the recent investigation into the consequences of simultaneous NBI and HHFW for scenario development on NSTX-U \cite{VanCompernolle2025PPCF}. 

Lastly, the transport models themselves could be improved to capture additional physics. For instance, the new GFS eigensolver \cite{Staebler2023POP} is expected to be more accurate than TGLF's current linear eigenmode solver for high $\beta$ spherical tokamaks \cite{Staebler2024SW,Kinsey2025POP}. More ambitiously, there is a longstanding observation of unexplained electron energy transport in NSTX, where increasing the NBI power often left the on-axis $\Te$ unchanged, instead broadening the temperature profile \cite{Stutman2009PRL}, creating anomalous $\Te$ flattening which could not be explained by gyrokinetic simulations of the near-axis region \cite{Ren2017NF}. Instead, mechanisms have been proposed that could generate anomalous electron energy transport via high frequency \Alfven eigenmodes ($f \lesssim \fci$) that become more unstable at higher $\W_\text{NBI}/\va^2 \propto n_i/B^2 \propto \beta$ (where $2\pi\fci = q_i B/m_i$ is the ion cyclotron frequency, $\W_\text{NBI}$ is the beam voltage and $\va = B/\sqrt{\mu_0 m_i n_i}$ is the \Alfven speed) \cite{Stutman2009PRL,Gorelenkov2010NF,Kolesnichenko2010PRL,Kolesnichenko2010NF,Belova2015PRL,Belova2017POP,Lestz2021NF} or ideal MHD modes known as infernal modes which can flatten profiles at high $\beta_N$ \cite{Jardin2022PRL,Jardin2023POP}. While both mechanisms have been demonstrated numerically in simulations, neither one has been shown to quantitatively explain the experimental observations. It remains to be seen if this unexplained electron energy transport will be exacerbated with the even greater NBI power available on NSTX-U, but if so, it will be important to understand how this transport channel can be accounted for in reduced models. 

\section{Acknowledgements}
\label{sec:acknowledgements}

The authors thank J. Abbate, X. Zhang, W. Choi, T. Rafiq, and X. Yuan for fruitful discussions. The TRANSP simulations reported here were performed with computing resources at the Princeton Plasma Physics Lab. The TGLF-NN and GKNN surrogate models described in \secref{sec:tglf_nn} were trained using computational resources of the National Energy Research Scientific Computing Center (NERSC), a Department of Energy User Facility using NERSC awards FES-ERCAP 30971 and FES-ERCAP 33303. Part of the data analysis was performed using the OMFIT integrated modeling framework \cite{Meneghini2015NF}. The data required to generate the figures in this paper are archived in the NSTX-U Data Repository ARK at the following address: (placeholder).
This research was supported by the U.S. Department of Energy (contracts DE-SC0021113, DE-AC02-09CH11466, DE-SC0018990, and DE-SC0024426).

\section{Disclaimer}
\label{sec:disclaimer}

\gadisclaimer

\appendix 

\section{TRANSP RunIDs}
\label{app:runids}

The NSTX discharges and corresponding TRANSP runIDs used for this study are listed in \tabref{tab:runids}, along with the category that each discharge was sorted into and previously published work that analyzed each discharge. The ``original'' interpretive ID is the TRANSP run where the equilibrium, profiles, input data, and most TRANSP settings were taken from, whereas the ``modernized'' interpretive ID is a recent rerun using the current version of TRANSP, with some settings tweaked for compatibility or consistency with other runs. The TRANSP IDs for the smaller subset of simulations that further varied the TGLF settings are listed in \tabref{tab:tglf_runs}. All runs are stored in MDSplus, access information available upon reasonable request.

\newcommand{\longpulse}{Long pulse\xspace}
\newcommand{\maxwmhd}{Max stored energy\xspace}
\newcommand{\highnustar}{High $\nu_e^*$\xspace}
\newcommand{\lownustar}{Low $\nu_e^*$\xspace}
\newcommand{\wphmode}{Wide pedestal\xspace}
\newcommand{\ephmode}{Enhanced pedestal\xspace}
\newcommand{\highbetap}{High $\beta_p$\xspace}
\newcommand{\lowbeta}{Low $\beta_e$\xspace}
\newcommand{\highbeta}{High $\beta_e$\xspace}
\newcommand{\hmode}{H mode\xspace}
\newcommand{\lmode}{L mode\xspace}
\newcommand{\stutman}{$\Te$ flattening\xspace}

\begin{table*}\centering
\begin{tabular}{ccccccccc}
\hline\hline
\twotab{NSTX}{Discharge} & \twotab{Original}{Interpretive} & \twotab{Modernized}{Interpretive} & MMM & \twotab{TGLF}{SAT0 EM} & \twotab{TGLF}{SAT1 ES} & \twotab{Category}{and Reference} & \twotab{Analysis}{Time (ms)} \\ 
\hline
116313 & G12 & N10 & N03 & N13 & N07 & \longpulse \cite{Gerhardt2011NFat} & 500 - 1050 \\ 
117707 & A04 & J11 & J03 & J15 & J17 & \maxwmhd \cite{Gerhardt2011NFat} & 650 - 950 \\
120967 & A03 & J02 & J05 & J18 & J17 & \highnustar \cite{Guttenfelder2013NF,Kaye2014POP} & 300 - 600 \\
120968 & A02 & L18 & L05 & L21 & L07 & \highnustar \cite{Guttenfelder2013NF,Rafiq2021POP,Clauser2022POP,Rafiq2024NF,Clauser2025POP} & 260 - 400 \\
120982 & A09 & J11 & J03 & J15 & J04 & \lownustar \cite{Clauser2022POP,Rafiq2024NF,Clauser2025POP} & 590 - 650 \\
121123 & A02 & J08 & J03 & J10 & J04 & \maxwmhd \cite{Gerhardt2011NFat} & 500 - 1000 \\
129016 & A03 & J11 & J02 & J12 & J04 & \highnustar \cite{Guttenfelder2013NF,Rafiq2021POP,Rafiq2024NF,Clauser2025POP} & 300 - 500 \\
129017 & A04 & K36 & K31 & K48 & K37 & \hmode \cite{Guttenfelder2013NF,Avdeeva2023NF,Avdeeva2024PPCF} & 320 - 500 \\
129039 & A05 & J14 & J06 & J15 & J07 & \lownustar \cite{Rafiq2021POP} & 300 - 340 \\
129041 & A10 & K03 & J02 & K05 & J03 & \lownustar \cite{Guttenfelder2013NF,Kaye2014POP,Clauser2022POP,Clauser2025POP} & 250 - 370 \\
129125 & B08 & J08 & J03 & J10 & J04 & \longpulse \cite{Gerhardt2011NFat} & 500 - 1100 \\
132588 & B01 & J08 & J03 & J09 & J04 & \wphmode \cite{Dominski2024POP} & 530 - 790 \\
132911 & A01 & J02 & J03 & J11 & J07 & \maxwmhd \cite{Gerhardt2011NFat,Gerhardt2011NFcur} & 500 - 700 \\
132913 & B01 & J02 & J03 & J10 & J07 & \maxwmhd \cite{Gerhardt2011NFat} & 640 - 720 \\
133958 & G63 & K02 & K03 & K11 & K10 & \highbetap \cite{Gerhardt2011NFcur} & 345 - 365 \\
133959 & D45 & K11 & K03 & K25 & K12 & \highbetap \cite{Gerhardt2011NFcur} & 600 - 900 \\
133964 & D05 & I26 & I20 & I85 & I63 & \highbetap \cite{Gerhardt2011NFat,Gerhardt2011NFcur,McClenaghan2023POP,McClenaghan2025PPCF} & 700 - 1050 \\
134767 & A05 & K09 & K03 & K10 & K06 & \hmode \cite{Gerhardt2011NFat} & 500 - 1200 \\
134837 & A11 & K09 & K03 & K10 & K06 & \maxwmhd \cite{Gerhardt2011NFat} & 500 - 950 \\
135117 & A02 & J11 & J03 & J12 & J08 & \maxwmhd \cite{Gerhardt2011NFcur} & 600 - 900 \\
135129 & A02 & J09 & J03 & J11 & J10 & \maxwmhd \cite{Gerhardt2011NFat} & 600 - 1050 \\
135440 & S05 & J14 & J05 & J15 & J07 & \longpulse \cite{Gerhardt2011NFcur} & 600 - 880 \\
135445 & A04 & J08 & J03 & J12 & J09 & \longpulse \cite{Gerhardt2011NFat,Gerhardt2011NFcur} & 600 - 1350 \\
138536 & J01 & J28 & J07 & J27 & J18 & \lownustar \cite{Rafiq2022POP,Rafiq2024NF} & 550 - 660 \\
139517 & A04 & J08 & J03 & J11 & J10 & \maxwmhd \cite{Gerhardt2011NFcur} & 550 - 850 \\
140035 & A05 & J02 & J03 & J13 & J12 & \maxwmhd \cite{Gerhardt2011NFat} & 500 - 1180 \\
141007 & A03 & J08 & J03 & J10 & J09 & \lowbeta, \lownustar \cite{Ren2012POP} & 300 - 430 \\
141031 & S05 & J02 & J03 & J09 & J04 & \lowbeta, \highnustar \cite{Ren2012POP,Guttenfelder2013NF} & 260 - 300 \\
141032 & A04 & J02 & J03 & J09 & J04 & \lowbeta, \highnustar \cite{Ren2012POP} & 260 - 300 \\
141040 & A02 & J02 & J03 & J09 & J04 & \lowbeta, \lownustar \cite{Ren2012POP,Guttenfelder2013NF} & 330 - 390 \\
141125 & D04 & J10 & J03 & J15 & J04 & \wphmode \cite{Battaglia2020POP} & 700 - 900 \\
141131 & D04 & J02 & J03 & J10 & J04 & \wphmode \cite{Battaglia2020POP} & 600 - 900 \\
141133 & B11 & K09 & K03 & K25 & K12 & \ephmode \cite{Battaglia2020POP,Gerhardt2014NF} & 750 - 1050 \\
141340 & B04 & J02 & J03 & J16 & J04 & \ephmode \cite{Gerhardt2014NF} & 300 - 400 \\
141623 & A11 & J02 & J03 & J10 & J09 & \highbetap \cite{Gerhardt2011NFat} & 450 - 900 \\
141633 & A11 & J02 & J03 & J13 & J05 & \highbetap \cite{Gerhardt2011NFat} & 300 - 600 \\
141767 & B01 & J02 & J03 & J08 & J04 & \hmode \cite{RuizRuiz2019PPCF,Ren2020NF} & 300 - 560 \\
\hline\hline
\end{tabular}
\caption{TRANSP runs for the full set of NSTX discharges analyzed in this work.}
\label{tab:runids}
\end{table*}

\begin{table*}\centering
\begin{tabular}{cccccccccc}
\hline\hline
\twotab{NSTX}{Discharge} & \twotab{ES}{SAT0} & \twotab{EM}{SAT0} & \threetab{EM}{SAT0}{$\dbpar = 0$} & \twotab{ES}{SAT1} & \twotab{EM}{SAT1} & \twotab{EM}{SAT2} & \twotab{EM}{SAT3} & \twotab{ES}{SAT1geo} & \twotab{EM}{SAT1geo} \\
\hline
129017 & K53 & K48 & K63 & K37 & K39 & K57 & K55 & K40 & K43 \\
133959 & K26 & K25 & K35 & K12 & K14 & K27 & K31 & K21 & K23 \\
133964 & I89 & I85 & H11 & I63 & I65 & I92 & I93 & --- & I79 \\
138536 & J46 & J27 & J54 & J18 & J20 & J47 & J48 & J21 & J23 \\
141133 & K34 & K25 & K45 & K12 & K21 & K38 & K41 & K24 & K23 \\
\hline\hline
\end{tabular}
\caption{TRANSP runs for the subset of simulations that varied the TGLF settings. ES denotes electrostatic, EM denotes fully electromagnetic, SATx denotes the saturation rule, and SAT1geo indicates the use of the recalibrated spectral shift model with SAT1. The missing entry is a simulation that was unable to converge within reasonable computation time.}
\label{tab:tglf_runs}
\end{table*}

\bibliography{all_bib} 

\end{document}